\documentclass{aa}  
\usepackage{graphicx}
\usepackage{txfonts}

\usepackage{xcolor}
\usepackage[colorlinks=true,linkcolor=blue,citecolor=blue,urlcolor=blue]{hyperref}
\usepackage{wasysym}

\usepackage[switch]{lineno} 
\usepackage{appendix}
\usepackage{color}
\usepackage{natbib}
\usepackage{pdflscape}
\usepackage{longtable}
\usepackage{pdflscape}
\usepackage{threeparttable} 
\usepackage{lineno}% Optional for footnotes
\usepackage{float} % in preamble
\usepackage{caption}
\usepackage{mdframed}

\newcommand{\code}[1]{\texttt{#1}\xspace}

\newcommand{\logg}{\ensuremath{\log\,g}\xspace}

\newcommand{\Angstrom}{\,{\AA}}

\newcommand{\objonename}{J0040$+$2729}

\newcommand{\objtwoname}{J0217$-$1903}

\usepackage{longtable}
\begin{document} 
%\linenumbers

   \title{The \emph{R}-Process Alliance: Exploring the cosmic scatter among ten \emph{r}-process sites with stellar abundances.\thanks{This paper includes data gathered with the 6.5~meter Magellan Telescopes located at Las Campanas Observatory, Chile and data taken at The McDonald Observatory of The University of Texas at Austin}}
%  \title{R-Process Alliance: an homogeneous chemical analysis of ten $r$-process enhanced stars}

   %\subtitle{??}

   \author{M. Racca\inst{1}
          \and
          T.T. Hansen\inst{1} 
          \and
          I.U. Roederer\inst{2,3} 
          \and
          V.M. Placco \inst{4}
          \and
          A. Frebel \inst{5,3}
          \and
          T.C. Beers \inst{6,3}
          \and
          R. Ezzeddine \inst{7,3}
          \and
           E.M. Holmbeck \inst{8,3}
          \and
          C.M. Sakari \inst{9}
          \and
          S. Monty \inst{10}
          \and
          Ø. Harket\inst{1}
          \and
          J.D. Simon \inst{11}
          \and
          C. Sneden \inst{12}
          \and
          I.B. Thompson \inst{11}
}

   \institute{
    Astronomy department, Stockholm University, Roslagstullsbacken 21, 114 21, Stockholm
    \and
     Department of Physics, North Carolina State University, 2401 Stinson Dr, Box 8202, Raleigh, NC 27695, USA
    \and
     Joint Institute for Nuclear Astrophysics – Center for the Evolution of the Elements (JINA-CEE), USA
    \and
     NSF NOIRLab, Tucson, AZ 85719, USA
    \and
    Department of Physics and Kavli Institute for Astrophysics and Space Research, Massachusetts Institute of Technology, Cambridge, MA 02139, USA
    \and
    Department of Physics and Astronomy, University of Notre Dame, Notre Dame, IN 46556, USA
    \and
     Department of Astronomy, University of Florida, Bryant Space Science Center, Gainesville, FL 32611, USA
    \and
     Lawrence Livermore National Laboratory, 7000 East Avenue, Livermore, CA 94550, USA
    \and
     Department of Physics and Astronomy, San Francisco State University, San Francisco, CA 94132, USA
    \and
     Institute of Astronomy, University of Cambridge, Madingley Rd, Cambridge, CB3 0HA, UK
    \and
     Observatories of the Carnegie Institution for Science, 813 Santa Barbara St., Pasadena, CA 91101, USA
    \and
     Department of Astronomy and McDonald Observatory, The University of Texas, Austin, TX, 78712, USA\\
    \email{mila.racca@astro.su.se}
             %\thanks{The university of heaven temporarily does not                     accept e-mails}
             }

   \date{Received xxx; accepted yyy}

% \abstract{}{}{}{}{} 
% 5 {} token are mandatory
 \abstract
  % context heading (optional), leave it empty if necessary  
  {The astrophysical origin of the rapid neutron-capture process ($r$-process), responsible for producing roughly half of the elements heavier than iron, remains uncertain. Detailed chemical signatures from the oldest, most metal-poor stars, which act as fossil records of the earliest nucleosynthesis events, can be used to identify the dominant $r$-process sites.}
  % aims heading (mandatory)
  {We present a homogeneous chemical abundance analysis of ten $r$-process element-enhanced stars. These old and metal-poor stars are strongly enriched in $r$-process elements with minimal contamination from other nucleosynthetic sources. By focusing on this chemically pure sample, we aim to investigate intrinsic variations in the $r$-process abundance patterns and explore their implications for the nature and potential diversity of $r$-process sites.}
  % methods heading (mandatory)
  {We performed a detailed chemical abundance analysis of high-resolution, high signal-to-noise spectra. For each star, we inspected over 1400 individual absorption lines using a combination of equivalent width measurements and spectral synthesis. The analysis was conducted under the assumption of one-dimensional local thermodynamic equilibrium, employing the MOOG radiative transfer code.}
  % results heading (mandatory)
  {We derived abundances for 54 chemical species, including 29 neutron-capture ($n$-capture) elements covering the full mass range of the \emph{r}-process abundance pattern. A kinematic analysis reveals that stars likely originated from ten kinematically distinct systems. Based on this assumption, we use the sample to probe the maximum variation expected from ten independent $r$-process nucleosynthesis events and compute the intrinsic dispersion of each element relative to Zr and Eu, for the light and heavy \emph{r}-process elements, respectively. This exercise results in a remarkably low cosmic scatter across the ten \emph{r}-process sites enriching these stars, for the rare earth and third peak elements, for example, we find $\sigma_{[\mathrm{La/Eu}]} = 0.08$ and $\sigma_{[\mathrm{Os/Eu}]} = 0.11$~dex while the scatter between light and heavy, $\sigma_{[\mathrm{Zr/Eu}]}$ is slightly higher at 0.18~dex.}
{The elemental abundance patterns across the ten independent \emph{r}-process sites show remarkably small cosmic dispersions. This minimal dispersion suggests a high degree of uniformity in \emph{r}-process yields across diverse astrophysical environments.}


   \maketitle

\section{Introduction}
The heaviest elements of the periodic table, those beyond the iron peak ($Z \gtrsim 30$), are primarily synthesized through neutron-capture processes during stellar evolution and explosive astrophysical events. Among these, the rapid neutron-capture process (\emph{r}-process) is a dominant mechanism for producing half of the isotopes and, in particular, the actinide elements thorium and uranium. The \emph{r}-process occurs under extreme conditions of high neutron densities ($10^{20}-10^{28}$ n/cm$^3$) \citep{Kratz_2007} and short timescales \citep{Burbidge_1957,Cameron1957AJ.....62....9C}, allowing atomic nuclei to rapidly capture neutrons before undergoing $\beta$-decay. Despite considerable investigations, the astrophysical sites of the \emph{r}-process remain an area of active investigation. Compact binary mergers, including both binary neutron star mergers (NSM) and neutron star-black hole systems, have long been proposed as prime candidates \citep{Lattimer1974ApJ...192L.145L, Eichler_2015}. The detection of the gravitational wave event GW170817 \citep{Abbott_2017, Abbott_2019}, along with its associated kilonova having emission consistent with the decay of heavy element isotopes \citep{Chornock_2017, Villar2017, Drout2017}, provided the first direct observational support for this scenario. Although NSMs are currently the leading candidates for the origin of \textit{r}-process elements, their contribution to Galactic chemical enrichment is still debated \citep{Tsujimoto2015, Cote2019ApJ...875..106C, Sku020A&A...634L...2S, Vanbeveren2024BSRSL..93..338V}. Factors such as merger rates, delay times, and the sensitivity of nucleosynthetic yields to system properties (e.g., neutron star mass) introduce uncertainties that are not yet fully captured in current chemical evolution models \citep{Holmbeck_2024, VdV2020MNRAS.494.4867V, VdV2022MNRAS.512.5258V}. On top of that, multiple recent studies demonstrated that at least two \emph{r-}process sites are required to explain the observed \emph{r-}process elements abundances in metal-poor stars \citep{Cote2019, Molero2023, Kuske2025}. For these reasons, alternative sites for \textit{r}-process nucleosynthesis have been proposed in addition to NSMs, these include magneto-rotationally supernovae \citep{Nishimura_2006, Winteler_2012, Cownan2021, Prasanna2024}, collapsars \citep{Barnes_2022}, common-envelope jets supernova (CEJSN) \citep{Grichener2022ApJ...926L...9G, Jin2024ApJ...971..189J, Soker2025OJAp....8E..67S} and the neutrino-driven wind emerging from the proto-neutron star in the aftermath of the explosion \citep{Arcones_2013, Wang2023}. For the latter, however, simulations have shown that under typical conditions, these winds tend to produce only a weak or limited \textit{r}-process, insufficient for forming the heaviest elements.
Additionally, certain classes of massive, fast-rotating, and highly magnetized stars, which lead to magnetars \citep{Patel_2025}, magnetohydrodynamic-jet supernovae, or black hole accretion disks \citep{Siegel2019Natur.569..241S}, are considered potential \textit{r}-process sites \citep{Halevi2018}. In these scenarios, the \textit{r}-process could occur in the ejecta from the central object, such as a neutron star or black hole. 

Accurately derived elemental abundances from old metal-poor stars can be used as an observational constraint for nucleosynthesis models and helps trace the chemical evolution of stars and galaxies \citep{sneden2008, Cescutti2022}.
However, the detailed determination of elemental abundances in metal-poor stars is challenging and strongly depends on the underlying assumptions of stellar atmosphere and radiative transfer models. The most physically realistic models, those incorporating three-dimensional (3D) hydrodynamics \citep[e.g.,][]{DiazLagae2024} and non-local thermodynamic equilibrium (NLTE) effects, better capture stellar surface inhomogeneities and radiative imbalances (for a detailed review of 3D-NLTE effects in late-type stars, see \citealp{LindAmarsi2024}). Yet, such models are not broadly available across the full range of stellar parameters required for the stars of interest in this kind of study, in which we analyze old, relatively cool ($<5500~K$) metal-poor red giants. In addition, atomic data for the development of model atoms for the heaviest elements remains sparse, further complicating accurate abundance determinations. For this reason, most of the abundances present in literature and in this work are determined under the assumption of 1D-LTE. However, despite these challenges, homogeneous abundance analyzes remain a powerful tool for minimizing systematic uncertainties and identifying meaningful trends in abundances, especially for stars with similar stellar parameters. 

In this work, we present a homogeneous chemical analysis of ten $r$-process enhanced stars with $\mathrm{[Eu/Fe]} > +0.3$ and $\mathrm{[Ba/Eu]} \leq -0.5$. 
These stars were discovered by the \emph{R}-Process Alliance (RPA)\footnote{\url{https://sites.google.com/view/rprocessalliance/home?authuser=0}} \citep{Hansen_2018} through their successful search for \emph{r}-process enriched stars \citep{Sakari2018ApJ...868..110S,Ezzeddine2020ApJ...898..150E, Holmbeck2020ApJS..249...30H, Bandyopadhyay2024}. With this sample, we investigate the cosmic spread among ten \emph{r}-process production sites that contributed to the enrichment of these stars.

This paper is outlined as follows: Information about the observations is presented in Section \ref{Data}. The derivation of the stellar parameters and the details about the abundance analysis are described in Sections \ref{StellarParameters} and \ref{AbundanceAnalysis}, respectively. The results are presented in Section \ref{Results} and discussed in Section \ref{Discussion}. A summary is provided in Section \ref{Summary}.

\section{Data}
\label{Data}
The ten stars analyzed in this paper are presented in Table \ref{tab:basicdata}, which lists the 2MASS identifier, right ascension (R.A.), declination (Decl.), photometric data ($V$, $G$ and $K$ magnitudes, and $BP$ and $RP$ colors), color excess ($E(B-V)$) and parallax ($\varpi$). For brevity, we adopt shortened source names constructed from the first four digits of the right ascension and the first four digits of the declination (including the sign), e.g., J00401252+2729247 becomes \objonename. These abbreviated names are used throughout the text and in the following tables. All ten stars were observed as part of the RPA "snapshot" survey, which uses relatively short, moderate signal-to-noise ratios ($\mathrm{SNR} \sim 30$ at 4100 \AA) high-resolution spectroscopy ($R \sim 30,000$) to efficiently identify \emph{r}-process enhanced stars \citep{Hansen_2018}. During this initial screening,  a set of key elemental abundances (specifically [Fe/H], [C/Fe], [Sr/Fe], [Ba/Fe], [Eu/Fe], [Ba/Eu], and [Sr/Ba]) are determined to categorize the stars. This approach allows the collaboration to quickly select promising targets for more detailed study. Eight of the stars analyzed here were discovered in the first data release from the RPA \citep{Hansen_2018}. Snapshot spectra for the remaining two stars, \objonename\ and \objtwoname, will be included in the forthcoming sixth RPA data release (T.T. Hansen et al., in prep).

Following this, all of them were selected for follow-up observations, having an $\mathrm{[Eu/Fe]>0.3}$ and $\mathrm{[Ba/Fe]<0.0}$. High-resolution, high signal-to-noise “portrait” spectra were obtained between May 2017 and July 2018 using the Magellan Inamori Kyocera Echelle (MIKE) spectrograph \citep{bernstein2003} on the Landon Clay (Magellan II) telescope at Las Campanas Observatory, Chile. 
The only exception is \objonename, which was observed in August 2020 with the TS23 echelle spectrograph \citep{Suntzeff1995} on the Harlan J. Smith 107-inch (2.7 m) telescope at McDonald Observatory. The higher SNR and resolution of the portrait observations facilitate a more detailed chemical abundance analysis of the identified \emph{r}-process enhanced stars. 
The MIKE spectra cover a wavelength range of 3350~\AA to 5000~\AA in the blue and 4900~\AA to 9500~\AA in the red.
The observations were taken with the $0.35\arcsec \times 5.00\arcsec$ slit using $2 \times 2$ binning and the $0.50\arcsec \times 5.00\arcsec$ slit using $2 \times 1$ binning. 
The corresponding resolving powers are $R \sim 56{,}000$ and $R \sim 54{,}000$ in the blue, and $R \sim 50{,}000$ and $R \sim 48{,}000$ in the red, respectively. 
The TS23 echelle spectrograph provides spectral coverage from 3400~\AA\ to 10{,}900~\AA, with a resolving power of $R \sim 43{,}000$ when using the $1.8\arcsec \times 8.00\arcsec$ slit and $1 \times 1$ binning.
Table \ref{tab:observinglog} summarizes the observing log, including the instrument setup (slit size and binning), exposure times, and SNR for each target.

\begin{table*}[htbp]
    \centering
    \renewcommand{\arraystretch}{1.2} 
    \caption{Properties of the stars.} 
    \resizebox{\textwidth}{!}{
    \begin{tabular}{r r r r r r r r r r}
    \hline
    \hline 
    2MASS stellar ID  & R.A. & Decl. & $V$ & $G$ & $BP$ & $RP$ & $K$ & $E(B-V)$ & $\varpi$ \\
        & & & mag & mag & mag & mag & mag & mag &  mas \\
    \hline
    J00401252+2729247  & 00 40 12.5  &  +27 29 24.7 & 11.13$\pm$0.11 & 10.81& 11.36& 10.10  & 8.52$\pm$0.02 &	0.0458$\pm$0.0026  & 0.35$\pm$0.02\\
    J02172993$-$1903583  & 02 17 29.9  &  $-$19 03 58.3 & 13.26$\pm$0.01 & 12.93  & 13.51  & 12.20  & 10.61$\pm$0.02 &	0.0238$\pm$0.0009  & 0.12$\pm$0.02 \\ 
    J02462013$-$1518419  & 02 46 20.1  & $-$15 18 41.9 & 12.44$\pm$0.01  & 12.18 & 12.62  & 11.57 & 10.20$\pm$0.02 & 0.0217$\pm$0.0003  & 0.34$\pm$0.02\\ 
    J14301385$-$2317388  & 14 30 13.9  &  $-$23 17 38.8 & 11.98$\pm$0.01 & 11.50 & 12.30  & 10.64  & 8.59$\pm$0.02  & 0.0870$\pm$0.0009  &0.14$\pm$0.02 \\ 
    J14325334$-$4125494  & 14 32 53.3  & $-$41 25 49.4 & 11.08$\pm$0.08 & 10.83& 11.29  & 10.20  & 8.81$\pm$0.02 &  0.1029$\pm$0.0046  & 0.91$\pm$0.02\\
    J19161821$-$5544454  & 19 16 18.2  & $-$55 44 45.4 & 11.45$\pm$0.09 & 11.07 & 11.74 & 10.29  & 8.53$\pm$0.02  & 0.0509$\pm$0.0007  & 0.17$\pm$0.02\\ 
    J20093393$-$3410273  & 20 09 33.9  &  $-$34 10 27.3 & 11.77$\pm$0.01 & 11.21  & 12.03 & 10.32  & 8.29$\pm$0.03 & 	0.0861$\pm$0.0017 	 & 0.13$\pm$0.02\\  
    J20492765$-$5124440  & 20 49 27.7  & $-$51 24 44.0 &  11.55$\pm$0.00 & 11.16& 11.79 & 10.40  & 8.73$\pm$0.02  & 0.0275$\pm$0.0017 & 0.16$\pm$0.02  \\ 
    J21064294$-$6828266  & 21 06 42.9  &  $-$68 28 26.6 & 12.80$\pm$0.01 & 12.58 & 12.99 & 12.00  &	10.73$\pm$0.02  & 0.0357$\pm$0.0008 & 0.47$\pm$0.01 \\
    J21091329$-$1310253 & 21 09 18.3  & $-$13 10 06.6 & 11.55$\pm$0.10 &  11.31 & 11.58 & 10.88  &	10.08$\pm$0.02 &  0.0466$\pm$0.0014 &2.91$\pm$0.03\\ 
    \hline 
    \end{tabular}}
    \label{tab:basicdata}
    \tablefoot{$V$ and $K$ magnitudes are taken respectively from \citet{Munari_2014} and \citet{2003yCat.2246....0C}; magnitude and colors in $G$, $BP$ and $RP$ bands come from the third Gaia data release (DR3; \citealt{Gaia2023}), and the parallax from \cite{Gaia2023}. Individual uncertainties for the Gaia photometry are not reported, as their typical values are on the order of $10^{-4}$~mag.}
\end{table*}

\begin{table*}[htbp]
    \centering
    \renewcommand{\arraystretch}{1.1} 
    \caption{Observing log.}
    \resizebox{0.78\textwidth}{!}{
    \begin{tabular}{l l r r r r r r r}
    \hline
    \hline 
    ID  &  Inst & Slit width & Bin & HJD & Exp. (s) & SNR &SNR & $RV_{helio}$\\
    & & (\arcsec)  & & & & $3800 $\AA & $4100 $\AA & (km s$^{-1}$)\\
    \hline
    J0040+2729  & McD & 1.80 & 1x1 &  2459087.88555 & 14x1800 & 11&	39 & $-91.70\pm0.49$\\
    J0217$-$1903 & MIKE & 0.50 & 2x1 & 2458093.58377 & 4x1800 & 32	& 100 & $-43.97\pm0.23$\\ 
    & MIKE & 0.50 & 2x1 & 2458438.14529& 5x1800 & &  & $-45.17\pm0.58$ \\
    J0246$-$1518  &  MIKE & 0.35 & 2x2  &  2457968.86570 &3x1800 & 56 & 85 & $+277.71\pm0.20$ \\ 
    J1430$-$2317 & MIKE & 0.35 & 2x2 &  2457882.55411 & 3x1800 & 20 & 51 & $+432.65\pm0.09$\\ 
    J1432$-$4125  &  MIKE & 0.50 & 2x1 & 2458323.44550 &3x1800 & 152 &	250 & $-229.78\pm0.30$\\
    & MIKE & 0.50 & 2x1 & 2458323.09657 & 4500 & & & $-231.80\pm0.41$ \\
    J1916$-$5544  & MIKE & 0.35 & 2x2 &  2457881.85935 &3x1200 & 44& 90 & $+49.43\pm0.17$  \\ 
    J2009$-$3410  & MIKE & 0.35& 2x2 &  2457968.59879 &5x1800 & 33 & 81 & $+27.80\pm0.51$ \\
    J2049$-$5124 &  MIKE & 0.35 &  2x2 &   2457882.78828 & 5400 &52	& 107 & $+24.31\pm0.17$\\   
    J2106$-$6828 & MIKE & 0.35 & 2x2 &  2457967.66552 & 4x1800 &48 & 83 & $-74.28\pm0.24$\\
    J2109$-$1310  &  MIKE & 0.35& 2x2 &   2457882.85526 &3x1200 & 107 & 183 & $-36.20\pm0.16$\\ 
    \hline  
    \end{tabular}}
    \label{tab:observinglog}
    \tablefoot{ For the stars \mbox{J0217$-$1903} and \mbox{J1432$-$4125}, multiple observations were taken at different epochs. These spectra were corrected for radial velocity shifts and subsequently coadded. We report the SNR of the final coadded spectrum only, as it is the version used for the analysis.
}

\end{table*}

\subsection{Data reduction}
The MIKE data were reduced using the Carnegie Python (CarPy) pipeline \citep{Kelson2000ApJ...531..184K,Kelson2003PASP..115..688K} and the TS23 echelle data was reduced using standard IRAF\footnote{NOIRLab IRAF is distributed by the Community Science and Data Center at NSF NOIRLab, which is managed by the Association of Universities for Research in Astronomy (AURA) under a cooperative agreement with the U.S. National Science Foundation.}  packages \citep{Tody1986SPIE..627..733T,Tody1993ASPC...52..173T,Fitzpatrick2024}. The data reduction process included corrections for bias, flat-fielding, and scattered light. For stars observed across multiple nights, the individual spectra were co-added. 

\subsection{Radial velocities}
To determine the heliocentric radial velocity (RV$_{\text{helio}}$) values, we performed an order-by-order cross-correlation using the IRAF's task \code{fxcor}\footnote{\url{https://articles.adsabs.harvard.edu/pdf/1979AJ.....84.1511T}}. The cross-correlation was carried out on 20 spectral orders centered around the Mg I triplet region, and the resulting velocity shift was then applied to all remaining orders. The standard stars used for calibration were HD~182488, with a heliocentric velocity of 
$v_{\mathrm{helio}} = 81.97~\mathrm{km\,s^{-1}}$, observed with the McD setup, and 
HD~211038 ($v_{\mathrm{helio}} = 10.44~\mathrm{km\,s^{-1}}$) together with 
HD~122563 ($v_{\mathrm{helio}} = -26.13~\mathrm{km\,s^{-1}}$), observed with the MIKE setup 
using slit widths of $0.50''$ and $0.35''$, respectively.
 Final values presented in Table \ref{tab:observinglog} are taken as the mean and standard deviation of the orders used for the cross-correlation. All the radial velocities are consistent with the values found by \citet{Hansen_2018} and by GaiaDR3 \citet{gaiadr3}, except for \mbox{J1916$-$5544}. For this star, we measure $+49.43\pm0.17$ km s$^{-1}$, in contrast to literature values of $+46.58\pm0.29$ km s$^{-1}$ \citep{Hansen_2018} and $+46.90\pm0.30$ km s$^{-1}$ \citep{Gaia2023}, suggesting the star belongs to a binary system. 

\section{Stellar parameter determination}
\label{StellarParameters}
Stellar parameters were derived using the methodology established by the RPA (see \citealt{roederer2018a,shah2024,Xylakis-Dornbusch2024A&A...688A.123X} for details). A short overview is provided below.
Effective temperatures ($T_{\mathrm{eff}}$) were determined photometrically using Gaia $G$, $BP$, $RP$-bands and 2MASS $K$-band magnitudes, listed in Table~\ref{tab:basicdata}. We adopted the color--metallicity--$T_{\mathrm{eff}}$ calibrations from \citet{Mucciarelli2021A&A...653A..90M}. 
Reddening corrections were applied to the Gaia magnitudes using the $E(B-V)$ values also reported in Table~\ref{tab:basicdata}, derived from the dust maps of \citet{schlafly2011}. Extinction coefficients were taken from \citet{McCall2004} for the $K$ filter and calculated following \cite{Gaia2018} for the Gaia filters, and applied individually.
Following the Monte Carlo approach by \citet{roederer2018a}, we estimated $T_{\mathrm{eff}}$ for each color combination by drawing $10^4$ realizations of the input parameters (magnitudes, reddening, metallicity) assuming Gaussian errors, and adopting the median of the resulting $T_{\mathrm{eff}}$ distribution. The final $T_{\mathrm{eff}}$ for each star was then computed as the weighted average of the values from all color indices. For comparison, we show the excitation imbalance in the left column of Figure~\ref{Ion_imbalance1}. The slight trend in the \ion{Fe}{I} and \ion{Fe}{II} line abundances as a function of excitation potential, $\chi$ reflects the difference between the photometric and the spectroscopic $T_{\mathrm{eff}}$. To reflect this uncertainty, as well as the limitations of the 1D model atmosphere, the total uncertainty includes the statistical weighted average plus 150~K added in quadrature \citep{Frebel_2013}. After determining $T_{\mathrm{eff}}$, the surface gravities (\logg) were calculated from the fundamental relation:
\begin{equation}
    \begin{split}
        \log g = 4 \log T_{\text{eff}} + \log \left(\frac{M}{M_{\odot}}\right) - 10.61 + 0.4 \cdot ( BC_V \\
        + m_V - 5 \log(d) + 5 - 3.1 \cdot E(B - V) - M_{\text{bol},\odot}),
    \end{split}
\end{equation}
where $M$ is the stellar mass, assumed to be $0.8 \pm 0.08 M_{\odot}$ for all the stars, being the canonical value for halo stars. $BC_V$ is the bolometric correction in the $V$ band, and  $m_V$ is the apparent $V$ band magnitude. The parameter $d$ denotes the distance in parsecs, and the Solar bolometric magnitude is fixed at $M_{\text{bol},\odot}=4.75$. The constant 10.61 is derived from the Solar reference values, $\log(T_{\mathrm{eff},\odot}) = 3.7617$ and $\log g_{\odot} = 4.438$.
To estimate $\log g$ and its associated uncertainty, we computed the median and standard deviation from $10^4$ Monte Carlo samples generated for each input parameter. An additional 0.3~dex was added in quadrature to the resulting standard deviation to account for systematic uncertainties. 

Metallicity, $\mathrm{[Fe/H]}$, and microturbulent velocity, $\xi$ were derived through equivalent width (EW) analysis of \ion{Fe}{i} and \ion{Fe}{ii} lines, where the EW is defined as the width of a rectangle, with the same height as the line depth relative to the continuum, and an area equal to that of the spectral line. The analysis was performed by fixing $T_{\mathrm{eff}}$ and $\log g$. The microturbulent velocity was obtained by removing any correlation between \ion{Fe}{i} line abundances and reduced equivalent width, while the model atmosphere metallicity was taken as the average of the $\mathrm{[Fe I/H]}$ and $\mathrm{[Fe II/H]}$ abundances, after manually removing outlier lines from the \ion{Fe}{i} and \ion{Fe}{ii} distributions that deviated by more than 2.5$\sigma$ from the mean abundance. This procedure results in a flat trend of \ion{Fe}{i} as a function of the line strength (REW), as shown in the central column of Figure~\ref{Ion_imbalance1}.
The final $\mathrm{[Fe \textsc{I}/H]}$ and $\mathrm{[Fe\textsc{II}/H]}$ values agree within 0.08 dex for all stars. Systematic uncertaintie on $\xi$ is estimated to be 0.2 km s$^{-1}$.

The final stellar parameters and their associated uncertainties for the ten stars are listed in Table~\ref{tab:Stellarparameters}.

\begin{table}[ht]
\centering
\caption{ Stellar parameters with systematic and statistical uncertainties.}
\begin{tabular}{lllllll}
\hline
Stellar ID & $T_{\text{eff}}$ [K] & $\log g$ & [Fe/H] & $\xi$ [km s$^{-1}$] \\
& & & $\pm 0.20$ & $\pm 0.20$ \\
\hline
J0040+2729  & $4695\pm154$ & $1.37\pm0.32$ & $-2.72 $ & $2.38 $ \\
J0217$-$1903  & $4609\pm154$ & $1.40\pm0.34$ & $-2.86 $ & $2.61 $ \\
J0246$-$1518  & $5019\pm155$ & $2.07\pm0.31$ & $-2.70 $ & $1.94$ \\
J1430$-$2317  & $4158\pm153$ & $0.49\pm0.33$ & $-1.83$ & $2.14$ \\
J1432$-$4125  & $5159\pm155$ & $2.25\pm0.31$ & $-2.76 $ & $1.63 $ \\
J1916$-$5544  & $4422\pm154$ & $0.70\pm0.33$ & $-2.39 $ & $2.31 $ \\
J2009$-$3410  & $4187\pm153$ & $0.33\pm0.30$ & $-2.32 $ & $2.89 $ \\
J2049$-$5124  & $4250\pm153$ & $0.83\pm0.30$ & $-2.61 $ & $2.29 $ \\
J2106$-$6828  & $5208\pm155$ & $2.51\pm0.31$ & $-2.72 $ & $1.50 $ \\
J2109$-$1310  & $4996\pm155$ & $1.58\pm0.31$ & $-2.45 $ & $2.03$ \\

\hline
\end{tabular}

    \label{tab:Stellarparameters}
\end{table}

\section{Abundance analysis}
\label{AbundanceAnalysis}
The abundance analysis was carried out using the software \texttt{SMHr}\footnote{https://github.com/eholmbeck/smhr-rpa/tree/smhr3-damping} \citep{casey2014}, which runs the 1D LTE radiative transfer code \texttt{MOOG}\footnote{https://github.com/alexji/moog17scat} \citep{Sneden1973,sobeck2011}. We utilized $\alpha$-enhanced ($\mathrm{[\alpha/Fe]} = +0.4$) \texttt{ATLAS9} model atmospheres \citep{castelli2003} and adopted Solar abundances from \cite{asplund2009}. We adopted the same line lists as those used in other RPA studies, ensuring methodological consistency across the collaboration. These were generated using \texttt{linemake}\footnote{https://github.com/vmplacco/linemake}\citep{placco2021} and include isotopic and hyperfine structure broadening where applicable, with $r$-process isotope ratios taken from \cite{sneden2008}. Elemental abundances were determined from a mixture of EW analysis and spectral synthesis. For the EW analysis, Gaussian or Voigt profiles were fitted to single, un-blended lines in the continuum-normalized spectra, while the spectral synthesis was used for blended features and lines affected by isotopic and/or hyperfine structure. The final abundances were computed as the average of individual line abundances. In Table \ref{tab:lines} in the appendix, we report the atomic data for the lines used in the analysis, with the measured EW and derived abundance of each line in J0040+2729. The same information for the other stars can be found in the CDS material.

To estimate abundance uncertainties, we applied the method described in \cite{ji2020a}, which performs propagation of stellar parameter uncertainties, incorporating statistical and systematic uncertainties on each spectral line. Moreover, a 0.1 dex uncertainty is included for all lines to account for systematic uncertainties like continuum placement and atomic data uncertainties. Finally, an additional uncertainty of 0.2 dex is applied in quadrature to all abundances derived from two or fewer lines.

\section{Results}
\label{Results}
\begin{figure}[hbt!]
\centering
\includegraphics[width=1.0\linewidth]{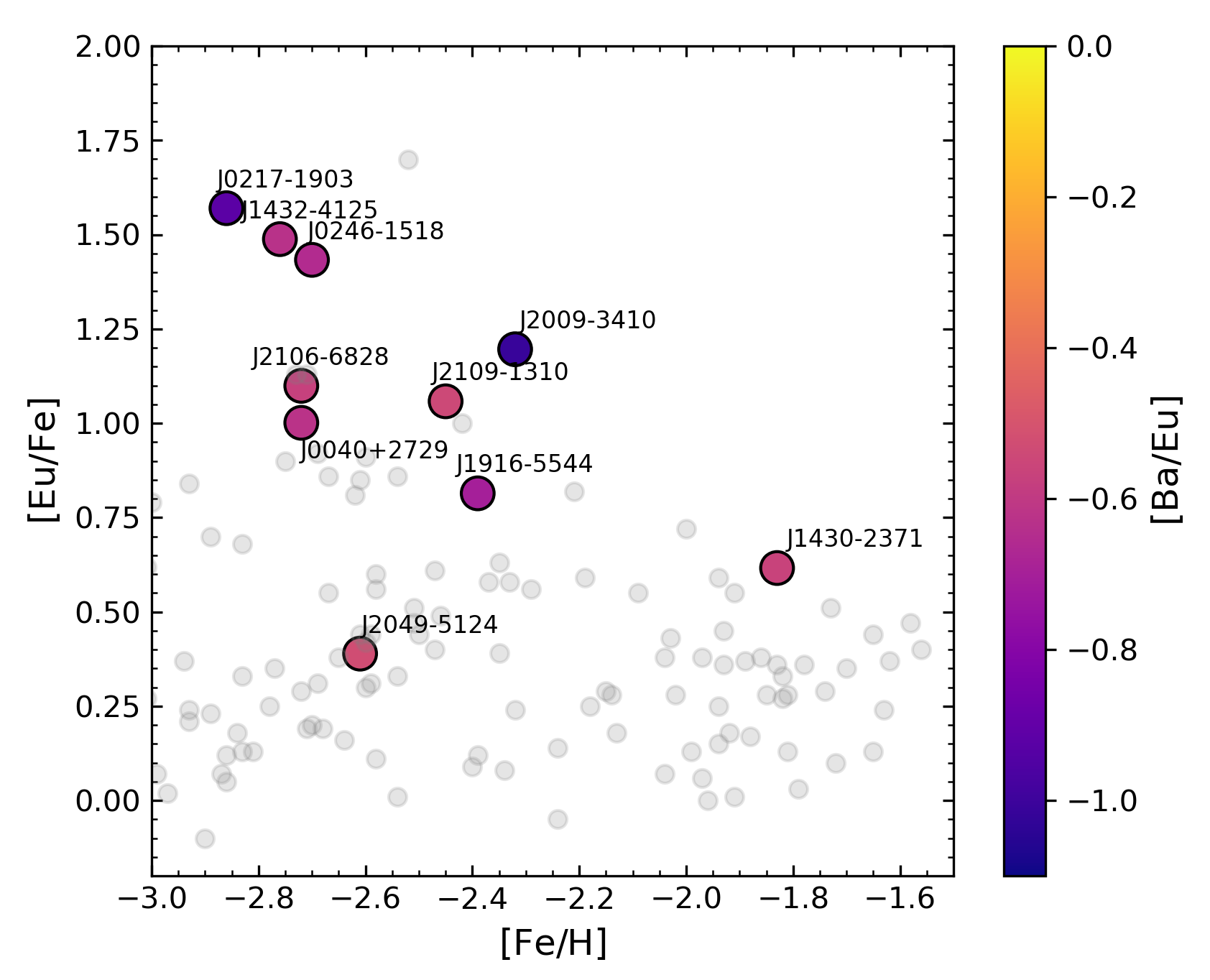}
\caption{$[\mathrm{Eu/Fe}]$ as a function of $[\mathrm{Fe/H}]$, with a color gradient representing $[\mathrm{Ba/Eu}]$ for the sample of ten stars. All stars lie well above the canonical $[\mathrm{Eu/Fe}] > +0.3$ threshold typically adopted for \emph{r}-process-enhanced stars, and their low $[\mathrm{Ba/Eu}]$ values further confirm that the enrichment is consistent with a pure \emph{r}-process origin. Grey stars in the background are taken from \cite{roederer2014a}.
\label{fig:rII}
}
\end{figure}

We inspected more than 1400 lines for each of the ten stars in our sample and derived abundances of 54 atomic species, including 29 neutron-capture elements. Abundances for \ion{Na}{i}, \ion{Mg}{i}, \ion{Al}{i}, \ion{Si}{i}, \ion{K}{i}, \ion{Ca}{i}, \ion{Ti}{i}, \ion{Ti}{ii}, \ion{Cr}{i}, \ion{Cr}{ii}, \ion{Fe}{i}, \ion{Fe}{ii}, \ion{Ni}{i} and \ion{Zn}{i} were determined measuring the EW of unblended lines, while abundances for all other species were derived via spectral synthesis. Final abundances and associated uncertainties for the ten stars are presented in Table~\ref{tab:Abundances}.
The stars exhibit enhanced $[\mathrm{Eu/Fe}]$ ratios, ranging from +0.39 to +1.60, and $[\mathrm{Ba/Eu}]$ between $-0.52$ and $-1.01$, as shown in Figure~\ref{fig:rII}. Such abundance ratios are indicative of dominant \emph{r}-process enrichment \citep{sneden2008}, making these stars excellent candidates for probing the nature of the \emph{r}-process nucleosynthesis sites.\\

\subsection{Carbon and Nitrogen}
Carbon abundances were determined for eight stars using spectral synthesis of the CH $G$-band at $4313 \text{\AA}$. For the stars J0246$-$1518 and J0217$-$1903, no CH absorption features were detected, and thus no carbon abundance could be determined.

Nitrogen abundances were estimated from the CN molecular bands at $3876\text{\AA}$ and $3589\text{\AA}$ were detected in six of the observed stars (\mbox{J0217$-$1903}, \mbox{J1430$-$2317}, \mbox{J1916$-$5544}, \mbox{J2009$-$3410}, \mbox{J2049$-$5124}, \mbox{J2109$-$1310}).
The NH band at 3369 was also investigated and detected in \mbox{J0217$-$1903}, \mbox{J2049$-$5124} and \mbox{J2109$-$1310}, for which we obtained a $1\sigma$ consistency with N abundance derived from CN, and in \mbox{J1432$-$4125} in which we have no CN detection instead.

Almost all the derived carbon and nitrogen abundances show no extreme enhancements across the sample. Carbon abundances, corrected for evolutionary effect according to \citet{Placco2014ApJ...797...21P}, spans the range \([\mathrm{C/Fe}] = -0.35\) to \(+0.55\), while nitrogen ranges from $[\mathrm{N/Fe}] = -0.23$ to $+0.56$. These values fall below the conventional thresholds for carbon-enhanced metal-poor (CEMP) and nitrogen-enhanced metal-poor (NEMP) of $[\mathrm{C/Fe}] > +0.7$ \citep{BeersChristlieb2005ARA&A..43..531B, Aoki2007}  and $[\mathrm{N/Fe}] = +1.00$ \citep{Pols2012}, respectively. The only exception is given by \mbox{J2019$-$1310}, for which the carbon value minimally exceeds the CEMP limit, being \([\mathrm{C/Fe}] = +0.74\)

\begin{figure*}[hbt!]
\hspace*{-0.02\linewidth}
\includegraphics[width=1.\linewidth]{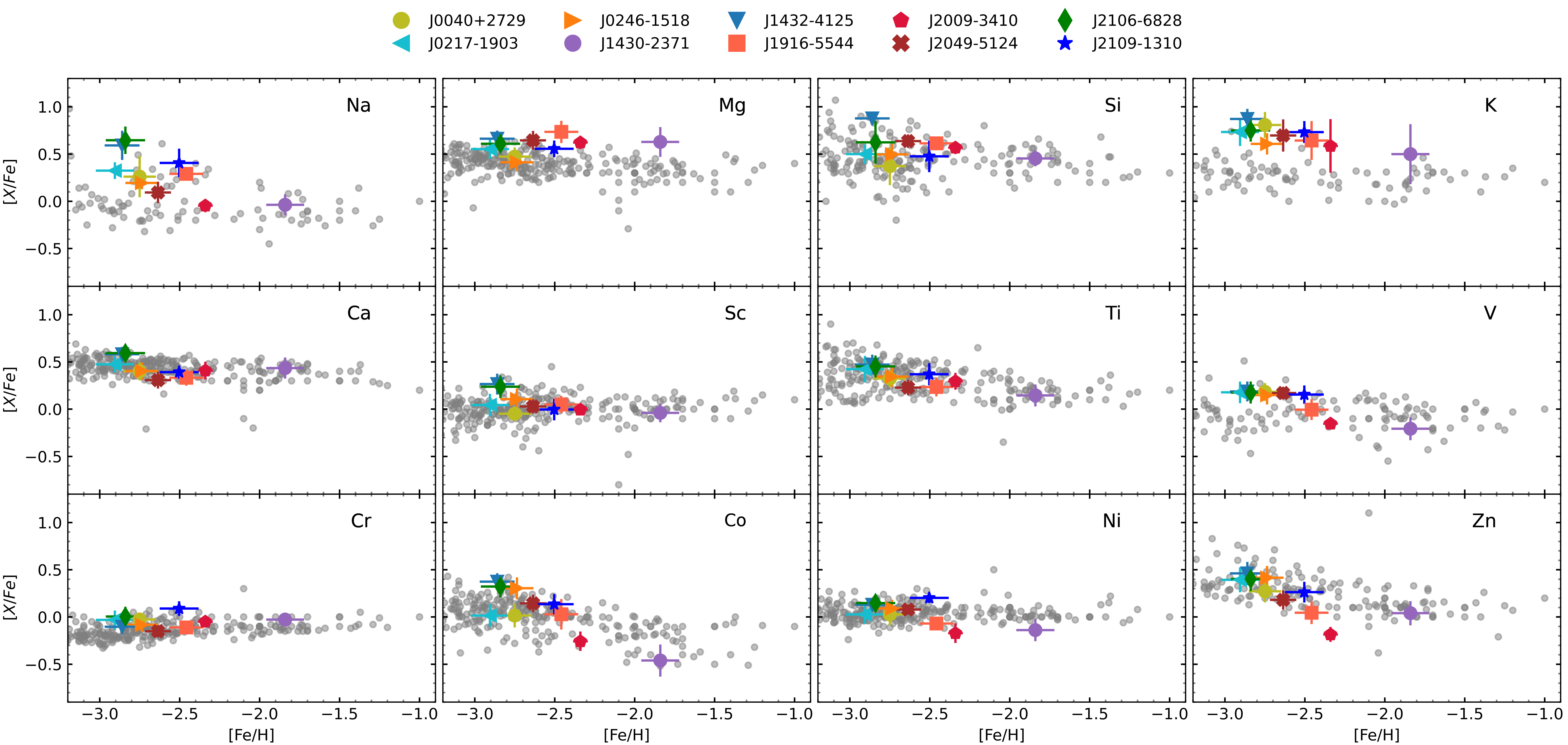}
\caption{$\mathrm{[X/Fe]}$ abundances of the light elements for the ten sample stars, compared to stellar abundances from the MW halo \citep[grey dots;][]{roederer2014a}.}
\label{light_ele}
\end{figure*}

\subsection{Elements from oxygen to zinc}
We determined abundances for 21 atomic species with $Z<30$ in all ten stars, including both neutral and singly ionized: \ion{O}{i}, \ion{Na}{i}, \ion{Mg}{i}, \ion{Al}{i}, \ion{Si}{i}, \ion{K}{i}, \ion{Ca}{i}, \ion{Sc}{ii}, \ion{Ti}{i}, \ion{Ti}{ii}, \ion{V}{i}, \ion{V}{ii}, \ion{Cr}{i}, \ion{Cr}{ii}, \ion{Mn}{i}, \ion{Mn}{ii}, \ion{Fe}{i}, \ion{Fe}{ii}, \ion{Co}{i}, \ion{Ni}{i}, and \ion{Zn}{i}. Additionally, we report the detection of \ion{Sc}{i} in one star and the detection of \ion{Cu}{i} in four stars in our sample.
The \ion{Cu}{i} abundance is based solely on the 5105~\AA line, arising from a high-excitation level ($\chi =1.39$~eV). Its intrinsic weakness and sensitivity to stellar parameters lead to non-detections in most cases. For \ion{Sc}{i}, the investigated lines have central wavelengths between 3900 and 4050~\AA, in the near-UV. For stars in our parameter range, this spectral region typically suffers from severe line blending, and \ion{Sc}{i} is a minority species in these stars, making detections difficult. The $\mathrm{[X/Fe]}$ abundance ratios for the light elements are shown in Figure \ref{light_ele}, where our sample is compared to Milky Way halo stars from \citet{roederer2014a}. In that study, NLTE corrections were applied to Na and K, resulting in systematically lower average abundances relative to our sample. For elements where abundances of both neutral and singly ionized species were derived, we plotted the average abundance derived from the two ionization states. Overall, the abundance ratios of the light elements in our stars are broadly consistent with those observed in other metal-poor stars, showing no significant anomalies.

\subsection{Neutron-capture Elements}
We analyzed spectral lines from 29 neutron-capture elements. A detailed review of each of these elements is provided in the following subsections.\\

\begin{figure*}[hbt!]
\hspace*{-0.02\linewidth}
\includegraphics[width=1.\linewidth]{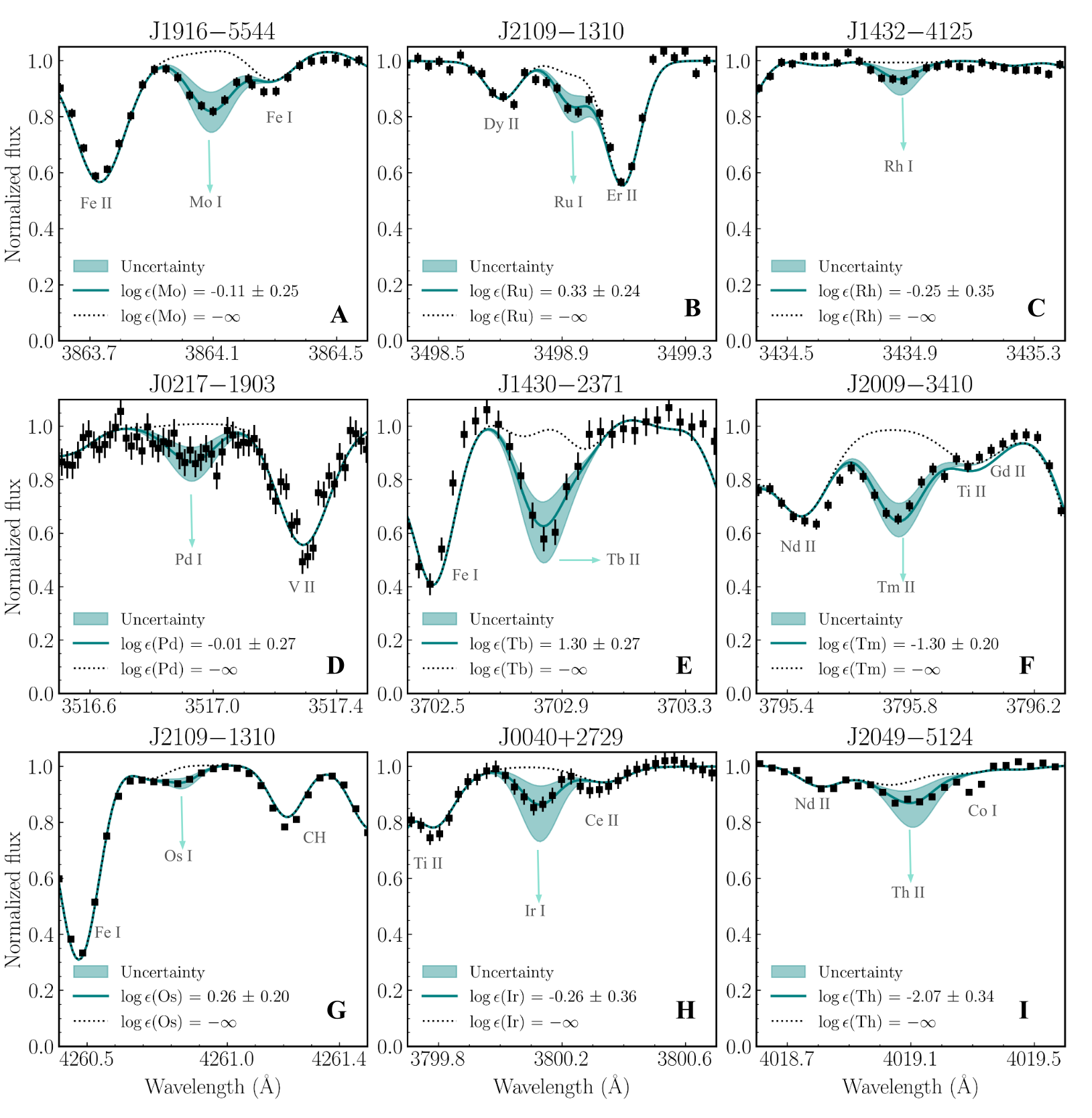}
    \caption{Examples of spectral synthesis fits used to derive elemental abundances. The observed data (black squares) are shown along with the best-fit synthetic spectra (solid teal lines) and associated uncertainties (shaded regions). The dotted lines correspond to synthetic spectra with no contribution from the indicated element (i.e., $\log \epsilon = -\infty$).}
    \label{fig:fit_examples}
\end{figure*}

\subsubsection{Rubidium}
We identify a tentative detection of rubidium (Z=37) in J1430$-$2371, the coolest and most metal-rich star in our sample, based on two \ion{Rb}{i} spectral lines, at 7800~\AA and 7947~\AA. The first one is a resonance feature, though partially blended on the blue wing with a nearby \ion{Si}{i} line, which complicates precise abundance determination. Fitting this feature yields an Rb abundance of $\log \epsilon(\mathrm{Rb}) = 0.82 \pm 0.20$. To support this result, we also examined another resonance line at 7947~\AA, from which we estimate an upper limit of $\log \epsilon(\mathrm{Rb}) < 0.91$. Since this upper limit lies only 0.09~dex above the value derived from the resonance line, we used this result as a validation of the previous estimate, interpreting it as a tentative detection of \ion{Rb}{i}.

\subsubsection{Elements between the first and second \emph{r}-process peaks}
\label{light_r}
Abundances for the light neutron-capture elements strontium (Sr, $Z=38$), yttrium (Y, $Z=39$), and zirconium (Zr, $Z=40$) were determined for all stars in the sample. In particular, Sr abundances were derived using the three \ion{Sr}{ii} lines located at 4077, 4161, and 4215~\AA. Y and Zr abundances were measured using line lists containing 41 and 51 transitions, respectively. For each star, the abundance determination relied on at least 10 lines, covering a wide wavelength range. 
We also report detections of molybdenum (Mo, $Z=42$), ruthenium (Ru, $Z=44$), rhodium (Rh, $Z=45$), and palladium (Pd, $Z=46$) in eight, eight, three, and five stars, respectively. Mo abundances are based primarily on the \ion{Mo}{i} line at 3864~\AA (shown in panel A of Figure~\ref{fig:fit_examples}), with the line at 5533~\AA also contributing in a few cases, showing agreement within 0.2 dex,  when both lines were detected. Ru abundances were derived using four \ion{Ru}{i} lines in the blue region of the spectrum at 3436, 3498 (see panel B in Figure~\ref{fig:fit_examples}), 3798, and 3799~\AA. Rh abundances were determined from the \ion{Rh}{i} line at 3434~\AA (see panel C in Figure~\ref{fig:fit_examples}) for stars J1432$-$4125, J1916$-$5544, and J2019$-$1310. For J2009$-$3410, an upper limit was derived using the same line. Pd abundances are based on the \ion{Pd}{i} lines at 3404, 3460, and 3517~\AA (panel D in Figure~\ref{fig:fit_examples}), with an upper limit derived for J1916$-$5544.

We report a detection of silver (Ag, Z=47) in only one star, J1432$-$4125, based on the \ion{Ag}{i} line at 3381~\AA, which is the only transition included in our linelist. This spectral feature is intrinsically weak and frequently blended with NH molecular lines, making it particularly challenging to detect in metal-poor stars. Due to these limitations, no Ag abundances could be determined for the other stars in our sample.\\

\subsubsection{Elements beyond the second \emph{r}-process peak}
The elements from barium (Ba, $Z=56$) to hafnium (Hf, $Z=72$) are called Lanthanides. Their abundances are generally well determined across our stellar sample.

For Ba, five strong \ion{Ba}{ii} transitions in the optical range were employed. 
Abundances for other heavy elements, namely lanthanum (La, $Z=57$), cerium (Ce, $Z=58$), praseodymium (Pr, $Z=59$), neodymium (Nd, $Z=60$), samarium (Sm, $Z=62$), gadolinium (Gd, $Z=64$), and dysprosium (Dy, $Z=66$), were derived from numerous lines arising from singly ionized atoms (typically more than 20 per element), ensuring robust abundance derivations.
Europium (Eu, $Z=63$) abundances were obtained using 10 \ion{Eu}{ii} lines, while terbium (Tb, $Z=65$), holmium (Ho, $Z=67$), and erbium (Er, $Z=68$) abundances were derived using up to five transitions, mostly located in the blue region of the spectrum. 
Thulium (Tm, $Z=69$), ytterbium (Yb, $Z=70$), and lutetium (Lu, $Z=71$) were detected in nine, ten, and nine stars, respectively. The Tm abundance is based on a maximum of six lines per star, all located below 4300~\AA (one example is shown in panel F of Figure~\ref{fig:fit_examples}), where line blending and continuum placement become increasingly uncertain. Similarly, Yb relies on a single Yb II line at 3694 \AA. 
In contrast, Lu abundances were derived from a single, relatively unblended line at 6220~\AA. 
Hafnium (Hf, $Z=72$) was detected in nine stars using five blue lines; although the features were relatively weak, they still allowed for reliable abundance estimates.\\
\subsubsection{Third peak elements}
\label{ThirdPeak}
Osmium (Os, $Z=76$) and iridium (Ir, $Z=77$) are among the so-called third $r$-process peak elements, corresponding to nuclei with mass numbers around $A=195$ near the closed neutron shell at $N=126$. We report respectively ten and nine new values of Os and Ir, and an upper limit for iridium.

In our analysis, we investigated two \ion{Os}{i} lines at 4135 and 4261 \AA\ (Figure~\ref{fig:fit_examples}, panel G). The absolute abundances derived from these two lines consistently disagreed, with the value derived from the first line being systematically higher than that obtained from the second by a minimum of 0.4 dex and up to 1 dex. Following the framework of “relative strength” as defined by \cite{Sneden_2009}, which under LTE conditions is expressed as $\log(\epsilon gf) - \chi \theta$ (where $\epsilon$ denotes the absolute abundance, $gf$ the oscillator strength, $\chi$ the excitation potential, and $\theta = 5040/T$ ), a line with lower relative strength is expected to be intrinsically weaker and consequently yield a lower abundance measurement.
The 4135 \AA\ line always exhibits a lower relative strength compared to the 4261 \AA\ line, and thus should produce correspondingly lower abundance values, contrary to what we observe. This systematic discrepancy, which becomes more pronounced in stars characterized by lower effective temperatures, lower surface gravities, and higher metallicities, strongly suggests the presence of an unidentified blend affecting the 4135 \AA\ feature. This interpretation is further supported by an anomalous radial velocity shift encountered during the fitting of the 4135 \AA\ line.
Given these considerations, we have elected to base our \ion{Os}{i} abundance determinations exclusively on the 4261 \AA\ line.

\begin{figure}[hbt!]
\centering
\includegraphics[width=1.00\linewidth]{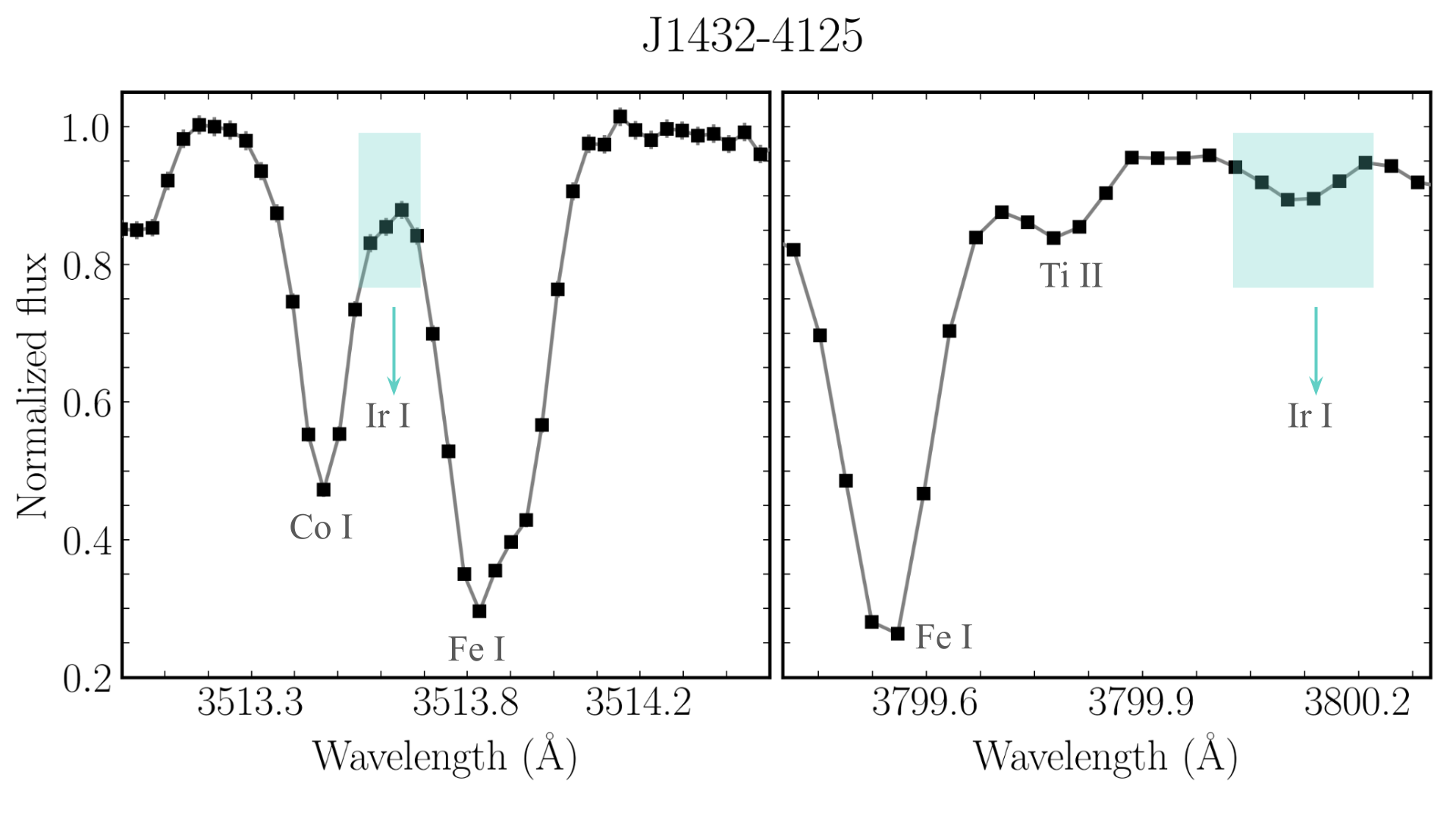}
\caption{
    Comparison of two \ion{Ir}{I} lines in the spectrum of J1432$-$4125. The black dots represent the observed spectrum, and the turquoise boxes identify the position of the two lines.
    \label{fig:irsyn}
}
\end{figure}

Similarly, we chose to base the Ir abundance solely on the \ion{Ir}{I} line at 3800~\AA (Figure \ref{fig:fit_examples}, panel H), excluding the one at 3513~\AA. To illustrate the reasoning for this, we show in Figure \ref{fig:irsyn} both \ion{Ir}{i} lines in the star J2109–1310. A closer inspection of this plot reveals that the 3800~\AA line is unblended. In contrast, the 3514~\AA feature is located between the nearby \ion{Co}{i} and \ion{Fe}{i} transitions, making its profile more challenging to model. The observed absorption lacks a clear Gaussian profile and appears instead as a weak bump between stronger features, undermining its reliability as an \ion{Ir}{i} abundance tracer. Such inconsistencies are easier to deal with for elements with more lines to choose from (e.g., \ion{La}{ii} or \ion{Ce}{ii}, which have 40–60 lines), but they become significant for elements like \ion{Ir}{i}, where only two lines are available.

\subsubsection{Beyond the third peak}
We derived lead (Pb, Z=82) abundance in two stars: J1430–2317 and J2009–3410. These are the two coolest stars in our sample, which allows for the detection of a neutral species like \ion{Pb}{I}. For all other, warmer stars, a larger fraction of Pb atoms are ionized, so such measurements are not feasible. Thorium (Th, Z=90) was detected in nine stars through three \ion{Th}{II} lines at 4019 (Figure~\ref{fig:fit_examples}, panel I), 4086, and 5989~$\text{\AA}$. In contrast, no uranium (U, $Z=92$) lines were detected. This is consistent with expectations, given the intrinsic weakness of \ion{U}{II} transitions. 
%(particularly the commonly used line at 3859~$\text{\AA}$).

\subsection{Notes on NLTE} 
All abundances presented in this work are derived under the assumption of LTE and using 1D model atmospheres. Whenever possible, we derived the abundances from ionized species, which are less sensitive to NLTE effects. In general, we did not apply NLTE corrections to maintain a homogeneous analysis, since 40\% of our sample stars have stellar parameters outside the ranges covered by the published NLTE grids. However, as we aim to investigate scatter in \emph{r}-process element abundances we report the NLTE-corrected abundances in Table \ref{tab:abundances_corrected} for the subset of stars that fall within the available grids for the \emph{r-}process species \ion{Sr}{ii}, \ion{Ba}{ii}, and \ion{Eu}{ii}, using corrections from \citet{Mashonkina2022}, \cite{Mashonkina2019}, and \cite{Mashonkina2000} respectively. As shown in Table \ref{tab:abundances_corrected}, the average NLTE correction is relatively small for $\mathrm{[Sr/Fe]}$ and $\mathrm{[Eu/Fe]}$ ($<0.1$ dex) but significantly larger for $\mathrm{[Ba/Fe]}$ ($\sim0.2$ dex). 
NLTE studies have also been performed for Nd, where \citet{Dixon2025} found corrections ranging from $-0.3$ to $+0.5$ dex, and Y, where \citet{Storm2023} found that \ion{Y}{ii} lines in metal-poor giants ($\mathrm{[Fe/H]} \simeq -3.0$, log\,g $\simeq 1.0$) can have NLTE corrections as large as 0.4 dex for low-excitation lines. However, none of these studies has provided correction grids.

For more information, we refer the reader to the following 3D NLTE grids present in the literature for O \citep{Amarsi2015_Ox}, C \citep{Amarsi2019}, Na \citep{Canocchi2024A&A...692A..43C}, Mg \citep{Matsuno2024}, Ca \citep{Lagae2025}, and Fe \citep{Amarsi2022A&A...668A..68A}. Correction in 1D NLTE are provided for Al \citep{Nordlander2017A&A}, and for Ti, Cr, Mn and Co for which we refer the reader to the MPIA NLTE database\footnote{\url{https://nlte.mpia.de/gui-siuAC_secE.php}} based on \cite{Bergemann2011MNRAS.413.2184B}, \cite{BergemannCescutti2010A&A...522A...9B}, \cite{BergemannGehren2008A&A...492..823B} and \cite{Bergemann2008PhST..133a4013B} respectively. For Ni, Cu and Zn corrections are provided in 1D by \cite{Eitner2023, Caliskan2025} and \cite{Sitnova2022}, respectively.  For more information on the above-mentioned elements see also the \code{NLiTE} website\footnote{\url{https://nlite.pythonanywhere.com/}}.

\subsection{Dynamical Origins and Orbital Properties}
\label{dynamic}
\begin{figure*}[h!]
\includegraphics[width=\linewidth]{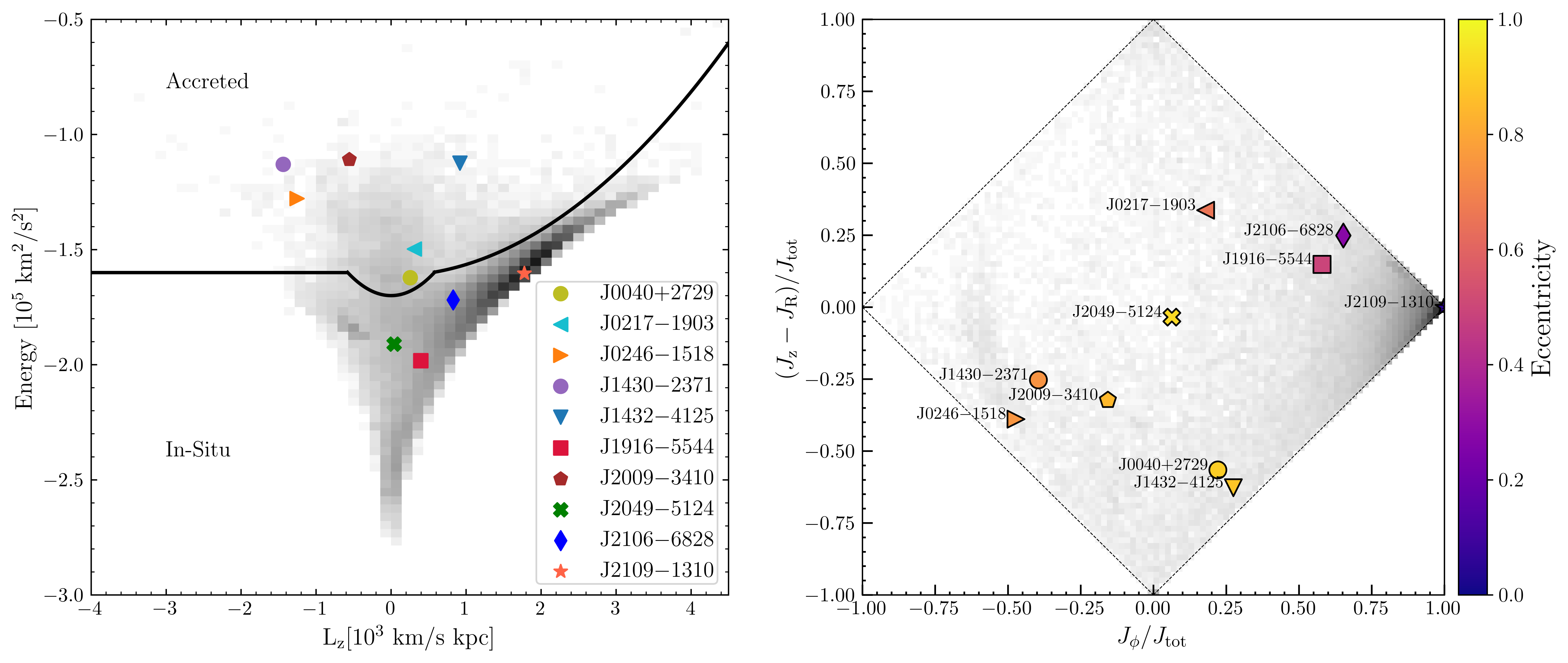}
\centering
\caption{
\textbf{Left panel:} Distribution of the sample stars in the orbital energy ($E$) versus angular momentum ($L_z$) plane. The greyscale background shows GALAH DR3 stars \citep{Buder2021} with metallicities in the range $-1.3 \leq \mathrm{[Fe/H]} \leq -0.9$. The thick black curve marks the boundary between dynamically in-situ stars (more bound, lower energy) and accreted populations (less bound, higher energy), following the prescription of \citet{Monty2024MNRAS.533.2420M}. Colored markers indicate the ten stars in our sample, as labeled in the legend. The grey contours identify the approximate region occupied by Gaia-Sausage-Enceladus (GSE) stars from \citet{Belokurov2023}, adapted to our adopted potential.
\textbf{Right panel:} Action-space diagram (''diamond diagram'') showing the normalized azimuthal action ($J_\phi / J_{\rm tot}$) on the x-axis and the normalized vertical-minus-radial action ($(J_z - J_R)/J_{\rm tot}$) on the y-axis. Colored markers correspond to the same stars as in the left panel, now color-coded by their orbital eccentricity. The greyscale background shows again the GALAH DR3 comparison sample. The grey rectangular region marks the approximate locations of the \textit{Gaia}-Sausage accreted substructure, following the definitions by \citet{Myeong2019MNRAS.488.1235M}.
}
\label{fig:cdtg_orbits}
\end{figure*}

To investigate the possible origins of our stars, we explored their orbital properties in the \(E\)--\(L_z\) space. The total orbital energy (E) and the vertical component of the angular momentum ($L_z$) were derived from the phase-space coordinates of the stars using the \texttt{galpy} package \citep{Bovy2015}, and their values are reported in Table~\ref{tab:orbits}. The results are visualized in the left panel of Figure~\ref{fig:cdtg_orbits}. This diagram displays the distribution of our sample stars in the orbital energy--angular momentum plane, where \(E\) is expressed in units of \(10^5\,\mathrm{km}^2\,\mathrm{s}^{-2}\) and \(L_z\) in \(10^3\,\mathrm{km\,s^{-1}\,kpc}\). The greyscale background represents the distribution of GALAH DR3 stars \citep{Buder2021} within the metallicity range $-1.3 \leq [\mathrm{Fe/H}] \leq -0.9$ dex. The thick black curve marks the dynamical boundary between in-situ stars (with higher binding energies) and accreted populations (less bound) originally calculated in \citet{Belokurov2023} and shifted to the energy scale of the \texttt{McMillan17} potential in \citet{Monty2024MNRAS.533.2420M}. The grey contours outline the region associated with the Gaia-Sausage-Enceladus (GSE) debris, as described by \citet{Belokurov2023} and adapted to our Galactic potential. In this space, two stars in our sample, \mbox{J0040+2729} and \mbox{J0217$-$1903}, appear to lie within the GSE-defined region.
\begin{table*}[ht!]
\centering
\caption{Orbital angular momentum (\(L_z\)), actions ($J_{\phi}$, $J_z$ and $J_r$), orbital energy (\(E\)) and eccentricity (Ecc.) for the stars in our sample.}
\resizebox{0.7\textwidth}{!}{
\begin{tabular}{lccccc}
\hline
Star ID &  $L_z (=J_{\phi})$  &$J_z$ & $J_r$ & E & Ecc.  \\
&$[10^3 \, \mathrm{km\,s^{-1}\,kpc}]$ & $[10^3 \, \mathrm{km\,s^{-1}\,kpc}]$ & $[10^3 \, \mathrm{km\,s^{-1}\,kpc}]$ & $[10^5 \, \mathrm{km}^2\,\mathrm{s}^{-2}]$& \\
\hline
J0040+2729 & +0.255  & +0.122 & +0.774 & $-$1.622 & 0.90 \\
J0217$-$1903 & +0.311 & +1.006 & +0.421 & $-$1.497 & 0.65\\
J0246$-$1518 & $-$1.253 & +0.183 &+1.214 & $-$1.278 & 0.75\\
J1430$-$2317 & $-$1.439 & +0.640 & +1.560&  $-$1.129 & 0.74 \\
J1432$-$4125 & +0.917 & +0.164 & +2.259 & $-$1.124 & 0.88\\
J1916$-$5544 & +0.396 & +0.195 & + 0.093&  $-$1.982 & 0.49\\
J2009$-$3410 & $-$0.558 & +0.924 & +2.073 & $-$1.109 & 0.85\\
J2049$-$5124 & +0.040 & +0.288 & + 0.311 & $-$1.910 & 0.92\\
J2106$-$6828 & +0.828 & +0.379& +0.062& $-$1.718 & 0.28\\
J2109$-$1310 & +1.776 & +0.002 & +0.003& $-$1.603 & 0.05\\
\hline
\end{tabular}}
\label{tab:orbits}
\end{table*}
To further test this hypothesis, we determined approximations for the orbital actions using the \texttt{Stäckel }fudge implemented in \texttt{galpy} \citep{Mackereth2018} to create the so-called "action diamond" \citep{Vasiliev2019, Myeong2019MNRAS.488.1235M}, shown in the right panel of Figure~\ref{fig:cdtg_orbits}. This diagram plots the normalized azimuthal action, \(J_\phi / J_\mathrm{tot}\), against the vertical-minus-radial action, \((J_z - J_R)/J_\mathrm{tot}\), providing a compact view of the orbital morphology. Our ten stars are shown as colored markers, with the color indicating their orbital eccentricity, obtained by computing the orbital integrations over 3~Gyr with 1000 time steps per Gyr, assuming the \texttt{McMillan17} Galactic potential \citep{McMillan2017} and  Local Standard of Rest as \citet{Monty2024MNRAS.533.2420M}. 
The background greyscale again shows the distribution of GALAH stars, while the grey box marks the highest density region of GSE members in \(E\)--\(L_z\) space, according to \citet{Myeong2019MNRAS.488.1235M}.

Interestingly, while \mbox{J0040+2729} and \mbox{J0217$-$1903} appear within the GSE region in the \(E\)--\(L_z\) plane, they sit at the edge of the typical boundary in \(J_\phi / J_\mathrm{tot}\) assumed for GSE members and well outside of the highest density region. This discrepancy indicates that their orbital morphologies are inconsistent with those of the GSE population, thereby weakening the case for their association with this major accretion event. Further support for this interpretation comes from their low metallicities of $-2.72$ and $-2.86$ dex, which are significantly below the characteristic GSE metallicity distribution, typically peaking around $\mathrm{[Fe/H]}\sim-1.5$ \citep{Myeong2019MNRAS.488.1235M, Monty2020MNRAS.497.1236M, Ceccarelli2024, Ou2024ApJ...974..232O}. Finally, the GSE is known to be depleted in $\mathrm{[Eu/Fe]}$ at low metallicities, with typical values of $\mathrm{[Eu/Fe]}$ being subsolar below $\mathrm{[Fe/H]} \sim -2.5$ \citep{Ou2024ApJ...974..232O, Monty2024MNRAS.533.2420M}.
These combined properties suggest that \mbox{J0040+2729} and \mbox{J0217$-$1903} may instead be associated with other halo substructures or represent remnants of distinct accretion events. 
The values of the actions and the eccentricity are reported in Table~\ref{tab:orbits} (note that with an axisymmetric potential like \citep{McMillan2017}, $J_\phi=L_z$).

In addition, we investigated whether any of the ten stars are associated with the chemodynamically tagged groups (CDTGs) identified by \citealt{Gudin_2021, Zepeda_2023, Shank_2023}. We find that only \mbox{J0246$-$5124} belongs to one of the 36 CDTGs identified by \citet{Shank_2023}, while the remaining stars do not appear in any of the groups analyzed in these studies. 

In summary, our dynamical analysis as mapped in Figure \ref{fig:cdtg_orbits}, indicates that the ten stars in our sample likely originate from ten distinct astrophysical environments, each tracing separate \emph{r}-process enrichment histories. Six of the stars (J0040+2729, J0217$-$1903, J0246$-$1518, J1430$-$2317, J1432$-$4125, J2009$-$3410)  lie above the dynamical boundary in the $E$--$L_z$ plane (left panel of Fig.~\ref{fig:cdtg_orbits}), consistent with an accreted origin, while three stars have prograde orbits and could either have formed in situ or been accreted, the last star have kinematics compatible with being a MW disk star. Working from the hypothesis that nine of our stars are accreted from now-disrupted dwarf galaxies, and one star formed in the MW we can use these stars to trace \emph{r}-process enrichment over a wide range of environments and estimate an upper limit on the scatter between them.

\section{Discussion}
\label{Discussion}
This work presents one of the largest homogeneously and comprehensively analyzed samples of $r$-process-enhanced stars to date. While most previous studies have been limited to small samples, often focusing on one or two stars \citep{sneden2003,  christlieb2004, aoki2010, Xylakis-Dornbusch2024A&A...688A.123X}, or relied on heterogeneous compilations from the literature, our study provides a systematic and self-consistent analysis of ten stars with robust abundance determinations. For each star, we derived abundances for over 50 species. Notably, this includes 29 neutron-capture elements, making our dataset uniquely suited to probe the detailed structure of the $r$-process abundance pattern.
In addition, our kinematic analysis (Section~\ref{dynamic}) shows the stars are not linked to the same accreted structure, suggesting that they each trace an independent enrichment history. This provides a valuable opportunity to investigate the diversity of $r$-process nucleosynthesis pathways across distinct progenitor systems.

\subsection{The \emph{r}-process element abundance pattern}
The complete \emph{r}-process element abundance patterns for the ten stars in our sample are presented in Figure~\ref{fig:abun_pattern}, separated into light ($30 \leq Z < 55$, left panel) and heavy ($55 \leq Z \leq 92$, right panel) $r$-process elements. The abundances for each star have been rescaled to the pattern of HD~222925 (grey dots and a solid line) by computing the average abundance offset with respect to this star \citep{roederer2018a, Roederer2022ApJ}, separately for light and heavy elements. We use HD~222925 as a reference, as this exhibits the most comprehensive $r$-process pattern observed to date. For comparison, the Solar $r$-process pattern from \citet{sneden2008} is shown as a grey dashed line. Residuals with respect to the average offset to HD~222925 for each star are shown in the lower panels. This re-scaling approach allows us to assess the scatter in the $r$-process element abundance pattern, with minimal influence from scatter induced by the analysis, enabling a clearer comparison across stars. The resulting trends and deviations will be discussed in detail in the following subsections.
Plots for the individual ten stars, each rescaled to the europium values of HD~222925, are shown in Figure \ref{fig:rescaled_Eu_10stars} in the Appendix.

\subsubsection{Light $r-$process elements}
\label{discussion:lightr}
The production of the lightest neutron capture (n-capture) elements in the \emph{r}-process has historically been seen to exhibit significant star-to-star variation when rescaled to Eu, leading to the view that their synthesis was not universal and potentially involved contributions from multiple astrophysical processes or sites \citep[e.g.,][]{Travaglio_2004,hansen2012,spite2018A&A...611A..30S}. This apparent lack of universality is contrasted with the relatively consistent patterns observed for the heavier lanthanides. However, recent work by \citet{Roederer2022ApJ} has challenged this picture by showing that several light trans-iron elements, including Se, Sr, Y, Nb, Mo, and Te, when scaled to Zr, follow a remarkably consistent abundance pattern in a sample of metal-poor, \emph{r}-process-enriched stars with a dispersion $\leq 0.13$ dex. These results suggest that, under certain astrophysical conditions, the production of light \emph{r}-process elements may also reflect a universal mechanism, potentially tied to a common nucleosynthetic site.\\
This picture of universality among light \emph{r}-process elements provides an important framework for interpreting our abundance measurements. In our sample, the Sr abundance presents a larger star-to-star variation compared to Y and Zr (see left panel of Figure~\ref{fig:abun_pattern}), for which the universality seems to hold.
However, Sr abundance was derived from the analysis of only three lines, compared to the other two elements for which over 20 clean transitions were investigated.

The limited number of Sr lines and their characteristics (See Section~\ref{light_r}) likely contribute to the larger observational spread.

The derived abundances of Mo, Ru, and Pd exhibit substantial star-to-star variation, as it is shown in Figure~\ref{fig:abun_pattern}. Such a spread is qualitatively consistent with the scenario proposed by \citet{Roederer2023}, in which the deposition of fission fragments contributes to the element-to-element abundance scatter in metal-poor stars. 
However, the reliability of these measurements is affected by the fact that they are based on weak and often blended lines located in the near-UV region (3400 - 4000~\AA). These observational challenges can introduce systematic uncertainties that may mimic or enhance real abundance variations. In particular, the Mo and Ru lines often lie in complex blended features, while the Pd lines are extremely weak and prone to contamination. As a result, it is difficult to determine whether the observed spread reflects true astrophysical scatter or is driven, at least in part, by measurement limitations. Alternative diagnostics would be necessary to disentangle observational effects from intrinsic abundance variations and to assess potential contributions from fission fragment yields in this mass region for this sample.

\subsubsection{Heavy $r-$process elements}
The heaviest \emph{r}-process elements, including the third \emph{r}-process peak and the actinides, provide essential constraints on the astrophysical conditions and robustness of the \emph{r}-process. Numerous studies have shown that, for a wide range of metal-poor stars enriched by the \emph{r}-process, the abundance pattern of heavy elements from the second peak ($Z\sim56$) to the third peak ($Z\sim76$)closely follows the scaled solar \emph{r}-process residuals \citep[e.g.,][]{Westin2000ApJ...530..783W, sneden2008}. This apparent invariance, often referred to as \textit{universality}, suggests that a single type of \emph{r}-process event (or at least a tightly constrained set of conditions) can robustly reproduce the heavy-element pattern over a wide range of progenitor environments. On the other end, the universality seems not to hold for the actinides, since some Milky Way halo stars show enhanced Th and U abundances relative to the rare earth elements \citep[the so-called "actinide boost";][]{Hill2002A&A...387..560H,Holmbeck2018ApJ...859L..24H, Placco2023APJ}. The other extreme in the shape of actinide deficiency has also been detected in one star in the \emph{r}-process enhanced ultra-faint dwarf galaxy Reticulum~II \citep{JiFrebel2018}. Therefore, examining the abundances of Pb, Th, and U in our sample offers a valuable test of the robustness and possible diversity of the \emph{r}-process at its heaviest end.

\begin{figure*}[hbt!]
\includegraphics[width=1.\linewidth]{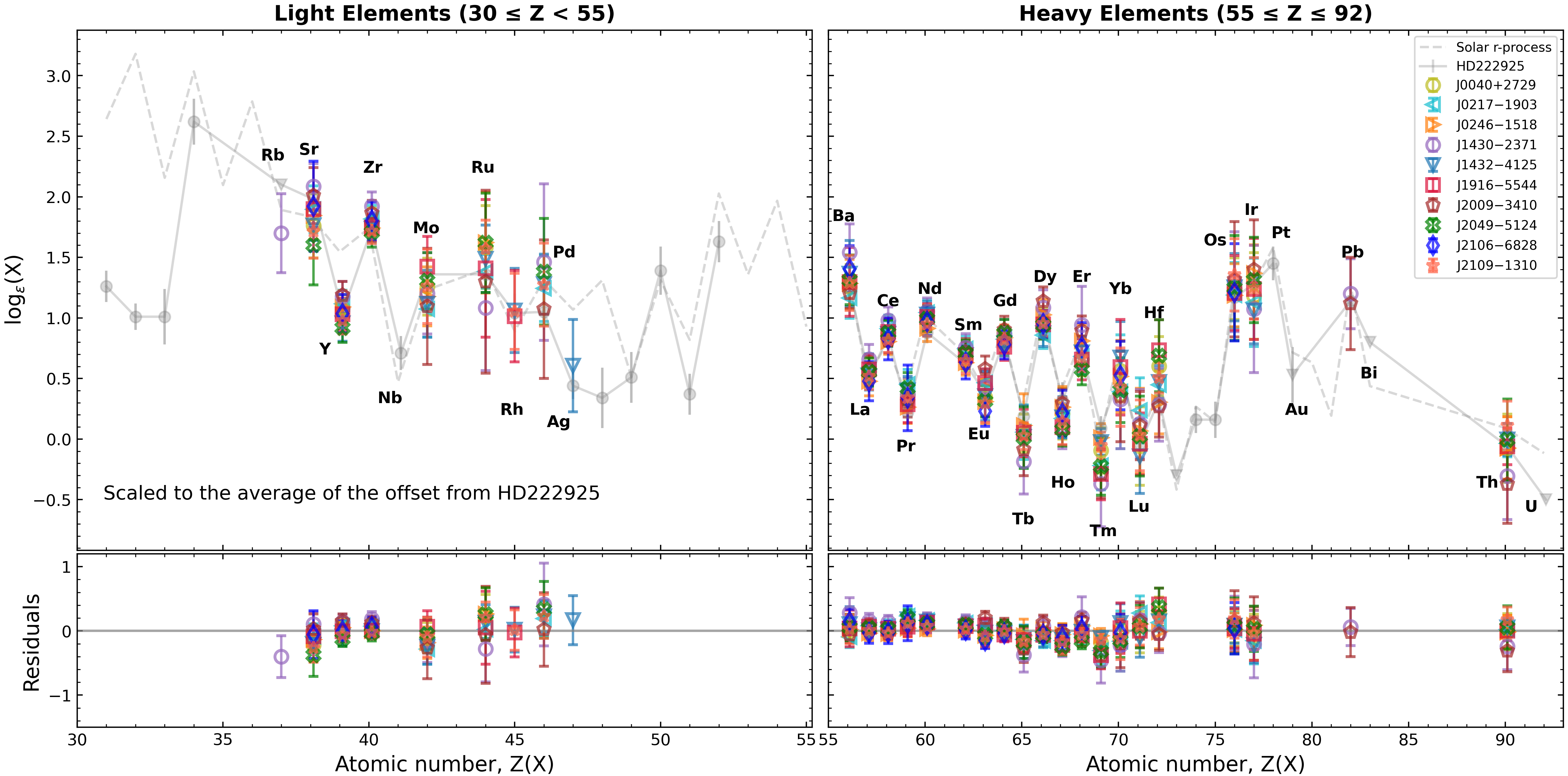}
\caption{Abundance patterns of light ($30\leq Z < 55$, left panel) and heavy neutron-capture elements ($55\leq Z\leq92$, right panel). The colored points represent derived abundances for the ten sample stars. The abundances for each star are rescaled by subtracting the mean offset from HD~222925 (\citealt{roederer2018a}). The pattern of HD~222925 is represented by the grey solid line and dots, upper limits are indicated with grey downwards triangles. The lower panel displays the residuals with respect to the reference star HD~222925. The dashed grey line is the solar \emph{r}-process pattern by \cite{sneden2008}.
\label{fig:abun_pattern}}
\end{figure*}

Across our sample, the abundances of elements from Ba to Er are consistently well determined and exhibit relatively small star-to-star scatter (See right panel in Figure~\ref{fig:abun_pattern}). This behavior aligns with the previous findings, where elements beyond the second \emph{r}-process peak are known to follow tight abundance trends, particularly in the highly \emph{r}-process-enhanced stars. These elements are typically associated with strong, relatively unblended absorption lines in the optical range, which facilitates robust abundance determination, in particular in high SNR spectra.
However, in the same plot (Figure~\ref{fig:abun_pattern}), is shown how, at higher atomic numbers ($Z \geq 68)$, the observed scatter increases, likely due to a combination of weaker lines and more limited line availability. 

Moving to the third peak, Os and Ir abundances can offer insight into the neutron density and entropy of the $r$-process environment, as well as potential signatures of fission recycling \citep{AlencastroPuls2025}. Despite their importance, Os and Ir abundances are notoriously difficult to derive in metal-poor stars due to the weakness of their lines, which are typically located in the crowded blue-UV region of the spectrum and often blended. For this reason, not many abundances of those elements were reported in literature before the recent work from \citet{AlencastroPuls2025} who published 33 \ion{Os}{i} abundances plus five upper limit values and 32 \ion{Ir}{i} plus eight upper limits. In our sample, Os and Ir show a larger dispersion, compared to the lanthanides, again likely due to the analysis of a single line as mentioned in Section~\ref{ThirdPeak}.

Finally, the actinide element Th displays remarkable consistency across stars in our sample (Figure~\ref{fig:abun_pattern}), with no significant star-to-star dispersion, except for two clear outliers: J1430$-$2371 and J2009$-$3410. These two stars exhibit significantly lower Th abundances, with a rescaled $\mathrm{\log{\epsilon}}$(Th/average of the offset from HD~222925) ratios of $-0.30$ and $-0.37$, respectively, compared to the average value of $-0.05$ of the other eight stars. This suggests a potential case of actinide depletion in these two stars, similar to what is observed in the Reticulum II star, contrasting with the otherwise uniform behavior of Th relative to Eu in the rest of the sample. However, we note that these two stars have the lowest effective temperatures in the sample. We therefore recommend caution in the interpretation of these abundances. 
\\

\subsection{Cosmic Abundance Variations Across Independent $r$-Process Sites}
\label{CosmicVariation}
As stated in Section~\ref{dynamic}, the ten stars analyzed in this study are chemically and dynamically untagged, and for the following further discussion, we therefore assume them to represent ten%. Among them, six stars were likely accreted from ultrafaint dwarfs or small system and probably enriched by a single site of \emph{r-}process production. Under this assumption, the star-to-star dispersion in elemental abundances offers a direct probe of the maximum possible cosmic variation between six}  
distinct $r$-process events. To provide a conservative upper limit, we assume that the measured abundance spread for each element arises entirely from astrophysical origins, neglecting any contribution from observational uncertainties. We then computed the standard deviation of the abundance of each $r$-process element across the ten stars, after normalizing the light ($Z < 55$) and heavy ($Z \geq 55$) elements to Zr and Eu, respectively. 
The results for elements detected in at least three stars are shown in Figure~\ref{fig:sigma_cosmic}, where the standard deviations $\sigma[\mathrm{X/Zr}]$ and $\sigma[\mathrm{X/Eu}]$ are plotted as a function of atomic number, and color-coded by the average number of lines used for abundance determinations. We also plot the standard deviation of the $\sigma_{[\mathrm{Zr/Eu}]}$ ratio itself, as a red horizontal line in the plot. This serves as a benchmark for the total variation between light and heavy $r$-process element production across the ten \emph{r}-process events. The numerical values of the standard deviations of all the elements are reported in Table~\ref{tab:variationssigma}. For most elements, the spread is remarkably small: we find $\sigma \approx 0.04-0.09$ dex (corresponding to fractional variations $<18\%$) for the light $r$-process elements, and $\sigma \approx 0.08-0.12$ dex ($<24\%$) for the heavier elements. A few notable exceptions exist, including Ru, Tm, Yb, and Hf, where the dispersion exceeds $\sigma_{[\mathrm{Zr/Eu}]}$. However, the scatter observed for these elements among the ten stars is most likely driven by observational limitations rather than genuine astrophysical variations. For these elements, we inspected up to six spectral lines, but in most cases, only three or fewer could be reliably analyzed. This is considerably less than what is typically available for other lanthanides, as also illustrated in Table~\ref{tab:Abundances}. As a result, while the line-to-line scatter within individual stars remains small (as it is clear from Table~\ref{tab:linelist}), the abundance uncertainty $\sigma$, in Table~\ref{tab:Abundances}, becomes larger whenever the determination relies on only few lines. A similar situation applies to Lu, Os, and Ir: although their dispersions across the stars appear smaller, the robustness of their abundances is likewise limited by the very small number of available lines. \\
The larger variation in the $\sigma_{[\mathrm{Zr/Eu}]}$ value is also expected, as this represents the scatter detected in numerous previous studies (see Sec \ref{discussion:lightr}). This variation in the ratio of the light to heavy \emph{r}-process elements can be directly compared to the lanthanide fraction of the \emph{r}-process element production site. For example, the ejecta of the GW170817 kilonova event \citep{Kasen2017} required a multi-component model including lanthanide-poor and lanthanide-rich material to explain both the blue and red parts of the light curve. The lanthanide fraction for this event was later compared to \emph{r}-process enriched stars by \cite{Ji_2019}, who found that this event did not produce enough lanthanide-rich material to match the stars. The $\sigma_{[\mathrm{Zr/Eu}]}$ value of 0.18 dex found for our ten stars thus provides a range of lanthanide fractions.
Since all ten stars in the sample presented here are $r$-process enhanced (to varying degrees, as reflected by their $\mathrm{[Eu/Fe]}$ and $\mathrm{[Ba/Eu]}$ ratios) and did not originate from one common environment, we interpret the measured $\sigma_{\rm cosmic}$ values as the maximum allowed variation in abundance ratios produced by potentially up to ten independent $r$-process enrichment paths operating in the distinct environments where these stars formed.
 
This means that, regardless of the specific astrophysical origin, whether NSMs, magnetorotational supernovae, collapsars, magnetars, or other channels, any viable $r$-process site must reproduce element ratios consistent with the $\sigma_{\rm cosmic}$ constraints established here. In this sense, our measurements provide robust empirical bounds on the diversity of $r$-process nucleosynthesis pathways. 

In order to investigate the potential effect of NLTE, we calculated the scatter among stars for which NLTE-corrected values for Sr, Ba, and Eu could be determined. The scatter with NLTE-corrected abundances are $\sigma_{[Sr/Zr],\rm NLTE} = 0.11$ and $\sigma_{[Ba/Eu],\rm NLTE} = 0.05$. For comparison, the scatter in LTE abundances for the same six stars are $\sigma_{[Sr/Zr],\rm LTE} = 0.06$ and $\sigma_{[Ba/Eu],\rm LTE} = 0.06$. While the comparison is not fully straightforward because only Sr was corrected in the [Sr/Zr] ratio, these results indicate that NLTE corrections tend to slightly reduce the scatter in [Ba/Eu] and increase it in [Sr/Zr], reflecting the differential impact of NLTE effects on different elements (see Table~\ref{tab:abundances_corrected}2).

\begin{figure*}
\hspace*{-0.0\linewidth}
\includegraphics[width=1.\linewidth]{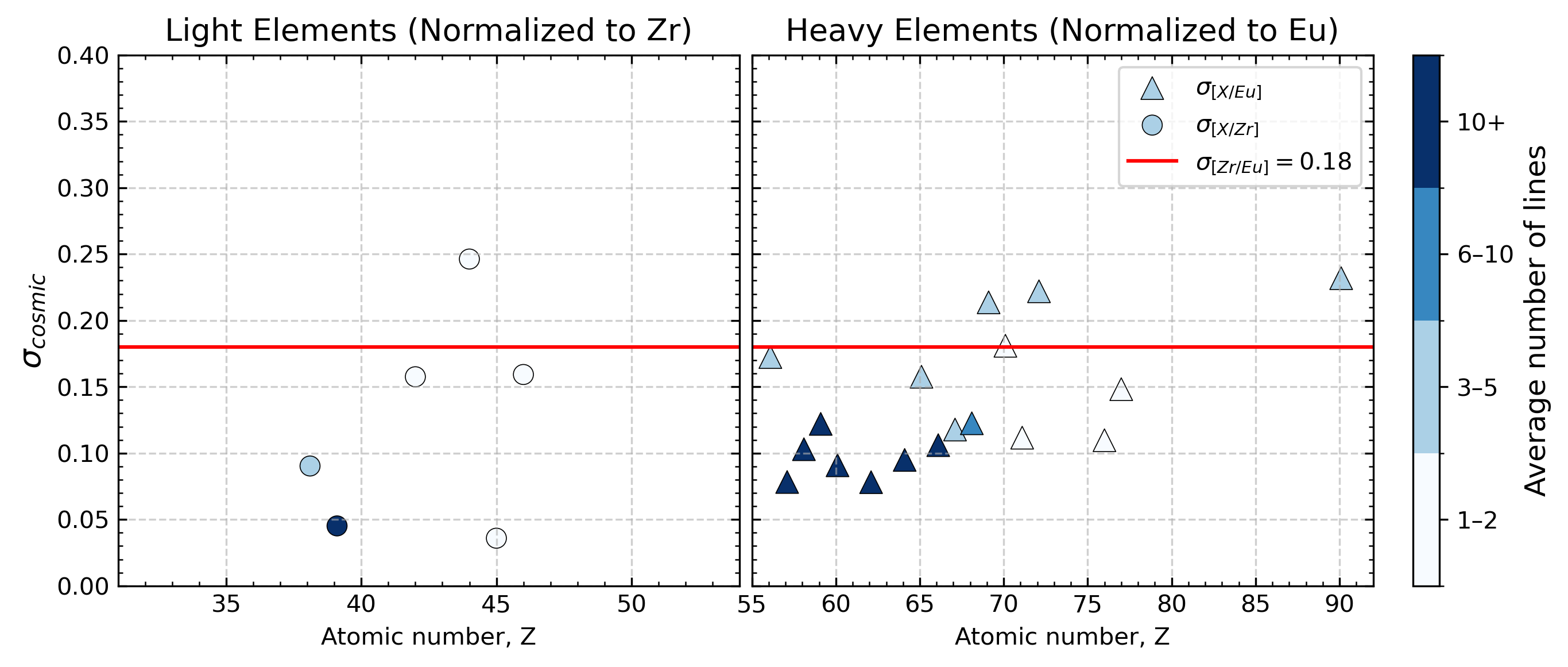}
\caption{
Cosmic standard deviation (\(\sigma_{\text{cosmic}}\)) of \(r\)-process element abundances across the stellar sample. 
\textbf{Left:} light \(r\)-process elements (normalized to Zr, shown as circles). 
\textbf{Right:} heavy \(r\)-process elements (normalized to Eu, shown as triangles). 
Only elements measured in at least three stars are included. 
Marker colors indicate the average number of spectral lines used for the abundance determination. 
The red horizontal line marks the standard deviation of \([\mathrm{Zr/Eu}]\), illustrating the overall variation between the light and heavy \(r\)-process components.
}

\end{figure*} 
\begin{table}[ht!]
\centering
\caption{Dispersion of light and heavy \emph{r}-process element abundances among ten stars.}
\label{fig:sigma_cosmic}
\begin{tabular}{lccc}
\hline
Element & Number of & Average number & $\sigma_{cosmic}$  \\
 & detections &  of lines & [dex]\\
\hline
Sr  & 10 & 2  & 0.09 \\
Y   & 10 & 20 & 0.05 \\
Zr  & 10 & 16 & --   \\
Mo  & 9  & 1  & 0.16 \\
Ru  & 9  & 1  & 0.25 \\
Rh  & 3  & 1  & 0.04 \\
Pd  & 6  & 1  & 0.16 \\
\hline
Ba  & 10 & 4  & 0.17 \\
La  & 10 & 26 & 0.08 \\
Ce  & 10 & 32 & 0.10 \\
Pr  & 10 & 10 & 0.12 \\
Nd  & 10 & 57 & 0.09 \\
Sm  & 10 & 48 & 0.08 \\
Eu  & 10 & 7  & --   \\
Gd  & 10 & 14 & 0.10 \\
Tb  & 10 & 2  & 0.16 \\
Dy  & 10 & 13 & 0.11 \\
Ho  & 10 & 4  & 0.12 \\
Er  & 10 & 6  & 0.12 \\
Tm  & 9  & 4  & 0.21 \\
Yb  & 10 & 1  & 0.18 \\
Lu  & 9  & 1  & 0.11 \\
Hf  & 9  & 2  & 0.22 \\
Os  & 10 & 1  & 0.11 \\
Ir  & 9  & 1  & 0.15 \\
Pb  & 2  & 1  & 0.16 \\
Th  & 9  & 2  & 0.23 \\

\hline
\end{tabular}
\tablefoot{Standard deviation ($\sigma$) of light ($Z \lesssim 50$) and heavy ($Z \gtrsim 56$) r-process element abundances. 
The second and the third columns contain the number of stars in which each element was detected and the average number of analyzed spectral lines per element, respectively. 
The fourth column abundance dispersion, expressed as $\mathrm{[X/Zr]}$ for light elements and $\mathrm{[X/Eu]}$ for heavy elements.
}
\label{tab:variationssigma}
\end{table}

\section{Summary}
\label{Summary}
In this work, we present a homogeneous and comprehensive chemical abundance analysis of a sample of ten $r$-process-enhanced stars. These stars are characterized by strong enhancements in $r$-process elements and no significant $s$-process contribution, having a $+0.39 \leq\mathrm{[Eu/Fe]} \leq +1.06$ and $ -1.01 \leq \mathrm{[Ba/Fe]} \leq -0.52$. For each star, we analyzed over 1400 spectral lines, deriving abundances for more than 50 species, including 29 neutron-capture elements from Rb to Th. The derived abundance patterns display the well-known universality for the heavy $r$-process elements (Ba to Ir), with only mild star-to-star variations. A slightly larger scatter is observed for Tm, Yb, and Hf, likely driven by increased measurement uncertainties due to line blending and continuum placement in the blue spectral region. A similar small scatter is seen for the light $r$-process elements in the range $38 \leq Z \leq 42$, while the elements in the range $ 43 < Z \lesssim 50$ exhibit a somewhat larger dispersion across the sample, possibly in accordance with the fission fragment deposition scenario, proposed by \citet{Roederer2023}. 

Kinematic analysis of the stars suggests they are not linked to a common birth environment but originated in distinct, unrelated progenitor systems. This sample thus offers valuable constraints on the diversity of ten individual astrophysical sites contributing to $r$-process enrichment. To quantify this, we calculated the cosmic standard deviation ($\sigma_{cosmic}$) for each element among the ten stars, representing ten different $r$-process nucleosynthesis sites. These values provide an upper limit to the possible variation introduced by different $r$-process nucleosynthesis environments across ten independent stellar sites and are remarkably small for the rare earth and third peak elements, for example, $\sigma_{[\mathrm{La/Eu}]} = 0.08$ dex and $\sigma_{[\mathrm{Os/Eu}]} = 0.11$ dex. A somewhat larger scatter of 0.18 dex is seen between the light and heavy parts of the $r$-process pattern ($\sigma_{[\mathrm{Zr/Eu}]}$) representing the ratio between lanthanide-rich and lanthanide-poor ejecta for the ten enrichment paths.

\begin{acknowledgements}
      The authors thank the referee for the insightful comments that helped improve the clarity of the paper. The authors also thank Alexander P. Ji for assistance with the observations.
      M.R. acknowledges support from the International Research Network for Nuclear Astrophysics (IReNA) through a Visiting Fellowship, which funded a research visit to North Carolina State University (U.S.) to collaborate with I.U.R. M.R. and T.T.H. acknowledge support from the Swedish Research Council (VR 2021-05556). 
      I.U.R. acknowledges support from the US National Science Foundation (NSF) grant AST~2205847.
      A.F. acknowledges support from NSF-AAG grant AST-2307436. 
      T.C.B. acknowledges partial support from grants PHY 14-30152; Physics Frontier Center/JINA Center for the Evolution of the Elements (JINA-CEE), and OISE-1927130; The International Research Network for Nuclear Astrophysics (IReNA), awarded by the US National Science Foundation, and DE-SC0023128CeNAM; the Center for Nuclear Astrophysics Across Messengers (CeNAM), awarded by the U.S. Department of Energy, Office of Science, Office of Nuclear Physics.
      E.M.H. acknowledges work performed under the auspices of the U.S. Department of Energy by Lawrence Livermore National Laboratory under Contract DE-AC52-07NA27344.
      This document has been approved for release under LLNL-JRNL-2008756.
      The work of V.M.P. is supported by NOIRLab, which is managed by the Association of Universities for Research in Astronomy (AURA) under a cooperative agreement with the U.S. National Science Foundation
      
\end{acknowledgements}

\bibliographystyle{aa_url}
\bibliography{bib}

\begin{thebibliography}{173}
\expandafter\ifx\csname natexlab\endcsname\relax\def\natexlab#1{#1}\fi

\bibitem[{Abbott {et~al.}(2019)Abbott, Abbott, Abbott, Abraham, Acernese, Ackley, Adams, Adhikari, Adya, Affeldt, Agathos, Agatsuma, Aggarwal, Aguiar, Aiello, Ain, Ajith, Allen, Allocca, Aloy, Altin, Amato, Ananyeva, Anderson, Anderson, Angelova, Antier, Appert, Arai, Araya, Areeda, Arène, Arnaud, Ascenzi, Ashton, Aston, Astone, Aubin, Aufmuth, AultONeal, Austin, Avendano, Avila-Alvarez, Babak, Bacon, Badaracco, Bader, Bae, Baker, Baldaccini, Ballardin, Ballmer, Banagiri, Barayoga, Barclay, Barish, Barker, Barkett, Barnum, Barone, Barr, Barsotti, Barsuglia, Barta, Bartlett, Bartos, Bassiri, Basti, Bawaj, Bayley, Bazzan, Bécsy, Bejger, Belahcene, Bell, Beniwal, Berger, Bergmann, Bernuzzi, Bero, Berry, Bersanetti, Bertolini, Betzwieser, Bhandare, Bidler, Bilenko, Bilgili, Billingsley, Birch, Birney, Birnholtz, Biscans, Biscoveanu, Bisht, Bitossi, Bizouard, Blackburn, Blair, Blair, Blair, Bloemen, Bode, Boer, Boetzel, Bogaert, Bondu, Bonilla, Bonnand, Booker, Boom, Booth, Bork, Boschi, Bose, Bossie, Bossilkov,
  Bosveld, Bouffanais, Bozzi, Bradaschia, Brady, Bramley, Branchesi, Brau, Briant, Briggs, Brighenti, Brillet, Brinkmann, Brisson, Brockill, Brooks, Brown, Brunett, Buikema, Bulik, Bulten, Buonanno, Buskulic, Buy, Byer, Cabero, Cadonati, Cagnoli, Cahillane, Bustillo, Callister, Calloni, Camp, Campbell, Canepa, Cannon, Cao, Cao, Capocasa, Carbognani, Caride, Carney, Carullo, Diaz, Casentini, Caudill, Cavaglià, Cavalier, Cavalieri, Cella, Cerdá-Durán, Cerretani, Cesarini, Chaibi, Chakravarti, Chamberlin, Chan, Chao, Charlton, Chase, Chassande-Mottin, Chatterjee, Chaturvedi, Cheeseboro, Chen, Chen, Chen, Cheng, Cheong, Chia, Chincarini, Chiummo, Cho, Cho, Cho, Christensen, Chu, Chua, Chung, Chung, Ciani, Ciobanu, Ciolfi, Cipriano, Cirone, Clara, Clark, Clearwater, Cleva, Cocchieri, Coccia, Cohadon, Cohen, Colgan, Colleoni, Collette, Collins, Cominsky, Constancio, Conti, Cooper, Corban, Corbitt, Cordero-Carrión, Corley, Cornish, Corsi, Cortese, Costa, Cotesta, Coughlin, Coughlin, Coulon, Countryman, Couvares,
  Covas, Cowan, Coward, Cowart, Coyne, Coyne, Creighton, Creighton, Cripe, Croquette, Crowder, Cullen, Cumming, Cunningham, Cuoco, Canton, Dálya, Danilishin, D’Antonio, Danzmann, Dasgupta, Da~Silva~Costa, Datrier, Dattilo, Dave, Davier, Davis, Daw, DeBra, Deenadayalan, Degallaix, De~Laurentis, Deléglise, Del~Pozzo, DeMarchi, Demos, Dent, De~Pietri, Derby, De~Rosa, De~Rossi, DeSalvo, de~Varona, Dhurandhar, Díaz, Dietrich, Di~Fiore, Di~Giovanni, Di~Girolamo, Di~Lieto, Ding, Di~Pace, Di~Palma, Di~Renzo, Dmitriev, Doctor, Donovan, Dooley, Doravari, Dorrington, Downes, Drago, Driggers, Du, Ducoin, Dupej, Dwyer, Easter, Edo, Edwards, Effler, Ehrens, Eichholz, Eikenberry, Eisenmann, Eisenstein, Essick, Estelles, Estevez, Etienne, Etzel, Evans, Evans, Fafone, Fair, Fairhurst, Fan, Farinon, Farr, Farr, Fauchon-Jones, Favata, Fays, Fazio, Fee, Feicht, Fejer, Feng, Fernandez-Galiana, Ferrante, Ferreira, Ferreira, Ferrini, Fidecaro, Fiori, Fiorucci, Fishbach, Fisher, Fishner, Fitz-Axen, Flaminio, Fletcher, Flynn,
  Fong, Font, Forsyth, Fournier, Frasca, Frasconi, Frei, Freise, Frey, Frey, Fritschel, Frolov, Fulda, Fyffe, Gabbard, Gadre, Gaebel, Gair, Gammaitoni, Ganija, Gaonkar, Garcia, García-Quirós, Garufi, Gateley, Gaudio, Gaur, Gayathri, Gemme, Genin, Gennai, George, George, Gergely, Germain, Ghonge, Ghosh, Ghosh, Ghosh, Giacomazzo, Giaime, Giardina, Giazotto, Gill, Giordano, Glover, Godwin, Goetz, Goetz, Goncharov, González, Gonzalez~Castro, Gopakumar, Gorodetsky, Gossan, Gosselin, Gouaty, Grado, Graef, Granata, Grant, Gras, Grassia, Gray, Gray, Greco, Green, Green, Gretarsson, Groot, Grote, Grunewald, Gruning, Guidi, Gulati, Guo, Gupta, Gupta, Gustafson, Gustafson, Haegel, Halim, Hall, Hall, Hamilton, Hammond, Haney, Hanke, Hanks, Hanna, Hannuksela, Hanson, Hardwick, Haris, Harms, Harry, Harry, Haster, Haughian, Hayes, Healy, Heidmann, Heintze, Heitmann, Hello, Hemming, Hendry, Heng, Hennig, Heptonstall, Vivanco, Heurs, Hild, Hinderer, Hoak, Hochheim, Hofman, Holgado, Holland, Holt, Holz, Hopkins, Horst,
  Hough, Howell, Hoy, Hreibi, Huerta, Huet, Hughey, Hulko, Husa, Huttner, Huynh-Dinh, Idzkowski, Iess, Ingram, Inta, Intini, Irwin, Isa, Isac, Isi, Iyer, Izumi, Jacqmin, Jadhav, Jani, Janthalur, Jaranowski, Jenkins, Jiang, Johnson, Jones, Jones, Jones, Jonker, Ju, Junker, Kalaghatgi, Kalogera, Kamai, Kandhasamy, Kang, Kanner, Kapadia, Karki, Karvinen, Kashyap, Kasprzack, Katsanevas, Katsavounidis, Katzman, Kaufer, Kawabe, Keerthana, Kéfélian, Keitel, Kennedy, Key, Khalili, Khan, Khan, Khan, Khan, Khazanov, Khursheed, Kijbunchoo, Kim, Kim, Kim, Kim, Kim, Kim, Kimball, King, King, Kinley-Hanlon, Kirchhoff, Kissel, Kleybolte, Klika, Klimenko, Knowles, Koch, Koehlenbeck, Koekoek, Koley, Kondrashov, Kontos, Koper, Korobko, Korth, Kowalska, Kozak, Kringel, Krishnendu, Królak, Kuehn, Kumar, Kumar, Kumar, Kumar, Kuo, Kutynia, Kwang, Lackey, Lai, Lam, Landry, Lane, Lang, Lange, Lantz, Lanza, Lartaux-Vollard, Lasky, Laxen, Lazzarini, Lazzaro, Leaci, Leavey, Lecoeuche, Lee, Lee, Lee, Lee, Lee, Lee, Lehmann, Lenon,
  Leroy, Letendre, Levin, Li, Li, Li, Li, Lin, Linde, Linker, Littenberg, Liu, Liu, Lo, Lockerbie, London, Longo, Lorenzini, Loriette, Lormand, Losurdo, Lough, Lousto, Lovelace, Lower, Lück, Lumaca, Lundgren, Lynch, Ma, Macas, Macfoy, MacInnis, Macleod, Macquet, Magaña-Sandoval, Zertuche, Magee, Majorana, Maksimovic, Malik, Man, Mandic, Mangano, Mansell, Manske, Mantovani, Marchesoni, Marion, Márka, Márka, Markakis, Markosyan, Markowitz, Maros, Marquina, Marsat, Martelli, Martin, Martin, Martynov, Mason, Massera, Masserot, Massinger, Masso-Reid, Mastrogiovanni, Matas, Matichard, Matone, Mavalvala, Mazumder, McCann, McCarthy, McClelland, McCormick, McCuller, McGuire, McIver, McManus, McRae, McWilliams, Meacher, Meadors, Mehmet, Mehta, Meidam, Melatos, Mendell, Mercer, Mereni, Merilh, Merzougui, Meshkov, Messenger, Messick, Metzdorff, Meyers, Miao, Michel, Middleton, Mikhailov, Milano, Miller, Miller, Millhouse, Mills, Milovich-Goff, Minazzoli, Minenkov, Mishkin, Mishra, Mistry, Mitra, Mitrofanov,
  Mitselmakher, Mittleman, Mo, Moffa, Mogushi, Mohapatra, Montani, Moore, Moraru, Moreno, Morisaki, Mours, Mow-Lowry, Mukherjee, Mukherjee, Mukherjee, Mukund, Mullavey, Munch, Muñiz, Muratore, Murray, Nagar, Nardecchia, Naticchioni, Nayak, Neilson, Nelemans, Nelson, Nery, Neunzert, Ng, Ng, Nguyen, Nichols, Nissanke, Nocera, North, Nuttall, Obergaulinger, Oberling, O’Brien, O’Dea, Ogin, Oh, Oh, Ohme, Ohta, Okada, Oliver, Oppermann, Oram, O’Reilly, Ormiston, Ortega, O’Shaughnessy, Ossokine, Ottaway, Overmier, Owen, Pace, Pagano, Page, Pai, Pai, Palamos, Palashov, Palomba, Pal-Singh, Pan, Pang, Pang, Pankow, Pannarale, Pant, Paoletti, Paoli, Parida, Parker, Pascucci, Pasqualetti, Passaquieti, Passuello, Patil, Patricelli, Pearlstone, Pedersen, Pedraza, Pedurand, Pele, Penn, Perez, Perreca, Pfeiffer, Phelps, Phukon, Piccinni, Pichot, Piergiovanni, Pillant, Pinard, Pirello, Pitkin, Poggiani, Pong, Ponrathnam, Popolizio, Porter, Powell, Prajapati, Prasad, Prasai, Prasanna, Pratten, Prestegard, Privitera,
  Prodi, Prokhorov, Puncken, Punturo, Puppo, Pürrer, Qi, Quetschke, Quinonez, Quintero, Quitzow-James, Raab, Radkins, Radulescu, Raffai, Raja, Rajan, Rajbhandari, Rakhmanov, Ramirez, Ramos-Buades, Rana, Rao, Rapagnani, Raymond, Razzano, Read, Regimbau, Rei, Reid, Reitze, Ren, Ricci, Richardson, Richardson, Ricker, Riles, Rizzo, Robertson, Robie, Robinet, Rocchi, Rolland, Rollins, Roma, Romanelli, Romano, Romel, Romie, Rose, Rosińska, Rosofsky, Ross, Rowan, Rüdiger, Ruggi, Rutins, Ryan, Sachdev, Sadecki, Sakellariadou, Salconi, Saleem, Samajdar, Sammut, Sanchez, Sanchez, Sanchis-Gual, Sandberg, Sanders, Santiago, Sarin, Sassolas, Saulson, Sauter, Savage, Schale, Scheel, Scheuer, Schmidt, Schnabel, Schofield, Schönbeck, Schreiber, Schulte, Schutz, Schwalbe, Scott, Scott, Seidel, Sellers, Sengupta, Sennett, Sentenac, Sequino, Sergeev, Setyawati, Shaddock, Shaffer, Shahriar, Shaner, Shao, Sharma, Shawhan, Shen, Shink, Shoemaker, Shoemaker, ShyamSundar, Siellez, Sieniawska, Sigg, Silva, Singer, Singh, Singhal,
  Sintes, Sitmukhambetov, Skliris, Slagmolen, Slaven-Blair, Smith, Smith, Somala, Son, Sorazu, Sorrentino, Souradeep, Sowell, Spencer, Srivastava, Srivastava, Staats, Stachie, Standke, Steer, Steinke, Steinlechner, Steinlechner, Steinmeyer, Stevenson, Stocks, Stone, Stops, Strain, Stratta, Strigin, Strunk, Sturani, Stuver, Sudhir, Summerscales, Sun, Sunil, Suresh, Sutton, Swinkels, Szczepańczyk, Tacca, Tait, Talbot, Talukder, Tanner, Tápai, Taracchini, Tasson, Taylor, Thies, Thomas, Thomas, Thondapu, Thorne, Thrane, Tiwari, Tiwari, Tiwari, Toland, Tonelli, Tornasi, Torres-Forné, Torrie, Töyrä, Travasso, Traylor, Tringali, Trovato, Trozzo, Trudeau, Tsang, Tse, Tso, Tsukada, Tsuna, Tuyenbayev, Ueno, Ugolini, Unnikrishnan, Urban, Usman, Vahlbruch, Vajente, Valdes, van Bakel, van Beuzekom, van~den Brand, Van Den~Broeck, Vander-Hyde, van Heijningen, van~der Schaaf, van Veggel, Vardaro, Varma, Vass, Vasúth, Vecchio, Vedovato, Veitch, Veitch, Venkateswara, Venugopalan, Verkindt, Vetrano, Viceré, Viets, Vine,
  Vinet, Vitale, Vo, Vocca, Vorvick, Vyatchanin, Wade, Wade, Wade, Walet, Walker, Wallace, Walsh, Wang, Wang, Wang, Wang, Wang, Ward, Warden, Warner, Was, Watchi, Weaver, Wei, Weinert, Weinstein, Weiss, Wellmann, Wen, Wessel, Weßels, Westhouse, Wette, Whelan, Whiting, Whittle, Wilken, Williams, Williamson, Willis, Willke, Wimmer, Winkler, Wipf, Wittel, Woan, Woehler, Wofford, Worden, Wright, Wu, Wysocki, Xiao, Yamamoto, Yancey, Yang, Yap, Yazback, Yeeles, Yu, Yu, Yuen, Yvert, Zadrożny, Zanolin, Zelenova, Zendri, Zevin, Zhang, Zhang, Zhang, Zhao, Zhou, Zhou, Zhu, Zucker, \& Zweizig}]{Abbott_2019}
Abbott, B.~P., Abbott, R., Abbott, T.~D., {et~al.} 2019, \href{http://dx.doi.org/10.3847/1538-4357/ab0e15}{\color{blue}APJ}, 874, 163

\bibitem[{Abbott {et~al.}(2017)Abbott, Abbott, Abbott, Acernese, Ackley, Adams, Adams, Addesso, Adhikari, Adya, Affeldt, Afrough, Agarwal, Agathos, Agatsuma, Aggarwal, Aguiar, Aiello, Ain, Ajith, Allen, Allen, Allocca, Aloy, Altin, Amato, Ananyeva, Anderson, Anderson, Angelova, Antier, Appert, Arai, Araya, Areeda, Arnaud, Arun, Ascenzi, Ashton, Ast, Aston, Astone, Atallah, Aufmuth, Aulbert, AultONeal, Austin, Avila-Alvarez, Babak, Bacon, Bader, Bae, Baker, Baldaccini, Ballardin, Ballmer, Banagiri, Barayoga, Barclay, Barish, Barker, Barkett, Barone, Barr, Barsotti, Barsuglia, Barta, Bartlett, Bartos, Bassiri, Basti, Batch, Bawaj, Bayley, Bazzan, Bécsy, Beer, Bejger, Belahcene, Bell, Berger, Bergmann, Bero, Berry, Bersanetti, Bertolini, Betzwieser, Bhagwat, Bhandare, Bilenko, Billingsley, Billman, Birch, Birney, Birnholtz, Biscans, Biscoveanu, Bisht, Bitossi, Biwer, Bizouard, Blackburn, Blackman, Blair, Blair, Blair, Bloemen, Bock, Bode, Boer, Bogaert, Bohe, Bondu, Bonilla, Bonnand, Boom, Bork, Boschi, Bose,
  Bossie, Bouffanais, Bozzi, Bradaschia, Brady, Branchesi, Brau, Briant, Brillet, Brinkmann, Brisson, Brockill, Broida, Brooks, Brown, Brown, Brunett, Buchanan, Buikema, Bulik, Bulten, Buonanno, Buskulic, Buy, Byer, Cabero, Cadonati, Cagnoli, Cahillane, Calderón~Bustillo, Callister, Calloni, Camp, Canepa, Canizares, Cannon, Cao, Cao, Capano, Capocasa, Carbognani, Caride, Carney, Diaz, Casentini, Caudill, Cavaglià, Cavalier, Cavalieri, Cella, Cepeda, Cerdá-Durán, Cerretani, Cesarini, Chamberlin, Chan, Chao, Charlton, Chase, Chassande-Mottin, Chatterjee, Chatziioannou, Cheeseboro, Chen, Chen, Chen, Cheng, Chia, Chincarini, Chiummo, Chmiel, Cho, Cho, Chow, Christensen, Chu, Chua, Chua, Chung, Chung, Ciani, Ciolfi, Cirelli, Cirone, Clara, Clark, Clearwater, Cleva, Cocchieri, Coccia, Cohadon, Cohen, Colla, Collette, Cominsky, Constancio~Jr., Conti, Cooper, Corban, Corbitt, Cordero-Carrión, Corley, Cornish, Corsi, Cortese, Costa, Coughlin, Coughlin, Coulon, Countryman, Couvares, Covas, Cowan, Coward, Cowart,
  Coyne, Coyne, Creighton, Creighton, Cripe, Crowder, Cullen, Cumming, Cunningham, Cuoco, Canton, Dálya, Danilishin, D’Antonio, Danzmann, Dasgupta, Costa, Dattilo, Dave, Davier, Davis, Daw, Day, De, DeBra, Degallaix, Laurentis, Deléglise, Pozzo, Demos, Denker, Dent, Pietri, Dergachev, Rosa, DeRosa, Rossi, DeSalvo, Varona, Devenson, Dhurandhar, Díaz, Fiore, Giovanni, Girolamo, Lieto, Pace, Palma, Renzo, Doctor, Dolique, Donovan, Dooley, Doravari, Dorrington, Douglas, Dovale~Álvarez, Downes, Drago, Dreissigacker, Driggers, Du, Ducrot, Dupej, Dwyer, Edo, Edwards, Effler, Eggenstein, Ehrens, Eichholz, Eikenberry, Eisenstein, Essick, Estevez, Etienne, Etzel, Evans, Evans, Factourovich, Fafone, Fair, Fairhurst, Fan, Farinon, Farr, Farr, Fauchon-Jones, Favata, Fays, Fee, Fehrmann, Feicht, Fejer, Fernandez-Galiana, Ferrante, Ferreira, Ferrini, Fidecaro, Finstad, Fiori, Fiorucci, Fishbach, Fisher, Fitz-Axen, Flaminio, Fletcher, Fong, Font, Forsyth, Forsyth, Fournier, Frasca, Frasconi, Frei, Freise, Frey, Frey,
  Fries, Fritschel, Frolov, Fulda, Fyffe, Gabbard, Gadre, Gaebel, Gair, Gammaitoni, Ganija, Gaonkar, Garcia-Quiros, Garufi, Gateley, Gaudio, Gaur, Gayathri, Gehrels, Gemme, Genin, Gennai, George, George, Gergely, Germain, Ghonge, Ghosh, Ghosh, Ghosh, Giaime, Giardina, Giazotto, Gill, Glover, Goetz, Goetz, Gomes, Goncharov, González, Castro, Gopakumar, Gorodetsky, Gossan, Gosselin, Gouaty, Grado, Graef, Granata, Grant, Gras, Gray, Greco, Green, Gretarsson, Groot, Grote, Grunewald, Gruning, Guidi, Guo, Gupta, Gupta, Gushwa, Gustafson, Gustafson, Halim, Hall, Hall, Hamilton, Hammond, Haney, Hanke, Hanks, Hanna, Hannam, Hannuksela, Hanson, Hardwick, Harms, Harry, Harry, Hart, Haster, Haughian, Healy, Heidmann, Heintze, Heitmann, Hello, Hemming, Hendry, Heng, Hennig, Heptonstall, Heurs, Hild, Hinderer, Hoak, Hofman, Holt, Holz, Hopkins, Horst, Hough, Houston, Howell, Hreibi, Hu, Huerta, Huet, Hughey, Husa, Huttner, Huynh-Dinh, Indik, Inta, Intini, Isa, Isac, Isi, Iyer, Izumi, Jacqmin, Jani, Jaranowski, Jawahar,
  Jiménez-Forteza, Johnson, Johnson-McDaniel, Jones, Jones, Jonker, Ju, Junker, Kalaghatgi, Kalogera, Kamai, Kandhasamy, Kang, Kanner, Kapadia, Karki, Karvinen, Kasprzack, Kastaun, Katolik, Katsavounidis, Katzman, Kaufer, Kawabe, Kéfélian, Keitel, Kemball, Kennedy, Kent, Key, Khalili, Khan, Khan, Khan, Khazanov, Kijbunchoo, Kim, Kim, Kim, Kim, Kim, Kim, Kimbrell, King, King, Kinley-Hanlon, Kirchhoff, Kissel, Kleybolte, Klimenko, Knowles, Koch, Koehlenbeck, Koley, Kondrashov, Kontos, Korobko, Korth, Kowalska, Kozak, Krämer, Kringel, Krishnan, Królak, Kuehn, Kumar, Kumar, Kumar, Kuo, Kutynia, Kwang, Lackey, Lai, Landry, Lang, Lange, Lantz, Lanza, Lartaux-Vollard, Lasky, Laxen, Lazzarini, Lazzaro, Leaci, Leavey, Lee, Lee, Lee, Lee, Lee, Lehmann, Lenon, Leonardi, Leroy, Letendre, Levin, Li, Linker, Littenberg, Liu, Lo, Lockerbie, London, Lord, Lorenzini, Loriette, Lormand, Losurdo, Lough, Lousto, Lovelace, Lück, Lumaca, Lundgren, Lynch, Ma, Macas, Macfoy, Machenschalk, MacInnis, Macleod, Magaña~Hernandez,
  Magaña-Sandoval, Magaña~Zertuche, Magee, Majorana, Maksimovic, Man, Mandic, Mangano, Mansell, Manske, Mantovani, Marchesoni, Marion, Márka, Márka, Markakis, Markosyan, Markowitz, Maros, Marquina, Martelli, Martellini, Martin, Martin, Martynov, Mason, Massera, Masserot, Massinger, Masso-Reid, Mastrogiovanni, Matas, Matichard, Matone, Mavalvala, Mazumder, McCarthy, McClelland, McCormick, McCuller, McGuire, McIntyre, McIver, McManus, McNeill, McRae, McWilliams, Meacher, Meadors, Mehmet, Meidam, Mejuto-Villa, Melatos, Mendell, Mercer, Merilh, Merzougui, Meshkov, Messenger, Messick, Metzdorff, Meyers, Miao, Michel, Middleton, Mikhailov, Milano, Miller, Miller, Miller, Millhouse, Milovich-Goff, Minazzoli, Minenkov, Ming, Mishra, Mitra, Mitrofanov, Mitselmakher, Mittleman, Moffa, Moggi, Mogushi, Mohan, Mohapatra, Montani, Moore, Moraru, Moreno, Morriss, Mours, Mow-Lowry, Mueller, Muir, Mukherjee, Mukherjee, Mukherjee, Mukund, Mullavey, Munch, Muñiz, Muratore, Murray, Napier, Nardecchia, Naticchioni, Nayak,
  Neilson, Nelemans, Nelson, Nery, Neunzert, Nevin, Newport, Newton, Ng, Nguyen, Nichols, Nielsen, Nissanke, Nitz, Noack, Nocera, Nolting, North, Nuttall, Oberling, O’Dea, Ogin, Oh, Oh, Ohme, Okada, Oliver, Oppermann, Oram, O’Reilly, Ormiston, Ortega, O’Shaughnessy, Ossokine, Ottaway, Overmier, Owen, Pace, Page, Page, Pai, Pai, Palamos, Palashov, Palomba, Pal-Singh, Pan, Pan, Pang, Pang, Pankow, Pannarale, Pant, Paoletti, Paoli, Papa, Parida, Parker, Pascucci, Pasqualetti, Passaquieti, Passuello, Patil, Patricelli, Pearlstone, Pedraza, Pedurand, Pekowsky, Pele, Penn, Perez, Perreca, Perri, Pfeiffer, Phelps, Piccinni, Pichot, Piergiovanni, Pierro, Pillant, Pinard, Pinto, Pirello, Pitkin, Poe, Poggiani, Popolizio, Porter, Post, Powell, Prasad, Pratt, Pratten, Predoi, Prestegard, Prijatelj, Principe, Privitera, Prodi, Prokhorov, Puncken, Punturo, Puppo, Pürrer, Qi, Quetschke, Quintero, Quitzow-James, Raab, Rabeling, Radkins, Raffai, Raja, Rajan, Rajbhandari, Rakhmanov, Ramirez, Ramos-Buades, Rapagnani,
  Raymond, Razzano, Read, Regimbau, Rei, Reid, Reitze, Ren, Reyes, Ricci, Ricker, Rieger, Riles, Rizzo, Robertson, Robie, Robinet, Rocchi, Rolland, Rollins, Roma, Romano, Romel, Romie, Rosińska, Ross, Rowan, Rüdiger, Ruggi, Rutins, Ryan, Sachdev, Sadecki, Sadeghian, Sakellariadou, Salconi, Saleem, Salemi, Samajdar, Sammut, Sampson, Sanchez, Sanchez, Sanchis-Gual, Sandberg, Sanders, Sassolas, Sathyaprakash, Saulson, Sauter, Savage, Sawadsky, Schale, Scheel, Scheuer, Schmidt, Schmidt, Schnabel, Schofield, Schönbeck, Schreiber, Schuette, Schulte, Schutz, Schwalbe, Scott, Scott, Seidel, Sellers, Sengupta, Sentenac, Sequino, Sergeev, Shaddock, Shaffer, Shah, Shahriar, Shaner, Shao, Shapiro, Shawhan, Sheperd, Shoemaker, Shoemaker, Siellez, Siemens, Sieniawska, Sigg, Silva, Singer, Singh, Singhal, Sintes, Slagmolen, Smith, Smith, Smith, Somala, Son, Sonnenberg, Sorazu, Sorrentino, Souradeep, Spencer, Srivastava, Staats, Staley, Steinke, Steinlechner, Steinlechner, Steinmeyer, Stevenson, Stone, Stops, Strain,
  Stratta, Strigin, Strunk, Sturani, Stuver, Summerscales, Sun, Sunil, Suresh, Sutton, Swinkels, Szczepańczyk, Tacca, Tait, Talbot, Talukder, Tanner, Tápai, Taracchini, Tasson, Taylor, Taylor, Tewari, Theeg, Thies, Thomas, Thomas, Thomas, Thorne, Thorne, Thrane, Tiwari, Tiwari, Tokmakov, Toland, Tonelli, Tornasi, Torres-Forné, Torrie, Töyrä, Travasso, Traylor, Trinastic, Tringali, Trozzo, Tsang, Tse, Tso, Tsukada, Tsuna, Tuyenbayev, Ueno, Ugolini, Unnikrishnan, Urban, Usman, Vahlbruch, Vajente, Valdes, Bakel, Beuzekom, Brand, Broeck, Vander-Hyde, Schaaf, Heijningen, Veggel, Vardaro, Varma, Vass, Vasúth, Vecchio, Vedovato, Veitch, Veitch, Venkateswara, Venugopalan, Verkindt, Vetrano, Viceré, Viets, Vinciguerra, Vine, Vinet, Vitale, Vo, Vocca, Vorvick, Vyatchanin, Wade, Wade, Wade, Walet, Walker, Wallace, Walsh, Wang, Wang, Wang, Wang, Wang, Ward, Warner, Was, Watchi, Weaver, Wei, Weinert, Weinstein, Weiss, Wen, Wessel, Weßels, Westerweck, Westphal, Wette, Whelan, Whitcomb, Whiting, Whittle, Wilken,
  Williams, Williams, Williamson, Willis, Willke, Wimmer, Winkler, Wipf, Wittel, Woan, Woehler, Wofford, Wong, Worden, Wright, Wu, Wysocki, Xiao, Yamamoto, Yancey, Yang, Yap, Yazback, Yu, Yu, Yvert, Zadrożny, Zanolin, Zelenova, Zendri, Zevin, Zhang, Zhang, Zhang, Zhang, Zhao, Zhou, Zhou, Zhu, Zhu, Zimmerman, Zucker, Zweizig, Collaboration, Collaboration), Burns, Veres, Kocevski, Racusin, Goldstein, Connaughton, Briggs, Blackburn, Hamburg, Hui, Kienlin, McEnery, Preece, Wilson-Hodge, Bissaldi, Cleveland, Gibby, Giles, Kippen, McBreen, Meegan, Paciesas, Poolakkil, Roberts, Stanbro, ray Burst~Monitor), Savchenko, Ferrigno, Kuulkers, Bazzano, Bozzo, Brandt, Chenevez, Courvoisier, Diehl, Domingo, Hanlon, Jourdain, Laurent, Lebrun, Lutovinov, Mereghetti, Natalucci, Rodi, Roques, Sunyaev, Ubertini, \& (INTEGRAL)}]{Abbott_2017}
Abbott, B.~P., Abbott, R., Abbott, T.~D., {et~al.} 2017, \href{http://dx.doi.org/10.3847/2041-8213/aa920c}{\color{blue}APJ Letters}, 848, L13

\bibitem[{{Aldenius} {et~al.}(2009){Aldenius}, {Lundberg}, \& {Blackwell-Whitehead}}]{aldenius2009}
{Aldenius}, M., {Lundberg}, H., \& {Blackwell-Whitehead}, R. 2009, \href{http://dx.doi.org/10.1051/0004-6361/200911844}{\color{blue}\aap}, \href{https://ui.adsabs.harvard.edu/abs/2009A\&A...502..989A}{502, 989}

\bibitem[{{Alencastro Puls} {et~al.}(2025){Alencastro Puls}, {Kuske}, {Hansen}, {Lombardo}, {Visentin}, {Arcones}, {Fernandes de Melo}, {Reichert}, {Bonifacio}, {Caffau}, \& {Fritzsche}}]{AlencastroPuls2025}
{Alencastro Puls}, A., {Kuske}, J., {Hansen}, C.~J., {et~al.} 2025, \href{http://dx.doi.org/10.1051/0004-6361/202452537}{\color{blue}\aap}, \href{https://ui.adsabs.harvard.edu/abs/2025A&A...693A.294A}{693, A294}

\bibitem[{Amarsi {et~al.}(2015)Amarsi, Asplund, Collet, \& Leenaarts}]{Amarsi2015_Ox}
Amarsi, A.~M., Asplund, M., Collet, R., \& Leenaarts, J. 2015, \href{http://dx.doi.org/10.1093/mnras/stv2608}{\color{blue}MNRAS}, 455, 3735

\bibitem[{{Amarsi} {et~al.}(2019){Amarsi}, {Barklem}, {Collet}, {Grevesse}, \& {Asplund}}]{Amarsi2019}
{Amarsi}, A.~M., {Barklem}, P.~S., {Collet}, R., {Grevesse}, N., \& {Asplund}, M. 2019, \href{http://dx.doi.org/10.1051/0004-6361/201833603}{\color{blue}\aap}, \href{https://ui.adsabs.harvard.edu/abs/2019A\&A...624A.111A}{624, A111}

\bibitem[{{Amarsi} {et~al.}(2022){Amarsi}, {Liljegren}, \& {Nissen}}]{Amarsi2022A&A...668A..68A}
{Amarsi}, A.~M., {Liljegren}, S., \& {Nissen}, P.~E. 2022, \href{http://dx.doi.org/10.1051/0004-6361/202244542}{\color{blue}\aap}, \href{https://ui.adsabs.harvard.edu/abs/2022A\&A...668A..68A}{668, A68}

\bibitem[{{Aoki} {et~al.}(2007){Aoki}, {Beers}, {Christlieb}, {Norris}, {Ryan}, \& {Tsangarides}}]{Aoki2007}
{Aoki}, W., {Beers}, T.~C., {Christlieb}, N., {et~al.} 2007, \href{http://dx.doi.org/10.1086/509817}{\color{blue}\apj}, \href{https://ui.adsabs.harvard.edu/abs/2007ApJ...655..492A}{655, 492}

\bibitem[{{Aoki} {et~al.}(2010){Aoki}, {Beers}, {Honda}, \& {Carollo}}]{aoki2010}
{Aoki}, W., {Beers}, T.~C., {Honda}, S., \& {Carollo}, D. 2010, \href{http://dx.doi.org/10.1088/2041-8205/723/2/L201}{\color{blue}\apjl}, \href{https://ui.adsabs.harvard.edu/abs/2010ApJ...723L.201A}{723, L201}

\bibitem[{Arcones \& Thielemann(2012)}]{Arcones_2013}
Arcones, A. \& Thielemann, F.-K. 2012, \href{http://dx.doi.org/10.1088/0954-3899/40/1/013201}{\color{blue}J. Phys. G}, 40, 013201

\bibitem[{{Asplund} {et~al.}(2009){Asplund}, {Grevesse}, {Sauval}, \& {Scott}}]{asplund2009}
{Asplund}, M., {Grevesse}, N., {Sauval}, A.~J., \& {Scott}, P. 2009, \href{http://dx.doi.org/10.1146/annurev.astro.46.060407.145222}{\color{blue}\araa}, \href{https://ui.adsabs.harvard.edu/abs/2009ARA\&A..47..481A}{47, 481}

\bibitem[{{Bandyopadhyay} {et~al.}(2024){Bandyopadhyay}, {Ezzeddine}, {Allende Prieto}, {Aria}, {Shah}, {Beers}, {Frebel}, {Hansen}, {Holmbeck}, {Placco}, {Roederer}, \& {Sakari}}]{Bandyopadhyay2024}
{Bandyopadhyay}, A., {Ezzeddine}, R., {Allende Prieto}, C., {et~al.} 2024, \href{http://dx.doi.org/10.3847/1538-4365/ad6f0f}{\color{blue}\apjs}, \href{https://ui.adsabs.harvard.edu/abs/2024ApJS..274...39B}{274, 39}

\bibitem[{Barnes \& Metzger(2022)}]{Barnes_2022}
Barnes, J. \& Metzger, B.~D. 2022, \href{http://dx.doi.org/10.3847/2041-8213/ac9b41}{\color{blue}APJ Letters}, 939, L29

\bibitem[{{Beers} \& {Christlieb}(2005)}]{BeersChristlieb2005ARA&A..43..531B}
{Beers}, T.~C. \& {Christlieb}, N. 2005, \href{http://dx.doi.org/10.1146/annurev.astro.42.053102.134057}{\color{blue}\araa}, \href{https://ui.adsabs.harvard.edu/abs/2005ARA\&A..43..531B}{43, 531}

\bibitem[{{Belmonte} {et~al.}(2017){Belmonte}, {Pickering}, {Ruffoni}, {Den Hartog}, {Lawler}, {Guzman}, \& {Heiter}}]{belmonte2017}
{Belmonte}, M.~T., {Pickering}, J.~C., {Ruffoni}, M.~P., {et~al.} 2017, \href{http://dx.doi.org/10.3847/1538-4357/aa8cd3}{\color{blue}\apj}, \href{https://ui.adsabs.harvard.edu/abs/2017ApJ...848..125B}{848, 125}

\bibitem[{Belokurov \& Kravtsov(2023)}]{Belokurov2023}
Belokurov, V. \& Kravtsov, A. 2023, \href{http://dx.doi.org/10.1093/mnras/stad2241}{\color{blue}MNRAS}, 525, 4456

\bibitem[{{Bergemann}(2008)}]{Bergemann2008PhST..133a4013B}
{Bergemann}, M. 2008, \href{http://dx.doi.org/10.1088/0031-8949/2008/T133/014013}{\color{blue}Physica Scripta Volume T}, \href{https://ui.adsabs.harvard.edu/abs/2008PhST..133a4013B}{133, 014013}

\bibitem[{{Bergemann}(2011)}]{Bergemann2011MNRAS.413.2184B}
{Bergemann}, M. 2011, \href{http://dx.doi.org/10.1111/j.1365-2966.2011.18295.x}{\color{blue}\mnras}, \href{https://ui.adsabs.harvard.edu/abs/2011MNRAS.413.2184B}{413, 2184}

\bibitem[{{Bergemann} \& {Cescutti}(2010)}]{BergemannCescutti2010A&A...522A...9B}
{Bergemann}, M. \& {Cescutti}, G. 2010, \href{http://dx.doi.org/10.1051/0004-6361/201014250}{\color{blue}\aap}, \href{https://ui.adsabs.harvard.edu/abs/2010A\&A...522A...9B}{522, A9}

\bibitem[{{Bergemann} \& {Gehren}(2008)}]{BergemannGehren2008A&A...492..823B}
{Bergemann}, M. \& {Gehren}, T. 2008, \href{http://dx.doi.org/10.1051/0004-6361:200810098}{\color{blue}\aap}, \href{https://ui.adsabs.harvard.edu/abs/2008A\&A...492..823B}{492, 823}

\bibitem[{{Bernstein} {et~al.}(2003){Bernstein}, {Shectman}, {Gunnels}, {Mochnacki}, \& {Athey}}]{bernstein2003}
{Bernstein}, R., {Shectman}, S.~A., {Gunnels}, S.~M., {Mochnacki}, S., \& {Athey}, A.~E. 2003, in Society of Photo-Optical Instrumentation Engineers (SPIE) Conference Series, Vol. 4841, Instrument Design and Performance for Optical/Infrared Ground-based Telescopes, ed. M.~{Iye} \& A.~F.~M. {Moorwood}, \href{https://ui.adsabs.harvard.edu/abs/2003SPIE.4841.1694B}{1694--1704}

\bibitem[{{Bi{\'e}mont} {et~al.}(2011){Bi{\'e}mont}, {Blagoev}, {Engstr{\"o}m}, {Hartman}, {Lundberg}, {Malcheva}, {Nilsson}, {Whitehead}, {Palmeri}, \& {Quinet}}]{biemont2011}
{Bi{\'e}mont}, {\'E}., {Blagoev}, K., {Engstr{\"o}m}, L., {et~al.} 2011, \href{http://dx.doi.org/10.1111/j.1365-2966.2011.18637.x}{\color{blue}\mnras}, \href{https://ui.adsabs.harvard.edu/abs/2011MNRAS.414.3350B}{414, 3350}

\bibitem[{Biémont {et~al.}(2000)Biémont, Garnir, Palmeri, Li, \& Svanberg}]{biemont2000}
Biémont, E., Garnir, H.~P., Palmeri, P., Li, Z.~S., \& Svanberg, S. 2000, \href{http://dx.doi.org/10.1046/j.1365-8711.2000.03094.x}{\color{blue}MNRAS}, 312, 116

\bibitem[{{Bovy}(2015)}]{Bovy2015}
{Bovy}, J. 2015, \href{http://dx.doi.org/10.1088/0067-0049/216/2/29}{\color{blue}\apjs}, \href{https://ui.adsabs.harvard.edu/abs/2015ApJS..216...29B}{216, 29}

\bibitem[{Buder {et~al.}(2021)Buder, Sharma, Kos, Amarsi, Nordlander, Lind, Martell, Asplund, Bland-Hawthorn, Casey, De Silva, D’Orazi, Freeman, Hayden, Lewis, Lin, Schlesinger, Simpson, Stello, Zucker, Zwitter, Beeson, Buck, Casagrande, Clark, Čotar, Da Costa, de Grijs, Feuillet, Horner, Kafle, Khanna, Kobayashi, Liu, Montet, Nandakumar, Nataf, Ness, Spina, Tepper-García, Ting(丁源森), Traven, Vogrinčič, Wittenmyer, Wyse, Žerjal, \& GALAH Collaboration}]{Buder2021}
Buder, S., Sharma, S., Kos, J., {et~al.} 2021, \href{http://dx.doi.org/10.1093/mnras/stab1242}{\color{blue}MNRAS}, 506, 150

\bibitem[{{Burbidge} {et~al.}(1957){Burbidge}, {Burbidge}, {Fowler}, \& {Hoyle}}]{Burbidge_1957}
{Burbidge}, E.~M., {Burbidge}, G.~R., {Fowler}, W.~A., \& {Hoyle}, F. 1957, \href{http://dx.doi.org/10.1103/RevModPhys.29.547}{\color{blue}RMP}, \href{https://ui.adsabs.harvard.edu/abs/1957RvMP...29..547B}{29, 547}

\bibitem[{{Caliskan} {et~al.}(2025){Caliskan}, {Amarsi}, {Racca}, {Koutsouridou}, {Barklem}, {Lind}, \& {Salvadori}}]{Caliskan2025}
{Caliskan}, S., {Amarsi}, A.~M., {Racca}, M., {et~al.} 2025, \href{http://dx.doi.org/10.1051/0004-6361/202554251}{\color{blue}\aap}, \href{https://ui.adsabs.harvard.edu/abs/2025A&A...696A.210C}{696, A210}

\bibitem[{{Cameron}(1957)}]{Cameron1957AJ.....62....9C}
{Cameron}, A.~G.~W. 1957, \href{http://dx.doi.org/10.1086/107435}{\color{blue}\aj}, \href{https://ui.adsabs.harvard.edu/abs/1957AJ.....62....9C}{62, 9}

\bibitem[{{Canocchi} {et~al.}(2024){Canocchi}, {Morello}, {Lind}, {Carleo}, {Stangret}, \& {Pall{\'e}}}]{Canocchi2024A&A...692A..43C}
{Canocchi}, G., {Morello}, G., {Lind}, K., {et~al.} 2024, \href{http://dx.doi.org/10.1051/0004-6361/202451972}{\color{blue}\aap}, \href{https://ui.adsabs.harvard.edu/abs/2024A\&A...692A..43C}{692, A43}

\bibitem[{{Casey}(2014)}]{casey2014}
{Casey}, A.~R. 2014, \href{https://ui.adsabs.harvard.edu/abs/2014PhDT.......394C}{{A Tale of Tidal Tales in the Milky Way}}, PhD thesis, Australian National University, Canberra

\bibitem[{{Castelli} \& {Kurucz}(2003)}]{castelli2003}
{Castelli}, F. \& {Kurucz}, R.~L. 2003, in Modelling of Stellar Atmospheres, ed. N.~{Piskunov}, W.~W. {Weiss}, \& D.~F. {Gray}, Vol. 210, \href{https://ui.adsabs.harvard.edu/abs/2003IAUS..210P.A20C}{A20}

\bibitem[{{Ceccarelli} {et~al.}(2024){Ceccarelli}, {Massari}, {Mucciarelli}, {Bellazzini}, {Nunnari}, {Cusano}, {Lardo}, {Romano}, {Ilyin}, \& {Stokholm}}]{Ceccarelli2024}
{Ceccarelli}, E., {Massari}, D., {Mucciarelli}, A., {et~al.} 2024, \href{http://dx.doi.org/10.1051/0004-6361/202348332}{\color{blue}\aap}, \href{https://ui.adsabs.harvard.edu/abs/2024A&A...684A..37C}{684, A37}

\bibitem[{{Cescutti} {et~al.}(2022){Cescutti}, {Bonifacio}, {Caffau}, {Monaco}, {Franchini}, {Lombardo}, {Matas Pinto}, {Lucertini}, {Fran{\c{c}}ois}, {Spitoni}, {Lallement}, {Sbordone}, {Mucciarelli}, {Spite}, {Hansen}, {Di Marcantonio}, {Ku{\v{c}}inskas}, {Dobrovolskas}, {Korn}, {Valentini}, {Magrini}, {Cristallo}, \& {Matteucci}}]{Cescutti2022}
{Cescutti}, G., {Bonifacio}, P., {Caffau}, E., {et~al.} 2022, \href{http://dx.doi.org/10.1051/0004-6361/202244515}{\color{blue}\aap}, \href{https://ui.adsabs.harvard.edu/abs/2022A\&A...668A.168C}{668, A168}

\bibitem[{Chornock {et~al.}(2017)Chornock, Berger, Kasen, Cowperthwaite, Nicholl, Villar, Alexander, Blanchard, Eftekhari, Fong, Margutti, Williams, Annis, Brout, Brown, Chen, Drout, Farr, Foley, Frieman, Fryer, Herner, Holz, Kessler, Matheson, Metzger, Quataert, Rest, Sako, Scolnic, Smith, \& Soares-Santos}]{Chornock_2017}
Chornock, R., Berger, E., Kasen, D., {et~al.} 2017, \href{http://dx.doi.org/10.3847/2041-8213/aa905c}{\color{blue}APJ Letters}, 848, L19

\bibitem[{{Christlieb} {et~al.}(2004){Christlieb}, {Gustafsson}, {Korn}, {Barklem}, {Beers}, {Bessell}, {Karlsson}, \& {Mizuno-Wiedner}}]{christlieb2004}
{Christlieb}, N., {Gustafsson}, B., {Korn}, A.~J., {et~al.} 2004, \href{http://dx.doi.org/10.1086/381237}{\color{blue}\apj}, \href{https://ui.adsabs.harvard.edu/abs/2004ApJ...603..708C}{603, 708}

\bibitem[{{C{\^o}t{\'e}} {et~al.}(2019{\natexlab{a}}){C{\^o}t{\'e}}, {Eichler}, {Arcones}, {Hansen}, {Simonetti}, {Frebel}, {Fryer}, {Pignatari}, {Reichert}, {Belczynski}, \& {Matteucci}}]{Cote2019ApJ...875..106C}
{C{\^o}t{\'e}}, B., {Eichler}, M., {Arcones}, A., {et~al.} 2019{\natexlab{a}}, \href{http://dx.doi.org/10.3847/1538-4357/ab10db}{\color{blue}\apj}, \href{https://ui.adsabs.harvard.edu/abs/2019ApJ...875..106C}{875, 106}

\bibitem[{{C{\^o}t{\'e}} {et~al.}(2019{\natexlab{b}}){C{\^o}t{\'e}}, {Eichler}, {Arcones}, {Hansen}, {Simonetti}, {Frebel}, {Fryer}, {Pignatari}, {Reichert}, {Belczynski}, \& {Matteucci}}]{Cote2019}
{C{\^o}t{\'e}}, B., {Eichler}, M., {Arcones}, A., {et~al.} 2019{\natexlab{b}}, \href{http://dx.doi.org/10.3847/1538-4357/ab10db}{\color{blue}\apj}, \href{https://ui.adsabs.harvard.edu/abs/2019ApJ...875..106C}{875, 106}

\bibitem[{{Cowan} {et~al.}(2005){Cowan}, {Sneden}, {Beers}, {Lawler}, {Simmerer}, {Truran}, {Primas}, {Collier}, \& {Burles}}]{cowan2005}
{Cowan}, J.~J., {Sneden}, C., {Beers}, T.~C., {et~al.} 2005, \href{http://dx.doi.org/10.1086/429952}{\color{blue}\apj}, \href{https://ui.adsabs.harvard.edu/abs/2005ApJ...627..238C}{627, 238}

\bibitem[{{Cowan} {et~al.}(2021){Cowan}, {Sneden}, {Lawler}, {Aprahamian}, {Wiescher}, {Langanke}, {Mart{\'\i}nez-Pinedo}, \& {Thielemann}}]{Cownan2021}
{Cowan}, J.~J., {Sneden}, C., {Lawler}, J.~E., {et~al.} 2021, \href{http://dx.doi.org/10.1103/RevModPhys.93.015002}{\color{blue}RMP}, \href{https://ui.adsabs.harvard.edu/abs/2021RvMP...93a5002C}{93, 015002}

\bibitem[{{Cutri} {et~al.}(2003){Cutri}, {Skrutskie}, {van Dyk}, {Beichman}, {Carpenter}, {Chester}, {Cambresy}, {Evans}, {Fowler}, {Gizis}, {Howard}, {Huchra}, {Jarrett}, {Kopan}, {Kirkpatrick}, {Light}, {Marsh}, {McCallon}, {Schneider}, {Stiening}, {Sykes}, {Weinberg}, {Wheaton}, {Wheelock}, \& {Zacarias}}]{2003yCat.2246....0C}
{Cutri}, R.~M., {Skrutskie}, M.~F., {van Dyk}, S., {et~al.} 2003, {VizieR Online Data Catalog: 2MASS All-Sky Catalog of Point Sources (Cutri+ 2003)}, VizieR On-line Data Catalog: II/246. Originally published in: University of Massachusetts and Infrared Processing and Analysis Center, (IPAC/California Institute of Technology) (2003)

\bibitem[{{Den Hartog} {et~al.}(2003){Den Hartog}, {Lawler}, {Sneden}, \& {Cowan}}]{denhartog2003}
{Den Hartog}, E.~A., {Lawler}, J.~E., {Sneden}, C., \& {Cowan}, J.~J. 2003, \href{http://dx.doi.org/10.1086/376940}{\color{blue}\apjs}, \href{https://ui.adsabs.harvard.edu/abs/2003ApJS..148..543D}{148, 543}

\bibitem[{{Den Hartog} {et~al.}(2006){Den Hartog}, {Lawler}, {Sneden}, \& {Cowan}}]{denhartog2006}
{Den Hartog}, E.~A., {Lawler}, J.~E., {Sneden}, C., \& {Cowan}, J.~J. 2006, \href{http://dx.doi.org/10.1086/508262}{\color{blue}\apjs}, \href{https://ui.adsabs.harvard.edu/abs/2006ApJS..167..292D}{167, 292}

\bibitem[{{Den Hartog} {et~al.}(2011){Den Hartog}, {Lawler}, {Sobeck}, {Sneden}, \& {Cowan}}]{denhartog2011}
{Den Hartog}, E.~A., {Lawler}, J.~E., {Sobeck}, J.~S., {Sneden}, C., \& {Cowan}, J.~J. 2011, \href{http://dx.doi.org/10.1088/0067-0049/194/2/35}{\color{blue}\apjs}, \href{https://ui.adsabs.harvard.edu/abs/2011ApJS..194...35D}{194, 35}

\bibitem[{{Den Hartog} {et~al.}(2014){Den Hartog}, {Ruffoni}, {Lawler}, {Pickering}, {Lind}, \& {Brewer}}]{denhartog2014}
{Den Hartog}, E.~A., {Ruffoni}, M.~P., {Lawler}, J.~E., {et~al.} 2014, \href{http://dx.doi.org/10.1088/0067-0049/215/2/23}{\color{blue}\apjs}, \href{https://ui.adsabs.harvard.edu/abs/2014ApJS..215...23D}{215, 23}

\bibitem[{{Dixon} {et~al.}(2025){Dixon}, {Ezzeddine}, {Li}, {Merle}, {Bautista}, \& {Guo}}]{Dixon2025}
{Dixon}, J.~D., {Ezzeddine}, R., {Li}, Y., {et~al.} 2025, \href{https://ui.adsabs.harvard.edu/abs/2025arXiv250922811D}{\href{http://dx.doi.org/10.48550/arXiv.2509.22811}{\color{blue}arXiv e-prints}, arXiv:2509.22811}

\bibitem[{{Drout} {et~al.}(2017){Drout}, {Piro}, {Shappee}, {Kilpatrick}, {Simon}, {Contreras}, {Coulter}, {Foley}, {Siebert}, {Morrell}, {Boutsia}, {Di Mille}, {Holoien}, {Kasen}, {Kollmeier}, {Madore}, {Monson}, {Murguia-Berthier}, {Pan}, {Prochaska}, {Ramirez-Ruiz}, {Rest}, {Adams}, {Alatalo}, {Ba{\~n}ados}, {Baughman}, {Beers}, {Bernstein}, {Bitsakis}, {Campillay}, {Hansen}, {Higgs}, {Ji}, {Maravelias}, {Marshall}, {Moni Bidin}, {Prieto}, {Rasmussen}, {Rojas-Bravo}, {Strom}, {Ulloa}, {Vargas-Gonz{\'a}lez}, {Wan}, \& {Whitten}}]{Drout2017}
{Drout}, M.~R., {Piro}, A.~L., {Shappee}, B.~J., {et~al.} 2017, \href{http://dx.doi.org/10.1126/science.aaq0049}{\color{blue}Science}, \href{https://ui.adsabs.harvard.edu/abs/2017Sci...358.1570D}{358, 1570}

\bibitem[{{Duquette} \& {Lawler}(1985)}]{duquette1985}
{Duquette}, D.~W. \& {Lawler}, J.~E. 1985, \href{http://dx.doi.org/10.1364/JOSAB.2.001948}{\color{blue}JOSA B}, \href{https://ui.adsabs.harvard.edu/abs/1985JOSAB...2.1948D}{2, 1948}

\bibitem[{Eichler {et~al.}(2015)Eichler, Arcones, Kelic, Korobkin, Langanke, Marketin, Martinez-Pinedo, Panov, Rauscher, Rosswog, Winteler, Zinner, \& Thielemann}]{Eichler_2015}
Eichler, M., Arcones, A., Kelic, A., {et~al.} 2015, \href{http://dx.doi.org/10.1088/0004-637X/808/1/30}{\color{blue}APJ}, 808, 30

\bibitem[{{Eitner} {et~al.}(2023){Eitner}, {Bergemann}, {Ruiter}, {Avril}, {Seitenzahl}, {Gent}, \& {C{\^o}t{\'e}}}]{Eitner2023}
{Eitner}, P., {Bergemann}, M., {Ruiter}, A.~J., {et~al.} 2023, \href{http://dx.doi.org/10.1051/0004-6361/202244286}{\color{blue}\aap}, \href{https://ui.adsabs.harvard.edu/abs/2023A&A...677A.151E}{677, A151}

\bibitem[{{Ezzeddine} {et~al.}(2020){Ezzeddine}, {Rasmussen}, {Frebel}, {Chiti}, {Hinojisa}, {Placco}, {Ji}, {Beers}, {Hansen}, {Roederer}, {Sakari}, \& {Melendez}}]{Ezzeddine2020ApJ...898..150E}
{Ezzeddine}, R., {Rasmussen}, K., {Frebel}, A., {et~al.} 2020, \href{http://dx.doi.org/10.3847/1538-4357/ab9d1a}{\color{blue}\apj}, \href{https://ui.adsabs.harvard.edu/abs/2020ApJ...898..150E}{898, 150}

\bibitem[{{Fitzpatrick} {et~al.}(2024){Fitzpatrick}, {Placco}, {Bolton}, {Merino}, {Ridgway}, \& {Stanghellini}}]{Fitzpatrick2024}
{Fitzpatrick}, M., {Placco}, V., {Bolton}, A., {et~al.} 2024, \href{https://ui.adsabs.harvard.edu/abs/2024arXiv240101982F}{\href{http://dx.doi.org/10.48550/arXiv.2401.01982}{\color{blue}arXiv e-prints}, arXiv:2401.01982}

\bibitem[{Frebel {et~al.}(2013)Frebel, Casey, Jacobson, \& Yu}]{Frebel_2013}
Frebel, A., Casey, A.~R., Jacobson, H.~R., \& Yu, Q. 2013, \href{http://dx.doi.org/10.1088/0004-637X/769/1/57}{\color{blue}APJ}, 769, 57

\bibitem[{{Gaia Collaboration} {et~al.}(2018){Gaia Collaboration}, {Babusiaux}, {van Leeuwen}, {Barstow}, {Jordi}, {Vallenari}, {Bossini}, {Bressan}, {Cantat-Gaudin}, {van Leeuwen}, {Brown}, {Prusti}, {de Bruijne}, {Bailer-Jones}, {Biermann}, {Evans}, {Eyer}, {Jansen}, {Klioner}, {Lammers}, {Lindegren}, {Luri}, {Mignard}, {Panem}, {Pourbaix}, {Randich}, {Sartoretti}, {Siddiqui}, {Soubiran}, {Walton}, {Arenou}, {Bastian}, {Cropper}, {Drimmel}, {Katz}, {Lattanzi}, {Bakker}, {Cacciari}, {Casta{\~n}eda}, {Chaoul}, {Cheek}, {De Angeli}, {Fabricius}, {Guerra}, {Holl}, {Masana}, {Messineo}, {Mowlavi}, {Nienartowicz}, {Panuzzo}, {Portell}, {Riello}, {Seabroke}, {Tanga}, {Th{\'e}venin}, {Gracia-Abril}, {Comoretto}, {Garcia-Reinaldos}, {Teyssier}, {Altmann}, {Andrae}, {Audard}, {Bellas-Velidis}, {Benson}, {Berthier}, {Blomme}, {Burgess}, {Busso}, {Carry}, {Cellino}, {Clementini}, {Clotet}, {Creevey}, {Davidson}, {De Ridder}, {Delchambre}, {Dell'Oro}, {Ducourant}, {Fern{\'a}ndez-Hern{\'a}ndez}, {Fouesneau},
  {Fr{\'e}mat}, {Galluccio}, {Garc{\'\i}a-Torres}, {Gonz{\'a}lez-N{\'u}{\~n}ez}, {Gonz{\'a}lez-Vidal}, {Gosset}, {Guy}, {Halbwachs}, {Hambly}, {Harrison}, {Hern{\'a}ndez}, {Hestroffer}, {Hodgkin}, {Hutton}, {Jasniewicz}, {Jean-Antoine-Piccolo}, {Jordan}, {Korn}, {Krone-Martins}, {Lanzafame}, {Lebzelter}, {L{\"o}ffler}, {Manteiga}, {Marrese}, {Mart{\'\i}n-Fleitas}, {Moitinho}, {Mora}, {Muinonen}, {Osinde}, {Pancino}, {Pauwels}, {Petit}, {Recio-Blanco}, {Richards}, {Rimoldini}, {Robin}, {Sarro}, {Siopis}, {Smith}, {Sozzetti}, {S{\"u}veges}, {Torra}, {van Reeven}, {Abbas}, {Abreu Aramburu}, {Accart}, {Aerts}, {Altavilla}, {{\'A}lvarez}, {Alvarez}, {Alves}, {Anderson}, {Andrei}, {Anglada Varela}, {Antiche}, {Antoja}, {Arcay}, {Astraatmadja}, {Bach}, {Baker}, {Balaguer-N{\'u}{\~n}ez}, {Balm}, {Barache}, {Barata}, {Barbato}, {Barblan}, {Barklem}, {Barrado}, {Barros}, {Bartholom{\'e} Mu{\~n}oz}, {Bassilana}, {Becciani}, {Bellazzini}, {Berihuete}, {Bertone}, {Bianchi}, {Bienaym{\'e}}, {Blanco-Cuaresma}, {Boch},
  {Boeche}, {Bombrun}, {Borrachero}, {Bouquillon}, {Bourda}, {Bragaglia}, {Bramante}, {Breddels}, {Brouillet}, {Br{\"u}semeister}, {Brugaletta}, {Bucciarelli}, {Burlacu}, {Busonero}, {Butkevich}, {Buzzi}, {Caffau}, {Cancelliere}, {Cannizzaro}, {Carballo}, {Carlucci}, {Carrasco}, {Casamiquela}, {Castellani}, {Castro-Ginard}, {Charlot}, {Chemin}, {Chiavassa}, {Cocozza}, {Costigan}, {Cowell}, {Crifo}, {Crosta}, {Crowley}, {Cuypers}, {Dafonte}, {Damerdji}, {Dapergolas}, {David}, {David}, \& {de Laverny}}]{Gaia2018}
{Gaia Collaboration}, {Babusiaux}, C., {van Leeuwen}, F., {et~al.} 2018, \href{http://dx.doi.org/10.1051/0004-6361/201832843}{\color{blue}\aap}, \href{https://ui.adsabs.harvard.edu/abs/2018A\&A...616A..10G}{616, A10}

\bibitem[{{Gaia Collaboration} {et~al.}(2023{\natexlab{a}}){Gaia Collaboration}, {Vallenari}, {Brown}, {Prusti}, {de Bruijne}, {Arenou}, {Babusiaux}, {Biermann}, {Creevey}, {Ducourant}, {Evans}, {Eyer}, {Guerra}, {Hutton}, {Jordi}, {Klioner}, {Lammers}, {Lindegren}, {Luri}, {Mignard}, {Panem}, {Pourbaix}, {Randich}, {Sartoretti}, {Soubiran}, {Tanga}, {Walton}, {Bailer-Jones}, {Bastian}, {Drimmel}, {Jansen}, {Katz}, {Lattanzi}, {van Leeuwen}, {Bakker}, {Cacciari}, {Casta{\~n}eda}, {De Angeli}, {Fabricius}, {Fouesneau}, {Fr{\'e}mat}, {Galluccio}, {Guerrier}, {Heiter}, {Masana}, {Messineo}, {Mowlavi}, {Nicolas}, {Nienartowicz}, {Pailler}, {Panuzzo}, {Riclet}, {Roux}, {Seabroke}, {Sordo}, {Th{\'e}venin}, {Gracia-Abril}, {Portell}, {Teyssier}, {Altmann}, {Andrae}, {Audard}, {Bellas-Velidis}, {Benson}, {Berthier}, {Blomme}, {Burgess}, {Busonero}, {Busso}, {C{\'a}novas}, {Carry}, {Cellino}, {Cheek}, {Clementini}, {Damerdji}, {Davidson}, {de Teodoro}, {Nu{\~n}ez Campos}, {Delchambre}, {Dell'Oro}, {Esquej},
  {Fern{\'a}ndez-Hern{\'a}ndez}, {Fraile}, {Garabato}, {Garc{\'\i}a-Lario}, {Gosset}, {Haigron}, {Halbwachs}, {Hambly}, {Harrison}, {Hern{\'a}ndez}, {Hestroffer}, {Hodgkin}, {Holl}, {Jan{\ss}en}, {Jevardat de Fombelle}, {Jordan}, {Krone-Martins}, {Lanzafame}, {L{\"o}ffler}, {Marchal}, {Marrese}, {Moitinho}, {Muinonen}, {Osborne}, {Pancino}, {Pauwels}, {Recio-Blanco}, {Reyl{\'e}}, {Riello}, {Rimoldini}, {Roegiers}, {Rybizki}, {Sarro}, {Siopis}, {Smith}, {Sozzetti}, {Utrilla}, {van Leeuwen}, {Abbas}, {{\'A}brah{\'a}m}, {Abreu Aramburu}, {Aerts}, {Aguado}, {Ajaj}, {Aldea-Montero}, {Altavilla}, {{\'A}lvarez}, {Alves}, {Anders}, {Anderson}, {Anglada Varela}, {Antoja}, {Baines}, {Baker}, {Balaguer-N{\'u}{\~n}ez}, {Balbinot}, {Balog}, {Barache}, {Barbato}, {Barros}, {Barstow}, {Bartolom{\'e}}, {Bassilana}, {Bauchet}, {Becciani}, {Bellazzini}, {Berihuete}, {Bernet}, {Bertone}, {Bianchi}, {Binnenfeld}, {Blanco-Cuaresma}, {Blazere}, {Boch}, {Bombrun}, {Bossini}, {Bouquillon}, {Bragaglia}, {Bramante}, {Breedt},
  {Bressan}, {Brouillet}, {Brugaletta}, {Bucciarelli}, {Burlacu}, {Butkevich}, {Buzzi}, {Caffau}, {Cancelliere}, {Cantat-Gaudin}, {Carballo}, {Carlucci}, {Carnerero}, {Carrasco}, {Casamiquela}, {Castellani}, {Castro-Ginard}, {Chaoul}, {Charlot}, {Chemin}, {Chiaramida}, {Chiavassa}, {Chornay}, {Comoretto}, {Contursi}, {Cooper}, {Cornez}, {Cowell}, {Crifo}, {Cropper}, {Crosta}, {Crowley}, {Dafonte}, {Dapergolas}, {David}, {David}, {de Laverny}, {De Luise}, {De March}, {De Ridder}, {de Souza}, {de Torres}, {del Peloso}, {del Pozo}, {Delbo}, {Delgado}, {Delisle}, {Demouchy}, {Dharmawardena}, {Di Matteo}, {Diakite}, {Diener}, {Distefano}, {Dolding}, {Edvardsson}, {Enke}, {Fabre}, {Fabrizio}, {Faigler}, {Fedorets}, {Fernique}, {Fienga}, {Figueras}, {Fournier}, {Fouron}, {Fragkoudi}, {Gai}, {Garcia-Gutierrez}, {Garcia-Reinaldos}, {Garc{\'\i}a-Torres}, {Garofalo}, {Gavel}, {Gavras}, {Gerlach}, {Geyer}, {Giacobbe}, {Gilmore}, {Girona}, {Giuffrida}, {Gomel}, {Gomez}, {Gonz{\'a}lez-N{\'u}{\~n}ez},
  {Gonz{\'a}lez-Santamar{\'\i}a}, {Gonz{\'a}lez-Vidal}, {Granvik}, {Guillout}, {Guiraud}, {Guti{\'e}rrez-S{\'a}nchez}, {Guy}, {Hatzidimitriou}, {Hauser}, {Haywood}, {Helmer}, {Helmi}, {Sarmiento}, {Hidalgo}, {Hilger}, {H{\l}adczuk}, {Hobbs}, {Holland}, {Huckle}, {Jardine}, {Jasniewicz}, {Jean-Antoine Piccolo}, {Jim{\'e}nez-Arranz}, {Jorissen}, {Juaristi Campillo}, {Julbe}, {Karbevska}, {Kervella}, {Khanna}, {Kontizas}, {Kordopatis}, {Korn}, {K{\'o}sp{\'a}l}, {Kostrzewa-Rutkowska}, {Kruszy{\'n}ska}, {Kun}, {Laizeau}, {Lambert}, {Lanza}, {Lasne}, {Le Campion}, {Lebreton}, {Lebzelter}, {Leccia}, {Leclerc}, {Lecoeur-Taibi}, {Liao}, {Licata}, {Lindstr{\o}m}, {Lister}, {Livanou}, {Lobel}, {Lorca}, {Loup}, {Madrero Pardo}, {Magdaleno Romeo}, {Managau}, {Mann}, {Manteiga}, {Marchant}, {Marconi}, {Marcos}, {Marcos Santos}, {Mar{\'\i}n Pina}, {Marinoni}, {Marocco}, {Marshall}, {Martin Polo}, {Mart{\'\i}n-Fleitas}, {Marton}, {Mary}, {Masip}, {Massari}, {Mastrobuono-Battisti}, {Mazeh}, {McMillan}, {Messina}, {Michalik},
  {Millar}, {Mints}, {Molina}, {Molinaro}, {Moln{\'a}r}, {Monari}, {Mongui{\'o}}, {Montegriffo}, {Montero}, {Mor}, {Mora}, {Morbidelli}, {Morel}, {Morris}, {Muraveva}, {Murphy}, {Musella}, {Nagy}, {Noval}, {Oca{\~n}a}, {Ogden}, {Ordenovic}, {Osinde}, {Pagani}, {Pagano}, {Palaversa}, {Palicio}, {Pallas-Quintela}, {Panahi}, {Payne-Wardenaar}, {Pe{\~n}alosa Esteller}, {Penttil{\"a}}, {Pichon}, {Piersimoni}, {Pineau}, {Plachy}, {Plum}, {Poggio}, {Pr{\v{s}}a}, {Pulone}, {Racero}, {Ragaini}, {Rainer}, {Raiteri}, {Rambaux}, {Ramos}, {Ramos-Lerate}, {Re Fiorentin}, {Regibo}, {Richards}, {Rios Diaz}, {Ripepi}, {Riva}, {Rix}, {Rixon}, {Robichon}, {Robin}, {Robin}, {Roelens}, {Rogues}, {Rohrbasser}, {Romero-G{\'o}mez}, {Rowell}, {Royer}, {Ruz Mieres}, {Rybicki}, {Sadowski}, {S{\'a}ez N{\'u}{\~n}ez}, {Sagrist{\`a} Sell{\'e}s}, {Sahlmann}, {Salguero}, {Samaras}, {Sanchez Gimenez}, {Sanna}, {Santove{\~n}a}, {Sarasso}, {Schultheis}, {Sciacca}, {Segol}, {Segovia}, {S{\'e}gransan}, {Semeux}, {Shahaf}, {Siddiqui}, {Siebert},
  {Siltala}, {Silvelo}, {Slezak}, {Slezak}, {Smart}, {Snaith}, {Solano}, {Solitro}, {Souami}, {Souchay}, {Spagna}, {Spina}, {Spoto}, {Steele}, {Steidelm{\"u}ller}, {Stephenson}, {S{\"u}veges}, {Surdej}, {Szabados}, {Szegedi-Elek}, {Taris}, {Taylor}, {Teixeira}, {Tolomei}, {Tonello}, {Torra}, {Torra}, {Torralba Elipe}, {Trabucchi}, {Tsounis}, {Turon}, {Ulla}, {Unger}, {Vaillant}, {van Dillen}, {van Reeven}, {Vanel}, {Vecchiato}, {Viala}, {Vicente}, {Voutsinas}, {Weiler}, {Wevers}, {Wyrzykowski}, {Yoldas}, {Yvard}, {Zhao}, {Zorec}, {Zucker}, \& {Zwitter}}]{gaiadr3}
{Gaia Collaboration}, {Vallenari}, A., {Brown}, A.~G.~A., {et~al.} 2023{\natexlab{a}}, \href{http://dx.doi.org/10.1051/0004-6361/202243940}{\color{blue}\aap}, \href{https://ui.adsabs.harvard.edu/abs/2023A&A...674A...1G}{674, A1}

\bibitem[{{Gaia Collaboration} {et~al.}(2023{\natexlab{b}}){Gaia Collaboration}, {Vallenari, A.}, {Brown, A. G. A.}, {Prusti, T.}, {de Bruijne, J. H. J.}, {Arenou, F.}, {Babusiaux, C.}, {Biermann, M.}, {Creevey, O. L.}, {Ducourant, C.}, {Evans, D. W.}, {Eyer, L.}, {Guerra, R.}, {Hutton, A.}, {Jordi, C.}, {Klioner, S. A.}, {Lammers, U. L.}, {Lindegren, L.}, {Luri, X.}, {Mignard, F.}, {Panem, C.}, {Pourbaix, D.}, {Randich, S.}, {Sartoretti, P.}, {Soubiran, C.}, {Tanga, P.}, {Walton, N. A.}, {Bailer-Jones, C. A. L.}, {Bastian, U.}, {Drimmel, R.}, {Jansen, F.}, {Katz, D.}, {Lattanzi, M. G.}, {van Leeuwen, F.}, {Bakker, J.}, {Cacciari, C.}, {Castañeda, J.}, {De Angeli, F.}, {Fabricius, C.}, {Fouesneau, M.}, {Frémat, Y.}, {Galluccio, L.}, {Guerrier, A.}, {Heiter, U.}, {Masana, E.}, {Messineo, R.}, {Mowlavi, N.}, {Nicolas, C.}, {Nienartowicz, K.}, {Pailler, F.}, {Panuzzo, P.}, {Riclet, F.}, {Roux, W.}, {Seabroke, G. M.}, {Sordo, R.}, {Thévenin, F.}, {Gracia-Abril, G.}, {Portell, J.}, {Teyssier, D.}, {Altmann,
  M.}, {Andrae, R.}, {Audard, M.}, {Bellas-Velidis, I.}, {Benson, K.}, {Berthier, J.}, {Blomme, R.}, {Burgess, P. W.}, {Busonero, D.}, {Busso, G.}, {Cánovas, H.}, {Carry, B.}, {Cellino, A.}, {Cheek, N.}, {Clementini, G.}, {Damerdji, Y.}, {Davidson, M.}, {de Teodoro, P.}, {Nuñez Campos, M.}, {Delchambre, L.}, {Dell’Oro, A.}, {Esquej, P.}, {Fernández-Hernández, J.}, {Fraile, E.}, {Garabato, D.}, {García-Lario, P.}, {Gosset, E.}, {Haigron, R.}, {Halbwachs, J.-L.}, {Hambly, N. C.}, {Harrison, D. L.}, {Hernández, J.}, {Hestroffer, D.}, {Hodgkin, S. T.}, {Holl, B.}, {Janßen, K.}, {Jevardat de Fombelle, G.}, {Jordan, S.}, {Krone-Martins, A.}, {Lanzafame, A. C.}, {Löffler, W.}, {Marchal, O.}, {Marrese, P. M.}, {Moitinho, A.}, {Muinonen, K.}, {Osborne, P.}, {Pancino, E.}, {Pauwels, T.}, {Recio-Blanco, A.}, {Reylé, C.}, {Riello, M.}, {Rimoldini, L.}, {Roegiers, T.}, {Rybizki, J.}, {Sarro, L. M.}, {Siopis, C.}, {Smith, M.}, {Sozzetti, A.}, {Utrilla, E.}, {van Leeuwen, M.}, {Abbas, U.}, {Ábrahám, P.}, {Abreu
  Aramburu, A.}, {Aerts, C.}, {Aguado, J. J.}, {Ajaj, M.}, {Aldea-Montero, F.}, {Altavilla, G.}, {Álvarez, M. A.}, {Alves, J.}, {Anders, F.}, {Anderson, R. I.}, {Anglada Varela, E.}, {Antoja, T.}, {Baines, D.}, {Baker, S. G.}, {Balaguer-Núñez, L.}, {Balbinot, E.}, {Balog, Z.}, {Barache, C.}, {Barbato, D.}, {Barros, M.}, {Barstow, M. A.}, {Bartolomé, S.}, {Bassilana, J.-L.}, {Bauchet, N.}, {Becciani, U.}, {Bellazzini, M.}, {Berihuete, A.}, {Bernet, M.}, {Bertone, S.}, {Bianchi, L.}, {Binnenfeld, A.}, {Blanco-Cuaresma, S.}, {Blazere, A.}, {Boch, T.}, {Bombrun, A.}, {Bossini, D.}, {Bouquillon, S.}, {Bragaglia, A.}, {Bramante, L.}, {Breedt, E.}, {Bressan, A.}, {Brouillet, N.}, {Brugaletta, E.}, {Bucciarelli, B.}, {Burlacu, A.}, {Butkevich, A. G.}, {Buzzi, R.}, {Caffau, E.}, {Cancelliere, R.}, {Cantat-Gaudin, T.}, {Carballo, R.}, {Carlucci, T.}, {Carnerero, M. I.}, {Carrasco, J. M.}, {Casamiquela, L.}, {Castellani, M.}, {Castro-Ginard, A.}, {Chaoul, L.}, {Charlot, P.}, {Chemin, L.}, {Chiaramida, V.},
  {Chiavassa, A.}, {Chornay, N.}, {Comoretto, G.}, {Contursi, G.}, {Cooper, W. J.}, {Cornez, T.}, {Cowell, S.}, {Crifo, F.}, {Cropper, M.}, {Crosta, M.}, {Crowley, C.}, {Dafonte, C.}, {Dapergolas, A.}, {David, M.}, {David, P.}, {de Laverny, P.}, {De Luise, F.}, {De March, R.}, {De Ridder, J.}, {de Souza, R.}, {de Torres, A.}, {del Peloso, E. F.}, {del Pozo, E.}, {Delbo, M.}, {Delgado, A.}, {Delisle, J.-B.}, {Demouchy, C.}, {Dharmawardena, T. E.}, {Di Matteo, P.}, {Diakite, S.}, {Diener, C.}, {Distefano, E.}, {Dolding, C.}, {Edvardsson, B.}, {Enke, H.}, {Fabre, C.}, {Fabrizio, M.}, {Faigler, S.}, {Fedorets, G.}, {Fernique, P.}, {Fienga, A.}, {Figueras, F.}, {Fournier, Y.}, {Fouron, C.}, {Fragkoudi, F.}, {Gai, M.}, {Garcia-Gutierrez, A.}, {Garcia-Reinaldos, M.}, {García-Torres, M.}, {Garofalo, A.}, {Gavel, A.}, {Gavras, P.}, {Gerlach, E.}, {Geyer, R.}, {Giacobbe, P.}, {Gilmore, G.}, {Girona, S.}, {Giuffrida, G.}, {Gomel, R.}, {Gomez, A.}, {González-Núñez, J.}, {González-Santamaría, I.}, {González-Vidal,
  J. J.}, {Granvik, M.}, {Guillout, P.}, {Guiraud, J.}, {Gutiérrez-Sánchez, R.}, {Guy, L. P.}, {Hatzidimitriou, D.}, {Hauser, M.}, {Haywood, M.}, {Helmer, A.}, {Helmi, A.}, {Sarmiento, M. H.}, {Hidalgo, S. L.}, {Hilger, T.}, {Hładczuk, N.}, {Hobbs, D.}, {Holland, G.}, {Huckle, H. E.}, {Jardine, K.}, {Jasniewicz, G.}, {Jean-Antoine Piccolo, A.}, {Jiménez-Arranz, Ó.}, {Jorissen, A.}, {Juaristi Campillo, J.}, {Julbe, F.}, {Karbevska, L.}, {Kervella, P.}, {Khanna, S.}, {Kontizas, M.}, {Kordopatis, G.}, {Korn, A. J.}, {Kóspál, Á}, {Kostrzewa-Rutkowska, Z.}, {Kruszyńska, K.}, {Kun, M.}, {Laizeau, P.}, {Lambert, S.}, {Lanza, A. F.}, {Lasne, Y.}, {Le Campion, J.-F.}, {Lebreton, Y.}, {Lebzelter, T.}, {Leccia, S.}, {Leclerc, N.}, {Lecoeur-Taibi, I.}, {Liao, S.}, {Licata, E. L.}, {Lindstrøm, H. E. P.}, {Lister, T. A.}, {Livanou, E.}, {Lobel, A.}, {Lorca, A.}, {Loup, C.}, {Madrero Pardo, P.}, {Magdaleno Romeo, A.}, {Managau, S.}, {Mann, R. G.}, {Manteiga, M.}, {Marchant, J. M.}, {Marconi, M.}, {Marcos, J.},
  {Marcos Santos, M. M. S.}, {Marín Pina, D.}, {Marinoni, S.}, {Marocco, F.}, {Marshall, D. J.}, {Martin Polo, L.}, {Martín-Fleitas, J. M.}, {Marton, G.}, {Mary, N.}, {Masip, A.}, {Massari, D.}, {Mastrobuono-Battisti, A.}, {Mazeh, T.}, {McMillan, P. J.}, {Messina, S.}, {Michalik, D.}, {Millar, N. R.}, {Mints, A.}, {Molina, D.}, {Molinaro, R.}, {Molnár, L.}, {Monari, G.}, {Monguió, M.}, {Montegriffo, P.}, {Montero, A.}, {Mor, R.}, {Mora, A.}, {Morbidelli, R.}, {Morel, T.}, {Morris, D.}, {Muraveva, T.}, {Murphy, C. P.}, {Musella, I.}, {Nagy, Z.}, {Noval, L.}, {Ocaña, F.}, {Ogden, A.}, {Ordenovic, C.}, {Osinde, J. O.}, {Pagani, C.}, {Pagano, I.}, {Palaversa, L.}, {Palicio, P. A.}, {Pallas-Quintela, L.}, {Panahi, A.}, {Payne-Wardenaar, S.}, {Peñalosa Esteller, X.}, {Penttilä, A.}, {Pichon, B.}, {Piersimoni, A. M.}, {Pineau, F.-X.}, {Plachy, E.}, {Plum, G.}, {Poggio, E.}, {Prša, A.}, {Pulone, L.}, {Racero, E.}, {Ragaini, S.}, {Rainer, M.}, {Raiteri, C. M.}, {Rambaux, N.}, {Ramos, P.}, {Ramos-Lerate, M.},
  {Re Fiorentin, P.}, {Regibo, S.}, {Richards, P. J.}, {Rios Diaz, C.}, {Ripepi, V.}, {Riva, A.}, {Rix, H.-W.}, {Rixon, G.}, {Robichon, N.}, {Robin, A. C.}, {Robin, C.}, {Roelens, M.}, {Rogues, H. R. O.}, {Rohrbasser, L.}, {Romero-Gómez, M.}, {Rowell, N.}, {Royer, F.}, {Ruz Mieres, D.}, {Rybicki, K. A.}, {Sadowski, G.}, {Sáez Núñez, A.}, {Sagristà Sellés, A.}, {Sahlmann, J.}, {Salguero, E.}, {Samaras, N.}, {Sanchez Gimenez, V.}, {Sanna, N.}, {Santoveña, R.}, {Sarasso, M.}, {Schultheis, M.}, {Sciacca, E.}, {Segol, M.}, {Segovia, J. C.}, {Ségransan, D.}, {Semeux, D.}, {Shahaf, S.}, {Siddiqui, H. I.}, {Siebert, A.}, {Siltala, L.}, {Silvelo, A.}, {Slezak, E.}, {Slezak, I.}, {Smart, R. L.}, {Snaith, O. N.}, {Solano, E.}, {Solitro, F.}, {Souami, D.}, {Souchay, J.}, {Spagna, A.}, {Spina, L.}, {Spoto, F.}, {Steele, I. A.}, {Steidelmüller, H.}, {Stephenson, C. A.}, {Süveges, M.}, {Surdej, J.}, {Szabados, L.}, {Szegedi-Elek, E.}, {Taris, F.}, {Taylor, M. B.}, {Teixeira, R.}, {Tolomei, L.}, {Tonello, N.},
  {Torra, F.}, {Torra, J.}, {Torralba Elipe, G.}, {Trabucchi, M.}, {Tsounis, A. T.}, {Turon, C.}, {Ulla, A.}, {Unger, N.}, {Vaillant, M. V.}, {van Dillen, E.}, {van Reeven, W.}, {Vanel, O.}, {Vecchiato, A.}, {Viala, Y.}, {Vicente, D.}, {Voutsinas, S.}, {Weiler, M.}, {Wevers, T.}, {Wyrzykowski, Ł.}, {Yoldas, A.}, {Yvard, P.}, {Zhao, H.}, {Zorec, J.}, {Zucker, S.}, \& {Zwitter, T.}}]{Gaia2023}
{Gaia Collaboration}, {Vallenari, A.}, {Brown, A. G. A.}, {et~al.} 2023{\natexlab{b}}, \href{http://dx.doi.org/10.1051/0004-6361/202243940}{\color{blue}A\&A}, 674, A1

\bibitem[{{Grichener} {et~al.}(2022){Grichener}, {Kobayashi}, \& {Soker}}]{Grichener2022ApJ...926L...9G}
{Grichener}, A., {Kobayashi}, C., \& {Soker}, N. 2022, \href{http://dx.doi.org/10.3847/2041-8213/ac4f68}{\color{blue}\apjl}, \href{https://ui.adsabs.harvard.edu/abs/2022ApJ...926L...9G}{926, L9}

\bibitem[{Gudin {et~al.}(2021)Gudin, Shank, Beers, Yuan, Limberg, Roederer, Placco, Holmbeck, Dietz, Rasmussen, Hansen, Sakari, Ezzeddine, \& Frebel}]{Gudin_2021}
Gudin, D., Shank, D., Beers, T.~C., {et~al.} 2021, \href{http://dx.doi.org/10.3847/1538-4357/abd7ed}{\color{blue}APJ}, 908, 79

\bibitem[{Halevi \& Mösta(2018)}]{Halevi2018}
Halevi, G. \& Mösta, P. 2018, \href{http://dx.doi.org/10.1093/mnras/sty797}{\color{blue}MNRAS}, 477, 2366

\bibitem[{{Hansen} {et~al.}(2012){Hansen}, {Primas}, {Hartman}, {Kratz}, {Wanajo}, {Leibundgut}, {Farouqi}, {Hallmann}, {Christlieb}, \& {Nilsson}}]{hansen2012}
{Hansen}, C.~J., {Primas}, F., {Hartman}, H., {et~al.} 2012, \href{http://dx.doi.org/10.1051/0004-6361/201118643}{\color{blue}\aap}, \href{https://ui.adsabs.harvard.edu/abs/2012A\&A...545A..31H}{545, A31}

\bibitem[{Hansen {et~al.}(2018)Hansen, Holmbeck, Beers, Placco, Roederer, Frebel, Sakari, Simon, \& Thompson}]{Hansen_2018}
Hansen, T.~T., Holmbeck, E.~M., Beers, T.~C., {et~al.} 2018, \href{http://dx.doi.org/10.3847/1538-4357/aabacc}{\color{blue}APJ}, 858, 92

\bibitem[{{Hill} {et~al.}(2002){Hill}, {Plez}, {Cayrel}, {Beers}, {Nordstr{\"o}m}, {Andersen}, {Spite}, {Spite}, {Barbuy}, {Bonifacio}, {Depagne}, {Fran{\c{c}}ois}, \& {Primas}}]{Hill2002A&A...387..560H}
{Hill}, V., {Plez}, B., {Cayrel}, R., {et~al.} 2002, \href{http://dx.doi.org/10.1051/0004-6361:20020434}{\color{blue}\aap}, \href{https://ui.adsabs.harvard.edu/abs/2002A\&A...387..560H}{387, 560}

\bibitem[{Holmbeck \& Andrews(2024)}]{Holmbeck_2024}
Holmbeck, E.~M. \& Andrews, J.~J. 2024, \href{http://dx.doi.org/10.3847/1538-4357/ad1e52}{\color{blue}APJ}, 963, 110

\bibitem[{{Holmbeck} {et~al.}(2018){Holmbeck}, {Beers}, {Roederer}, {Placco}, {Hansen}, {Sakari}, {Sneden}, {Liu}, {Lee}, {Cowan}, \& {Frebel}}]{Holmbeck2018ApJ...859L..24H}
{Holmbeck}, E.~M., {Beers}, T.~C., {Roederer}, I.~U., {et~al.} 2018, \href{http://dx.doi.org/10.3847/2041-8213/aac722}{\color{blue}\apjl}, \href{https://ui.adsabs.harvard.edu/abs/2018ApJ...859L..24H}{859, L24}

\bibitem[{{Holmbeck} {et~al.}(2020){Holmbeck}, {Hansen}, {Beers}, {Placco}, {Whitten}, {Rasmussen}, {Roederer}, {Ezzeddine}, {Sakari}, {Frebel}, {Drout}, {Simon}, {Thompson}, {Bland-Hawthorn}, {Gibson}, {Grebel}, {Kordopatis}, {Kunder}, {Mel{\'e}ndez}, {Navarro}, {Reid}, {Seabroke}, {Steinmetz}, {Watson}, \& {Wyse}}]{Holmbeck2020ApJS..249...30H}
{Holmbeck}, E.~M., {Hansen}, T.~T., {Beers}, T.~C., {et~al.} 2020, \href{http://dx.doi.org/10.3847/1538-4365/ab9c19}{\color{blue}\apjs}, \href{https://ui.adsabs.harvard.edu/abs/2020ApJS..249...30H}{249, 30}

\bibitem[{{Ivans} {et~al.}(2006){Ivans}, {Simmerer}, {Sneden}, {Lawler}, {Cowan}, {Gallino}, \& {Bisterzo}}]{ivans2006}
{Ivans}, I.~I., {Simmerer}, J., {Sneden}, C., {et~al.} 2006, \href{http://dx.doi.org/10.1086/504069}{\color{blue}\apj}, \href{https://ui.adsabs.harvard.edu/abs/2006ApJ...645..613I}{645, 613}

\bibitem[{{Ivarsson} {et~al.}(2001){Ivarsson}, {Litz{\'e}n}, \& {Wahlgren}}]{ivarsson2001}
{Ivarsson}, S., {Litz{\'e}n}, U., \& {Wahlgren}, G.~M. 2001, \href{http://dx.doi.org/10.1238/Physica.Regular.064a00455}{\color{blue}\physscr}, \href{https://ui.adsabs.harvard.edu/abs/2001PhyS...64..455I}{64, 455}

\bibitem[{Ji {et~al.}(2019)Ji, Drout, \& Hansen}]{Ji_2019}
Ji, A.~P., Drout, M.~R., \& Hansen, T.~T. 2019, \href{http://dx.doi.org/10.3847/1538-4357/ab3291}{\color{blue}APJ}, 882, 40

\bibitem[{{Ji} \& {Frebel}(2018)}]{JiFrebel2018}
{Ji}, A.~P. \& {Frebel}, A. 2018, \href{http://dx.doi.org/10.3847/1538-4357/aab14a}{\color{blue}\apj}, \href{https://ui.adsabs.harvard.edu/abs/2018ApJ...856..138J}{856, 138}

\bibitem[{{Ji} {et~al.}(2020){Ji}, {Li}, {Simon}, {Marshall}, {Vivas}, {Pace}, {Bechtol}, {Drlica-Wagner}, {Koposov}, {Hansen}, {Allam}, {Gruendl}, {Johnson}, {McNanna}, {No{\"e}l}, {Tucker}, \& {Walker}}]{ji2020a}
{Ji}, A.~P., {Li}, T.~S., {Simon}, J.~D., {et~al.} 2020, \href{http://dx.doi.org/10.3847/1538-4357/ab6213}{\color{blue}\apj}, \href{https://ui.adsabs.harvard.edu/abs/2020ApJ...889...27J}{889, 27}

\bibitem[{{Jin} \& {Soker}(2024)}]{Jin2024ApJ...971..189J}
{Jin}, S. \& {Soker}, N. 2024, \href{http://dx.doi.org/10.3847/1538-4357/ad5f8e}{\color{blue}\apj}, \href{https://ui.adsabs.harvard.edu/abs/2024ApJ...971..189J}{971, 189}

\bibitem[{{Kasen} {et~al.}(2017){Kasen}, {Metzger}, {Barnes}, {Quataert}, \& {Ramirez-Ruiz}}]{Kasen2017}
{Kasen}, D., {Metzger}, B., {Barnes}, J., {Quataert}, E., \& {Ramirez-Ruiz}, E. 2017, \href{http://dx.doi.org/10.1038/nature24453}{\color{blue}\nat}, \href{https://ui.adsabs.harvard.edu/abs/2017Natur.551...80K}{551, 80}

\bibitem[{{Kelson}(2003)}]{Kelson2003PASP..115..688K}
{Kelson}, D.~D. 2003, \href{http://dx.doi.org/10.1086/375502}{\color{blue}\pasp}, \href{https://ui.adsabs.harvard.edu/abs/2003PASP..115..688K}{115, 688}

\bibitem[{{Kelson} {et~al.}(2000){Kelson}, {Illingworth}, {van Dokkum}, \& {Franx}}]{Kelson2000ApJ...531..184K}
{Kelson}, D.~D., {Illingworth}, G.~D., {van Dokkum}, P.~G., \& {Franx}, M. 2000, \href{http://dx.doi.org/10.1086/308440}{\color{blue}\apj}, \href{https://ui.adsabs.harvard.edu/abs/2000ApJ...531..184K}{531, 184}

\bibitem[{{Kramida} {et~al.}(2018){Kramida}, {Ralchenko}, {Nave}, \& {Reader}}]{kramida2018}
{Kramida}, A., {Ralchenko}, Y., {Nave}, G., \& {Reader}, J. 2018, in APS Meeting Abstracts, Vol. 2018, APS Division of Atomic, Molecular and Optical Physics Meeting Abstracts, \href{https://ui.adsabs.harvard.edu/abs/2018APS..DMPM01004K}{M01.004}

\bibitem[{Kratz {et~al.}(2007)Kratz, Farouqi, Pfeiffer, Truran, Sneden, \& Cowan}]{Kratz_2007}
Kratz, K.-L., Farouqi, K., Pfeiffer, B., {et~al.} 2007, \href{http://dx.doi.org/10.1086/517495}{\color{blue}APJ}, 662, 39

\bibitem[{{Kurucz} \& {Bell}(1995)}]{kurucz1995}
{Kurucz}, R. \& {Bell}, B. 1995, Atomic Line Data (R.L. Kurucz and B. Bell) Kurucz CD-ROM No. 23. Cambridge, \href{https://ui.adsabs.harvard.edu/abs/1995KurCD..23.....K}{23}

\bibitem[{{Kuske} {et~al.}(2025){Kuske}, {Arcones}, \& {Reichert}}]{Kuske2025}
{Kuske}, J., {Arcones}, A., \& {Reichert}, M. 2025, \href{http://dx.doi.org/10.3847/1538-4357/adf0f7}{\color{blue}\apj}, \href{https://ui.adsabs.harvard.edu/abs/2025ApJ...990...37K}{990, 37}

\bibitem[{{Lagae} {et~al.}(2025){Lagae}, {Amarsi}, \& {Lind}}]{Lagae2025}
{Lagae}, C., {Amarsi}, A.~M., \& {Lind}, K. 2025, \href{http://dx.doi.org/10.1051/0004-6361/202452874}{\color{blue}\aap}, \href{https://ui.adsabs.harvard.edu/abs/2025A&A...697A..60L}{697, A60}

\bibitem[{{Lattimer} \& {Schramm}(1974)}]{Lattimer1974ApJ...192L.145L}
{Lattimer}, J.~M. \& {Schramm}, D.~N. 1974, \href{http://dx.doi.org/10.1086/181612}{\color{blue}\apjl}, \href{https://ui.adsabs.harvard.edu/abs/1974ApJ...192L.145L}{192, L145}

\bibitem[{{Lawler} {et~al.}(2001{\natexlab{a}}){Lawler}, {Bonvallet}, \& {Sneden}}]{lawler2001a}
{Lawler}, J.~E., {Bonvallet}, G., \& {Sneden}, C. 2001{\natexlab{a}}, \href{http://dx.doi.org/10.1086/321549}{\color{blue}\apj}, \href{https://ui.adsabs.harvard.edu/abs/2001ApJ...556..452L}{556, 452}

\bibitem[{{Lawler} \& {Dakin}(1989)}]{lawler1989}
{Lawler}, J.~E. \& {Dakin}, J.~T. 1989, \href{http://dx.doi.org/10.1364/JOSAB.6.001457}{\color{blue}JOSA B}, \href{https://ui.adsabs.harvard.edu/abs/1989JOSAB...6.1457L}{6, 1457}

\bibitem[{{Lawler} {et~al.}(2007){Lawler}, {den Hartog}, {Labby}, {Sneden}, {Cowan}, \& {Ivans}}]{lawler2007}
{Lawler}, J.~E., {den Hartog}, E.~A., {Labby}, Z.~E., {et~al.} 2007, \href{http://dx.doi.org/10.1086/510368}{\color{blue}\apjs}, \href{https://ui.adsabs.harvard.edu/abs/2007ApJS..169..120L}{169, 120}

\bibitem[{{Lawler} {et~al.}(2006){Lawler}, {Den Hartog}, {Sneden}, \& {Cowan}}]{lawler2006}
{Lawler}, J.~E., {Den Hartog}, E.~A., {Sneden}, C., \& {Cowan}, J.~J. 2006, \href{http://dx.doi.org/10.1086/498213}{\color{blue}\apjs}, \href{https://ui.adsabs.harvard.edu/abs/2006ApJS..162..227L}{162, 227}

\bibitem[{{Lawler} {et~al.}(2013){Lawler}, {Guzman}, {Wood}, {Sneden}, \& {Cowan}}]{lawler2013}
{Lawler}, J.~E., {Guzman}, A., {Wood}, M.~P., {Sneden}, C., \& {Cowan}, J.~J. 2013, \href{http://dx.doi.org/10.1088/0067-0049/205/2/11}{\color{blue}\apjs}, \href{https://ui.adsabs.harvard.edu/abs/2013ApJS..205...11L}{205, 11}

\bibitem[{{Lawler} {et~al.}(2004){Lawler}, {Sneden}, \& {Cowan}}]{lawler2004}
{Lawler}, J.~E., {Sneden}, C., \& {Cowan}, J.~J. 2004, \href{http://dx.doi.org/10.1086/382068}{\color{blue}\apj}, \href{https://ui.adsabs.harvard.edu/abs/2004ApJ...604..850L}{604, 850}

\bibitem[{{Lawler} {et~al.}(2015){Lawler}, {Sneden}, \& {Cowan}}]{lawler2015}
{Lawler}, J.~E., {Sneden}, C., \& {Cowan}, J.~J. 2015, \href{http://dx.doi.org/10.1088/0067-0049/220/1/13}{\color{blue}\apjs}, \href{https://ui.adsabs.harvard.edu/abs/2015ApJS..220...13L}{220, 13}

\bibitem[{{Lawler} {et~al.}(2009){Lawler}, {Sneden}, {Cowan}, {Ivans}, \& {Den Hartog}}]{lawler2009}
{Lawler}, J.~E., {Sneden}, C., {Cowan}, J.~J., {Ivans}, I.~I., \& {Den Hartog}, E.~A. 2009, \href{http://dx.doi.org/10.1088/0067-0049/182/1/51}{\color{blue}\apjs}, \href{https://ui.adsabs.harvard.edu/abs/2009ApJS..182...51L}{182, 51}

\bibitem[{{Lawler} {et~al.}(2008){Lawler}, {Sneden}, {Cowan}, {Wyart}, {Ivans}, {Sobeck}, {Stockett}, \& {Den Hartog}}]{lawler2008}
{Lawler}, J.~E., {Sneden}, C., {Cowan}, J.~J., {et~al.} 2008, \href{http://dx.doi.org/10.1086/589834}{\color{blue}\apjs}, \href{https://ui.adsabs.harvard.edu/abs/2008ApJS..178...71L}{178, 71}

\bibitem[{{Lawler} {et~al.}(2017){Lawler}, {Sneden}, {Nave}, {Den Hartog}, {Emraho{\u{g}}lu}, \& {Cowan}}]{lawler2017}
{Lawler}, J.~E., {Sneden}, C., {Nave}, G., {et~al.} 2017, \href{http://dx.doi.org/10.3847/1538-4365/228/1/10}{\color{blue}\apjs}, \href{https://ui.adsabs.harvard.edu/abs/2017ApJS..228...10L}{228, 10}

\bibitem[{{Lawler} {et~al.}(2001{\natexlab{b}}){Lawler}, {Wickliffe}, {Cowley}, \& {Sneden}}]{lawler2001b}
{Lawler}, J.~E., {Wickliffe}, M.~E., {Cowley}, C.~R., \& {Sneden}, C. 2001{\natexlab{b}}, \href{http://dx.doi.org/10.1086/323001}{\color{blue}\apjs}, \href{https://ui.adsabs.harvard.edu/abs/2001ApJS..137..341L}{137, 341}

\bibitem[{{Lawler} {et~al.}(2001{\natexlab{c}}){Lawler}, {Wickliffe}, {den Hartog}, \& {Sneden}}]{lawler2001c}
{Lawler}, J.~E., {Wickliffe}, M.~E., {den Hartog}, E.~A., \& {Sneden}, C. 2001{\natexlab{c}}, \href{http://dx.doi.org/10.1086/323407}{\color{blue}\apj}, \href{https://ui.adsabs.harvard.edu/abs/2001ApJ...563.1075L}{563, 1075}

\bibitem[{{Lawler} {et~al.}(2014){Lawler}, {Wood}, {Den Hartog}, {Feigenson}, {Sneden}, \& {Cowan}}]{lawler2014a}
{Lawler}, J.~E., {Wood}, M.~P., {Den Hartog}, E.~A., {et~al.} 2014, \href{http://dx.doi.org/10.1088/0067-0049/215/2/20}{\color{blue}\apjs}, \href{https://ui.adsabs.harvard.edu/abs/2014ApJS..215...20L}{215, 20}

\bibitem[{{Lawler} {et~al.}(2001{\natexlab{d}}){Lawler}, {Wyart}, \& {Blaise}}]{lawler2001d}
{Lawler}, J.~E., {Wyart}, J.~F., \& {Blaise}, J. 2001{\natexlab{d}}, \href{http://dx.doi.org/10.1086/323000}{\color{blue}\apjs}, \href{https://ui.adsabs.harvard.edu/abs/2001ApJS..137..351L}{137, 351}

\bibitem[{{Li} {et~al.}(2007){Li}, {Chatelain}, {Holt}, {Rehse}, {Rosner}, \& {Scholl}}]{li2007}
{Li}, R., {Chatelain}, R., {Holt}, R.~A., {et~al.} 2007, \href{http://dx.doi.org/10.1088/0031-8949/76/5/028}{\color{blue}\physscr}, \href{https://ui.adsabs.harvard.edu/abs/2007PhyS...76..577L}{76, 577}

\bibitem[{{Lind} \& {Amarsi}(2024)}]{LindAmarsi2024}
{Lind}, K. \& {Amarsi}, A.~M. 2024, \href{http://dx.doi.org/10.1146/annurev-astro-052722-103557}{\color{blue}\araa}, \href{https://ui.adsabs.harvard.edu/abs/2024ARA\&A..62..475L}{62, 475}

\bibitem[{{Ljung} {et~al.}(2006){Ljung}, {Nilsson}, {Asplund}, \& {Johansson}}]{ljung2006}
{Ljung}, G., {Nilsson}, H., {Asplund}, M., \& {Johansson}, S. 2006, \href{http://dx.doi.org/10.1051/0004-6361:20065212}{\color{blue}\aap}, \href{https://ui.adsabs.harvard.edu/abs/2006A\&A...456.1181L}{456, 1181}

\bibitem[{{Mackereth} \& {Bovy}(2018)}]{Mackereth2018}
{Mackereth}, J.~T. \& {Bovy}, J. 2018, \href{http://dx.doi.org/10.1088/1538-3873/aadcdd}{\color{blue}\pasp}, \href{https://ui.adsabs.harvard.edu/abs/2018PASP..130k4501M}{130, 114501}

\bibitem[{{Mashonkina} \& {Gehren}(2000)}]{Mashonkina2000}
{Mashonkina}, L. \& {Gehren}, T. 2000, \aap, \href{https://ui.adsabs.harvard.edu/abs/2000A&A...364..249M}{364, 249}

\bibitem[{{Mashonkina} {et~al.}(2022){Mashonkina}, {Pakhomov}, {Sitnova}, {Jablonka}, {Yakovleva}, \& {Belyaev}}]{Mashonkina2022}
{Mashonkina}, L., {Pakhomov}, Y.~V., {Sitnova}, T., {et~al.} 2022, \href{http://dx.doi.org/10.1093/mnras/stab3189}{\color{blue}\mnras}, \href{https://ui.adsabs.harvard.edu/abs/2022MNRAS.509.3626M}{509, 3626}

\bibitem[{{Mashonkina} \& {Belyaev}(2019)}]{Mashonkina2019}
{Mashonkina}, L.~I. \& {Belyaev}, A.~K. 2019, \href{http://dx.doi.org/10.1134/S1063773719060033}{\color{blue}Astron. Lett}, \href{https://ui.adsabs.harvard.edu/abs/2019AstL...45..341M}{45, 341}

\bibitem[{{Matsuno} {et~al.}(2024){Matsuno}, {Amarsi}, {Carlos}, \& {Nissen}}]{Matsuno2024}
{Matsuno}, T., {Amarsi}, A.~M., {Carlos}, M., \& {Nissen}, P.~E. 2024, {VizieR Online Data Catalog: 3D non-LTE Mg abundance (Matsuno+, 2024)}, VizieR On-line Data Catalog: J/A+A/688/A72. Originally published in: 2024A\&A...688A..72M

\bibitem[{{McCall}(2004)}]{McCall2004}
{McCall}, M.~L. 2004, \href{http://dx.doi.org/10.1086/424933}{\color{blue}\aj}, \href{https://ui.adsabs.harvard.edu/abs/2004AJ....128.2144M}{128, 2144}

\bibitem[{{McMillan}(2017)}]{McMillan2017}
{McMillan}, P.~J. 2017, \href{http://dx.doi.org/10.1093/mnras/stw2759}{\color{blue}\mnras}, \href{https://ui.adsabs.harvard.edu/abs/2017MNRAS.465...76M}{465, 76}

\bibitem[{{McWilliam}(1998)}]{mcwilliam1998}
{McWilliam}, A. 1998, \href{http://dx.doi.org/10.1086/300289}{\color{blue}\aj}, \href{https://ui.adsabs.harvard.edu/abs/1998AJ....115.1640M}{115, 1640}

\bibitem[{Molero {et~al.}(2023)Molero, Magrini, Matteucci, Romano, Palla, Cescutti, Vázquez, \& Spitoni}]{Molero2023}
Molero, M., Magrini, L., Matteucci, F., {et~al.} 2023, \href{http://dx.doi.org/10.1093/mnras/stad1577}{\color{blue}MNRAS}, 523, 2974

\bibitem[{{Monty} {et~al.}(2024){Monty}, {Belokurov}, {Sanders}, {Hansen}, {Sakari}, {McKenzie}, {Myeong}, {Davies}, {Ardern-Arentsen}, \& {Massari}}]{Monty2024MNRAS.533.2420M}
{Monty}, S., {Belokurov}, V., {Sanders}, J.~L., {et~al.} 2024, \href{http://dx.doi.org/10.1093/mnras/stae1895}{\color{blue}\mnras}, \href{https://ui.adsabs.harvard.edu/abs/2024MNRAS.533.2420M}{533, 2420}

\bibitem[{{Monty} {et~al.}(2020){Monty}, {Venn}, {Lane}, {Lokhorst}, \& {Yong}}]{Monty2020MNRAS.497.1236M}
{Monty}, S., {Venn}, K.~A., {Lane}, J. M.~M., {Lokhorst}, D., \& {Yong}, D. 2020, \href{http://dx.doi.org/10.1093/mnras/staa1995}{\color{blue}\mnras}, \href{https://ui.adsabs.harvard.edu/abs/2020MNRAS.497.1236M}{497, 1236}

\bibitem[{{Morton}(2000)}]{morton2000}
{Morton}, D.~C. 2000, \href{http://dx.doi.org/10.1086/317349}{\color{blue}\apjs}, \href{https://ui.adsabs.harvard.edu/abs/2000ApJS..130..403M}{130, 403}

\bibitem[{{Mucciarelli} {et~al.}(2021){Mucciarelli}, {Bellazzini}, \& {Massari}}]{Mucciarelli2021A&A...653A..90M}
{Mucciarelli}, A., {Bellazzini}, M., \& {Massari}, D. 2021, \href{http://dx.doi.org/10.1051/0004-6361/202140979}{\color{blue}\aap}, \href{https://ui.adsabs.harvard.edu/abs/2021A\&A...653A..90M}{653, A90}

\bibitem[{Munari {et~al.}(2014)Munari, Henden, Frigo, Zwitter, Bienaymé, Bland-Hawthorn, Boeche, Freeman, Gibson, Gilmore, Grebel, Helmi, Kordopatis, Levine, Navarro, Parker, Reid, Seabroke, Siebert, Siviero, Smith, Steinmetz, Templeton, Terrell, Welch, Williams, \& Wyse}]{Munari_2014}
Munari, U., Henden, A., Frigo, A., {et~al.} 2014, \href{http://dx.doi.org/10.1088/0004-6256/148/5/81}{\color{blue}AJ}, 148, 81

\bibitem[{{Myeong} {et~al.}(2019){Myeong}, {Vasiliev}, {Iorio}, {Evans}, \& {Belokurov}}]{Myeong2019MNRAS.488.1235M}
{Myeong}, G.~C., {Vasiliev}, E., {Iorio}, G., {Evans}, N.~W., \& {Belokurov}, V. 2019, \href{http://dx.doi.org/10.1093/mnras/stz1770}{\color{blue}\mnras}, \href{https://ui.adsabs.harvard.edu/abs/2019MNRAS.488.1235M}{488, 1235}

\bibitem[{{Nilsson} \& {Ivarsson}(2008)}]{nilsson2008}
{Nilsson}, H. \& {Ivarsson}, S. 2008, \href{http://dx.doi.org/10.1051/0004-6361:200811019}{\color{blue}\aap}, \href{https://ui.adsabs.harvard.edu/abs/2008A\&A...492..609N}{492, 609}

\bibitem[{{Nilsson} {et~al.}(2002{\natexlab{a}}){Nilsson}, {Ivarsson}, {Johansson}, \& {Lundberg}}]{nilsson2002b}
{Nilsson}, H., {Ivarsson}, S., {Johansson}, S., \& {Lundberg}, H. 2002{\natexlab{a}}, \href{http://dx.doi.org/10.1051/0004-6361:20011540}{\color{blue}\aap}, \href{https://ui.adsabs.harvard.edu/abs/2002A\&A...381.1090N}{381, 1090}

\bibitem[{{Nilsson} {et~al.}(2002{\natexlab{b}}){Nilsson}, {Zhang}, {Lundberg}, {Johansson}, \& {Nordstr{\"o}m}}]{nilsson2002a}
{Nilsson}, H., {Zhang}, Z.~G., {Lundberg}, H., {Johansson}, S., \& {Nordstr{\"o}m}, B. 2002{\natexlab{b}}, \href{http://dx.doi.org/10.1051/0004-6361:20011597}{\color{blue}\aap}, \href{https://ui.adsabs.harvard.edu/abs/2002A\&A...382..368N}{382, 368}

\bibitem[{Nishimura {et~al.}(2006)Nishimura, Kotake, Hashimoto, Yamada, Nishimura, Fujimoto, \& Sato}]{Nishimura_2006}
Nishimura, S., Kotake, K., Hashimoto, M.-a., {et~al.} 2006, \href{http://dx.doi.org/10.1086/500786}{\color{blue}APJ}, 642, 410

\bibitem[{{Nordlander} \& {Lind}(2017)}]{Nordlander2017A&A}
{Nordlander}, T. \& {Lind}, K. 2017, \href{http://dx.doi.org/10.1051/0004-6361/201730427}{\color{blue}\aap}, \href{https://ui.adsabs.harvard.edu/abs/2017A\&A...607A..75N}{607, A75}

\bibitem[{{O'Brian} {et~al.}(1991){O'Brian}, {Wickliffe}, {Lawler}, {Whaling}, \& {Brault}}]{obrian1991}
{O'Brian}, T.~R., {Wickliffe}, M.~E., {Lawler}, J.~E., {Whaling}, W., \& {Brault}, J.~W. 1991, \href{http://dx.doi.org/10.1364/JOSAB.8.001185}{\color{blue}JOSA B}, \href{https://ui.adsabs.harvard.edu/abs/1991JOSAB...8.1185O}{8, 1185}

\bibitem[{{Ou} {et~al.}(2024){Ou}, {Ji}, {Frebel}, {Naidu}, \& {Limberg}}]{Ou2024ApJ...974..232O}
{Ou}, X., {Ji}, A.~P., {Frebel}, A., {Naidu}, R.~P., \& {Limberg}, G. 2024, \href{http://dx.doi.org/10.3847/1538-4357/ad6f9b}{\color{blue}\apj}, \href{https://ui.adsabs.harvard.edu/abs/2024ApJ...974..232O}{974, 232}

\bibitem[{Patel {et~al.}(2025)Patel, Metzger, Cehula, Burns, Goldberg, \& Thompson}]{Patel_2025}
Patel, A., Metzger, B.~D., Cehula, J., {et~al.} 2025, \href{http://dx.doi.org/10.3847/2041-8213/adc9b0}{\color{blue}APJ Letters}, 984, L29

\bibitem[{{Pehlivan Rhodin} {et~al.}(2017){Pehlivan Rhodin}, {Hartman}, {Nilsson}, \& {J{\"o}nsson}}]{pehlivan2017}
{Pehlivan Rhodin}, A., {Hartman}, H., {Nilsson}, H., \& {J{\"o}nsson}, P. 2017, \href{http://dx.doi.org/10.1051/0004-6361/201629849}{\color{blue}\aap}, \href{https://ui.adsabs.harvard.edu/abs/2017A\&A...598A.102P}{598, A102}

\bibitem[{{Placco} {et~al.}(2023){Placco}, {Almeida-Fernandes}, {Holmbeck}, {Roederer}, {Mardini}, {Hayes}, {Venn}, {Chiboucas}, {Deibert}, {Gamen}, {Heo}, {Jeong}, {Kalari}, {Martioli}, {Xu}, {Diaz}, {Gomez-Jimenez}, {Henderson}, {Prado}, {Quiroz}, {Ruiz-Carmona}, {Simpson}, {Urrutia}, {McConnachie}, {Pazder}, {Burley}, {Ireland}, {Waller}, {Berg}, {Robertson}, {Hartman}, {Jones}, {Labrie}, {Perez}, {Ridgway}, \& {Thomas-Osip}}]{Placco2023APJ}
{Placco}, V.~M., {Almeida-Fernandes}, F., {Holmbeck}, E.~M., {et~al.} 2023, \href{http://dx.doi.org/10.3847/1538-4357/ad077e}{\color{blue}\apj}, \href{https://ui.adsabs.harvard.edu/abs/2023ApJ...959...60P}{959, 60}

\bibitem[{{Placco} {et~al.}(2014){Placco}, {Frebel}, {Beers}, \& {Stancliffe}}]{Placco2014ApJ...797...21P}
{Placco}, V.~M., {Frebel}, A., {Beers}, T.~C., \& {Stancliffe}, R.~J. 2014, \href{http://dx.doi.org/10.1088/0004-637X/797/1/21}{\color{blue}\apj}, \href{https://ui.adsabs.harvard.edu/abs/2014ApJ...797...21P}{797, 21}

\bibitem[{{Placco} {et~al.}(2021){Placco}, {Sneden}, {Roederer}, {Lawler}, {Den Hartog}, {Hejazi}, {Maas}, \& {Bernath}}]{placco2021}
{Placco}, V.~M., {Sneden}, C., {Roederer}, I.~U., {et~al.} 2021, \href{http://dx.doi.org/10.3847/2515-5172/abf651}{\color{blue}RNAAS}, \href{https://ui.adsabs.harvard.edu/abs/2021RNAAS...5...92P}{5, 92}

\bibitem[{{Pols} {et~al.}(2012){Pols}, {Izzard}, {Stancliffe}, \& {Glebbeek}}]{Pols2012}
{Pols}, O.~R., {Izzard}, R.~G., {Stancliffe}, R.~J., \& {Glebbeek}, E. 2012, \href{http://dx.doi.org/10.1051/0004-6361/201219597}{\color{blue}\aap}, \href{https://ui.adsabs.harvard.edu/abs/2012A\&A...547A..76P}{547, A76}

\bibitem[{{Prasanna} {et~al.}(2024){Prasanna}, {Coleman}, \& {Thompson}}]{Prasanna2024}
{Prasanna}, T., {Coleman}, M. S.~B., \& {Thompson}, T.~A. 2024, \href{http://dx.doi.org/10.3847/1538-4357/ad4d90}{\color{blue}\apj}, \href{https://ui.adsabs.harvard.edu/abs/2024ApJ...973...91P}{973, 91}

\bibitem[{{Quinet} {et~al.}(2006){Quinet}, {Palmeri}, {Bi{\'e}mont}, {Jorissen}, {van Eck}, {Svanberg}, {Xu}, \& {Plez}}]{quinet2006}
{Quinet}, P., {Palmeri}, P., {Bi{\'e}mont}, {\'E}., {et~al.} 2006, \href{http://dx.doi.org/10.1051/0004-6361:20053852}{\color{blue}\aap}, \href{https://ui.adsabs.harvard.edu/abs/2006A\&A...448.1207Q}{448, 1207}

\bibitem[{{Rodríguez Díaz} {et~al.}(2024){Rodríguez Díaz}, {Lagae, Cis}, {Amarsi, Anish M.}, {Bigot, Lionel}, {Zhou, Yixiao}, {Aguirre Børsen-Koch, Víctor}, {Lind, Karin}, {Trampedach, Regner}, \& {Collet, Remo}}]{DiazLagae2024}
{Rodríguez Díaz}, {Lagae, Cis}, {Amarsi, Anish M.}, {et~al.} 2024, \href{http://dx.doi.org/10.1051/0004-6361/202348480}{\color{blue}A\&A}, 688, A212

\bibitem[{{Roederer} {et~al.}(2022){Roederer}, {Cowan}, {Pignatari}, {Beers}, {Den Hartog}, {Ezzeddine}, {Frebel}, {Hansen}, {Holmbeck}, {Mumpower}, {Placco}, {Sakari}, {Surman}, \& {Vassh}}]{Roederer2022ApJ}
{Roederer}, I.~U., {Cowan}, J.~J., {Pignatari}, M., {et~al.} 2022, \href{http://dx.doi.org/10.3847/1538-4357/ac85bc}{\color{blue}\apj}, \href{https://ui.adsabs.harvard.edu/abs/2022ApJ...936...84R}{936, 84}

\bibitem[{{Roederer} \& {Lawler}(2012)}]{roedererlawler2012}
{Roederer}, I.~U. \& {Lawler}, J.~E. 2012, \href{http://dx.doi.org/10.1088/0004-637X/750/1/76}{\color{blue}\apj}, \href{https://ui.adsabs.harvard.edu/abs/2012ApJ...750...76R}{750, 76}

\bibitem[{{Roederer} {et~al.}(2008){Roederer}, {Lawler}, {Sneden}, {Cowan}, {Sobeck}, \& {Pilachowski}}]{roederer2008}
{Roederer}, I.~U., {Lawler}, J.~E., {Sneden}, C., {et~al.} 2008, \href{http://dx.doi.org/10.1086/526452}{\color{blue}\apj}, \href{https://ui.adsabs.harvard.edu/abs/2008ApJ...675..723R}{675, 723}

\bibitem[{{Roederer} {et~al.}(2012){Roederer}, {Lawler}, {Sobeck}, {Beers}, {Cowan}, {Frebel}, {Ivans}, {Schatz}, {Sneden}, \& {Thompson}}]{roederer2012b}
{Roederer}, I.~U., {Lawler}, J.~E., {Sobeck}, J.~S., {et~al.} 2012, \href{http://dx.doi.org/10.1088/0067-0049/203/2/27}{\color{blue}\apjs}, \href{https://ui.adsabs.harvard.edu/abs/2012ApJS..203...27R}{203, 27}

\bibitem[{{Roederer} {et~al.}(2014){Roederer}, {Preston}, {Thompson}, {Shectman}, {Sneden}, {Burley}, \& {Kelson}}]{roederer2014a}
{Roederer}, I.~U., {Preston}, G.~W., {Thompson}, I.~B., {et~al.} 2014, \href{http://dx.doi.org/10.1088/0004-6256/147/6/136}{\color{blue}\aj}, \href{https://ui.adsabs.harvard.edu/abs/2014AJ....147..136R}{147, 136}

\bibitem[{{Roederer} {et~al.}(2018){Roederer}, {Sakari}, {Placco}, {Beers}, {Ezzeddine}, {Frebel}, \& {Hansen}}]{roederer2018a}
{Roederer}, I.~U., {Sakari}, C.~M., {Placco}, V.~M., {et~al.} 2018, \href{http://dx.doi.org/10.3847/1538-4357/aadd92}{\color{blue}\apj}, \href{https://ui.adsabs.harvard.edu/abs/2018ApJ...865..129R}{865, 129}

\bibitem[{Roederer {et~al.}(2023)Roederer, Vassh, Holmbeck, Mumpower, Surman, Cowan, Beers, Ezzeddine, Frebel, Hansen, Placco, \& Sakari}]{Roederer2023}
Roederer, I.~U., Vassh, N., Holmbeck, E.~M., {et~al.} 2023, \href{http://dx.doi.org/10.1126/science.adf1341}{\color{blue}Science}, 382, 1177

\bibitem[{{Ruffoni} {et~al.}(2014){Ruffoni}, {Den Hartog}, {Lawler}, {Brewer}, {Lind}, {Nave}, \& {Pickering}}]{ruffoni2014}
{Ruffoni}, M.~P., {Den Hartog}, E.~A., {Lawler}, J.~E., {et~al.} 2014, \href{http://dx.doi.org/10.1093/mnras/stu780}{\color{blue}\mnras}, \href{https://ui.adsabs.harvard.edu/abs/2014MNRAS.441.3127R}{441, 3127}

\bibitem[{{Sakari} {et~al.}(2018){Sakari}, {Placco}, {Farrell}, {Roederer}, {Wallerstein}, {Beers}, {Ezzeddine}, {Frebel}, {Hansen}, {Holmbeck}, {Sneden}, {Cowan}, {Venn}, {Davis}, {Matijevi{\v{c}}}, {Wyse}, {Bland-Hawthorn}, {Chiappini}, {Freeman}, {Gibson}, {Grebel}, {Helmi}, {Kordopatis}, {Kunder}, {Navarro}, {Reid}, {Seabroke}, {Steinmetz}, \& {Watson}}]{Sakari2018ApJ...868..110S}
{Sakari}, C.~M., {Placco}, V.~M., {Farrell}, E.~M., {et~al.} 2018, \href{http://dx.doi.org/10.3847/1538-4357/aae9df}{\color{blue}\apj}, \href{https://ui.adsabs.harvard.edu/abs/2018ApJ...868..110S}{868, 110}

\bibitem[{{Schlafly} \& {Finkbeiner}(2011)}]{schlafly2011}
{Schlafly}, E.~F. \& {Finkbeiner}, D.~P. 2011, \href{http://dx.doi.org/10.1088/0004-637X/737/2/103}{\color{blue}\apj}, \href{https://ui.adsabs.harvard.edu/abs/2011ApJ...737..103S}{737, 103}

\bibitem[{{Shah} {et~al.}(2024){Shah}, {Ezzeddine}, {Roederer}, {Hansen}, {Placco}, {Beers}, {Frebel}, {Ji}, {Holmbeck}, {Marshall}, \& {Sakari}}]{shah2024}
{Shah}, S.~P., {Ezzeddine}, R., {Roederer}, I.~U., {et~al.} 2024, \href{http://dx.doi.org/10.1093/mnras/stae255}{\color{blue}\mnras}, \href{https://ui.adsabs.harvard.edu/abs/2024MNRAS.529.1917S}{529, 1917}

\bibitem[{Shank {et~al.}(2023)Shank, Beers, Placco, Gudin, Catapano, Holmbeck, Ezzeddine, Roederer, Sakari, Frebel, \& Hansen}]{Shank_2023}
Shank, D., Beers, T.~C., Placco, V.~M., {et~al.} 2023, \href{http://dx.doi.org/10.3847/1538-4357/aca322}{\color{blue}APJ}, 943, 23

\bibitem[{{Siegel} {et~al.}(2019){Siegel}, {Barnes}, \& {Metzger}}]{Siegel2019Natur.569..241S}
{Siegel}, D.~M., {Barnes}, J., \& {Metzger}, B.~D. 2019, \href{http://dx.doi.org/10.1038/s41586-019-1136-0}{\color{blue}\nat}, \href{https://ui.adsabs.harvard.edu/abs/2019Natur.569..241S}{569, 241}

\bibitem[{{Sitnova} {et~al.}(2022){Sitnova}, {Yakovleva}, {Belyaev}, \& {Mashonkina}}]{Sitnova2022}
{Sitnova}, T.~M., {Yakovleva}, S.~A., {Belyaev}, A.~K., \& {Mashonkina}, L.~I. 2022, \href{http://dx.doi.org/10.1093/mnras/stac1813}{\color{blue}\mnras}, \href{https://ui.adsabs.harvard.edu/abs/2022MNRAS.515.1510S}{515, 1510}

\bibitem[{{Sk{\'u}lad{\'o}ttir} \& {Salvadori}(2020)}]{Sku020A&A...634L...2S}
{Sk{\'u}lad{\'o}ttir}, {\'A}. \& {Salvadori}, S. 2020, \href{http://dx.doi.org/10.1051/0004-6361/201937293}{\color{blue}\aap}, \href{https://ui.adsabs.harvard.edu/abs/2020A&A...634L...2S}{634, L2}

\bibitem[{{Sneden} {et~al.}(2008){Sneden}, {Cowan}, \& {Gallino}}]{sneden2008}
{Sneden}, C., {Cowan}, J.~J., \& {Gallino}, R. 2008, \href{http://dx.doi.org/10.1146/annurev.astro.46.060407.145207}{\color{blue}\araa}, \href{https://ui.adsabs.harvard.edu/abs/2008ARA\&A..46..241S}{46, 241}

\bibitem[{{Sneden} {et~al.}(2003){Sneden}, {Cowan}, {Lawler}, {Ivans}, {Burles}, {Beers}, {Primas}, {Hill}, {Truran}, {Fuller}, {Pfeiffer}, \& {Kratz}}]{sneden2003}
{Sneden}, C., {Cowan}, J.~J., {Lawler}, J.~E., {et~al.} 2003, \href{http://dx.doi.org/10.1086/375491}{\color{blue}\apj}, \href{https://ui.adsabs.harvard.edu/abs/2003ApJ...591..936S}{591, 936}

\bibitem[{Sneden {et~al.}(2009)Sneden, Lawler, Cowan, Ivans, \& Hartog}]{Sneden_2009}
Sneden, C., Lawler, J.~E., Cowan, J.~J., Ivans, I.~I., \& Hartog, E. A.~D. 2009, \href{http://dx.doi.org/10.1088/0067-0049/182/1/80}{\color{blue}APJ Supplement Series}, 182, 80

\bibitem[{{Sneden}(1973)}]{Sneden1973}
{Sneden}, C.~A. 1973, \href{https://ui.adsabs.harvard.edu/abs/1973PhDT.......180S}{{Carbon and Nitrogen Abundances in Metal-Poor Stars.}}, PhD thesis, University of Texas, Austin

\bibitem[{{Sobeck} {et~al.}(2011){Sobeck}, {Kraft}, {Sneden}, {Preston}, {Cowan}, {Smith}, {Thompson}, {Shectman}, \& {Burley}}]{sobeck2011}
{Sobeck}, J.~S., {Kraft}, R.~P., {Sneden}, C., {et~al.} 2011, \href{http://dx.doi.org/10.1088/0004-6256/141/6/175}{\color{blue}\aj}, \href{https://ui.adsabs.harvard.edu/abs/2011AJ....141..175S}{141, 175}

\bibitem[{{Sobeck} {et~al.}(2007){Sobeck}, {Lawler}, \& {Sneden}}]{sobeck2007}
{Sobeck}, J.~S., {Lawler}, J.~E., \& {Sneden}, C. 2007, \href{http://dx.doi.org/10.1086/519987}{\color{blue}\apj}, \href{https://ui.adsabs.harvard.edu/abs/2007ApJ...667.1267S}{667, 1267}

\bibitem[{{Soker}(2025)}]{Soker2025OJAp....8E..67S}
{Soker}, N. 2025, \href{http://dx.doi.org/10.33232/001c.138777}{\color{blue}OJAp}, \href{https://ui.adsabs.harvard.edu/abs/2025OJAp....8E..67S}{8, 67}

\bibitem[{{Spite} {et~al.}(2018){Spite}, {Spite}, {Barbuy}, {Bonifacio}, {Caffau}, \& {Fran{\c{c}}ois}}]{spite2018A&A...611A..30S}
{Spite}, F., {Spite}, M., {Barbuy}, B., {et~al.} 2018, \href{http://dx.doi.org/10.1051/0004-6361/201732096}{\color{blue}\aap}, \href{https://ui.adsabs.harvard.edu/abs/2018A&A...611A..30S}{611, A30}

\bibitem[{{Storm} \& {Bergemann}(2023)}]{Storm2023}
{Storm}, N. \& {Bergemann}, M. 2023, \href{http://dx.doi.org/10.1093/mnras/stad2488}{\color{blue}\mnras}, \href{https://ui.adsabs.harvard.edu/abs/2023MNRAS.525.3718S}{525, 3718}

\bibitem[{{Suntzeff}(1995)}]{Suntzeff1995}
{Suntzeff}, N.~B. 1995, \href{http://dx.doi.org/10.1086/133649}{\color{blue}\pasp}, \href{https://ui.adsabs.harvard.edu/abs/1995PASP..107..990S}{107, 990}

\bibitem[{{Tody}(1986)}]{Tody1986SPIE..627..733T}
{Tody}, D. 1986, in Society of Photo-Optical Instrumentation Engineers (SPIE) Conference Series, Vol. 627, Instrumentation in astronomy VI, ed. D.~L. {Crawford}, \href{https://ui.adsabs.harvard.edu/abs/1986SPIE..627..733T}{733}

\bibitem[{{Tody}(1993)}]{Tody1993ASPC...52..173T}
{Tody}, D. 1993, in Astronomical Society of the Pacific Conference Series, Vol.~52, Astronomical Data Analysis Software and Systems II, ed. R.~J. {Hanisch}, R.~J.~V. {Brissenden}, \& J.~{Barnes}, \href{https://ui.adsabs.harvard.edu/abs/1993ASPC...52..173T}{173}

\bibitem[{Travaglio {et~al.}(2004)Travaglio, Gallino, Arnone, Cowan, Jordan, \& Sneden}]{Travaglio_2004}
Travaglio, C., Gallino, R., Arnone, E., {et~al.} 2004, \href{http://dx.doi.org/10.1086/380507}{\color{blue}APJ}, 601, 864

\bibitem[{{Tsujimoto, T.} \& {Shigeyama, T.}(2014)}]{Tsujimoto2015}
{Tsujimoto, T.} \& {Shigeyama, T.} 2014, \href{http://dx.doi.org/10.1051/0004-6361/201423751}{\color{blue}A\&A}, 565, L5

\bibitem[{{van de Voort} {et~al.}(2022){van de Voort}, {Pakmor}, {Bieri}, \& {Grand}}]{VdV2022MNRAS.512.5258V}
{van de Voort}, F., {Pakmor}, R., {Bieri}, R., \& {Grand}, R. J.~J. 2022, \href{http://dx.doi.org/10.1093/mnras/stac710}{\color{blue}\mnras}, \href{https://ui.adsabs.harvard.edu/abs/2022MNRAS.512.5258V}{512, 5258}

\bibitem[{{van de Voort} {et~al.}(2020){van de Voort}, {Pakmor}, {Grand}, {Springel}, {G{\'o}mez}, \& {Marinacci}}]{VdV2020MNRAS.494.4867V}
{van de Voort}, F., {Pakmor}, R., {Grand}, R. J.~J., {et~al.} 2020, \href{http://dx.doi.org/10.1093/mnras/staa754}{\color{blue}\mnras}, \href{https://ui.adsabs.harvard.edu/abs/2020MNRAS.494.4867V}{494, 4867}

\bibitem[{{Vanbeveren} \& {Mennekens}(2024)}]{Vanbeveren2024BSRSL..93..338V}
{Vanbeveren}, D. \& {Mennekens}, N. 2024, \href{http://dx.doi.org/10.25518/0037-9565.12403}{\color{blue}Bull. Soc. R. Sci. Liège}, \href{https://ui.adsabs.harvard.edu/abs/2024BSRSL..93..338V}{93, 338}

\bibitem[{{Vasiliev}(2019)}]{Vasiliev2019}
{Vasiliev}, E. 2019, \href{http://dx.doi.org/10.1093/mnras/stz171}{\color{blue}\mnras}, \href{https://ui.adsabs.harvard.edu/abs/2019MNRAS.484.2832V}{484, 2832}

\bibitem[{{Villar} {et~al.}(2017){Villar}, {Guillochon}, {Berger}, {Metzger}, {Cowperthwaite}, {Nicholl}, {Alexander}, {Blanchard}, {Chornock}, {Eftekhari}, {Fong}, {Margutti}, \& {Williams}}]{Villar2017}
{Villar}, V.~A., {Guillochon}, J., {Berger}, E., {et~al.} 2017, \href{http://dx.doi.org/10.3847/2041-8213/aa9c84}{\color{blue}\apjl}, \href{https://ui.adsabs.harvard.edu/abs/2017ApJ...851L..21V}{851, L21}

\bibitem[{{Wang} \& {Burrows}(2023)}]{Wang2023}
{Wang}, T. \& {Burrows}, A. 2023, \href{http://dx.doi.org/10.3847/1538-4357/ace7b2}{\color{blue}\apj}, \href{https://ui.adsabs.harvard.edu/abs/2023ApJ...954..114W}{954, 114}

\bibitem[{{Westin} {et~al.}(2000){Westin}, {Sneden}, {Gustafsson}, \& {Cowan}}]{Westin2000ApJ...530..783W}
{Westin}, J., {Sneden}, C., {Gustafsson}, B., \& {Cowan}, J.~J. 2000, \href{http://dx.doi.org/10.1086/308407}{\color{blue}\apj}, \href{https://ui.adsabs.harvard.edu/abs/2000ApJ...530..783W}{530, 783}

\bibitem[{{Wickliffe} \& {Lawler}(1997)}]{wickliffe1997}
{Wickliffe}, M.~E. \& {Lawler}, J.~E. 1997, \href{http://dx.doi.org/10.1364/JOSAB.14.000737}{\color{blue}JOSA B}, \href{https://ui.adsabs.harvard.edu/abs/1997JOSAB..14..737W}{14, 737}

\bibitem[{{Wickliffe} {et~al.}(2000){Wickliffe}, {Lawler}, \& {Nave}}]{wickliffe2000}
{Wickliffe}, M.~E., {Lawler}, J.~E., \& {Nave}, G. 2000, \href{http://dx.doi.org/10.1016/S0022-4073(99)00173-9}{\color{blue}\jqsrt}, \href{https://ui.adsabs.harvard.edu/abs/2000JQSRT..66..363W}{66, 363}

\bibitem[{{Wickliffe} {et~al.}(1994){Wickliffe}, {Salih}, \& {Lawler}}]{wickliffe1994}
{Wickliffe}, M.~E., {Salih}, S., \& {Lawler}, J.~E. 1994, \href{http://dx.doi.org/10.1016/0022-4073(94)90108-2}{\color{blue}\jqsrt}, \href{https://ui.adsabs.harvard.edu/abs/1994JQSRT..51..545W}{51, 545}

\bibitem[{Winteler {et~al.}(2012)Winteler, Käppeli, Perego, Arcones, Vasset, Nishimura, Liebendörfer, \& Thielemann}]{Winteler_2012}
Winteler, C., Käppeli, R., Perego, A., {et~al.} 2012, \href{http://dx.doi.org/10.1088/2041-8205/750/1/L22}{\color{blue}APJ Letters}, 750, L22

\bibitem[{{Wood} {et~al.}(2014{\natexlab{a}}){Wood}, {Lawler}, {Den Hartog}, {Sneden}, \& {Cowan}}]{wood2014a}
{Wood}, M.~P., {Lawler}, J.~E., {Den Hartog}, E.~A., {Sneden}, C., \& {Cowan}, J.~J. 2014{\natexlab{a}}, \href{http://dx.doi.org/10.1088/0067-0049/214/2/18}{\color{blue}\apjs}, \href{https://ui.adsabs.harvard.edu/abs/2014ApJS..214...18W}{214, 18}

\bibitem[{{Wood} {et~al.}(2013){Wood}, {Lawler}, {Sneden}, \& {Cowan}}]{wood2013}
{Wood}, M.~P., {Lawler}, J.~E., {Sneden}, C., \& {Cowan}, J.~J. 2013, \href{http://dx.doi.org/10.1088/0067-0049/208/2/27}{\color{blue}\apjs}, \href{https://ui.adsabs.harvard.edu/abs/2013ApJS..208...27W}{208, 27}

\bibitem[{{Wood} {et~al.}(2014{\natexlab{b}}){Wood}, {Lawler}, {Sneden}, \& {Cowan}}]{wood2014b}
{Wood}, M.~P., {Lawler}, J.~E., {Sneden}, C., \& {Cowan}, J.~J. 2014{\natexlab{b}}, \href{http://dx.doi.org/10.1088/0067-0049/211/2/20}{\color{blue}\apjs}, \href{https://ui.adsabs.harvard.edu/abs/2014ApJS..211...20W}{211, 20}

\bibitem[{{Xu} {et~al.}(2007){Xu}, {Svanberg}, {Quinet}, {Palmeri}, \& {Bi{\'e}mont}}]{xu2007}
{Xu}, H.~L., {Svanberg}, S., {Quinet}, P., {Palmeri}, P., \& {Bi{\'e}mont}, {\'E}. 2007, \href{http://dx.doi.org/10.1016/j.jqsrt.2006.08.010}{\color{blue}\jqsrt}, \href{https://ui.adsabs.harvard.edu/abs/2007JQSRT.104...52X}{104, 52}

\bibitem[{{Xylakis-Dornbusch} {et~al.}(2024){Xylakis-Dornbusch}, {Hansen}, {Beers}, {Christlieb}, {Ezzeddine}, {Frebel}, {Holmbeck}, {Placco}, {Roederer}, {Sakari}, \& {Sneden}}]{Xylakis-Dornbusch2024A&A...688A.123X}
{Xylakis-Dornbusch}, T., {Hansen}, T.~T., {Beers}, T.~C., {et~al.} 2024, \href{http://dx.doi.org/10.1051/0004-6361/202449376}{\color{blue}\aap}, \href{https://ui.adsabs.harvard.edu/abs/2024A\&A...688A.123X}{688, A123}

\bibitem[{Zepeda {et~al.}(2023)Zepeda, Beers, Placco, Shank, Gudin, Hirai, Mardini, Pifer, Catapano, \& Calagna}]{Zepeda_2023}
Zepeda, J., Beers, T.~C., Placco, V.~M., {et~al.} 2023, \href{http://dx.doi.org/10.3847/1538-4357/acbbcc}{\color{blue}APJ}, 947, 23

\end{thebibliography}

\clearpage
\begin{appendix} 
\label{appendix}
\section{Stellar parameters}

\begin{center}
\begin{mdframed}[linewidth=0.pt,roundcorner=5pt,innermargin=0pt,outermargin=0pt]
    \centering
    \includegraphics[width=2\linewidth]{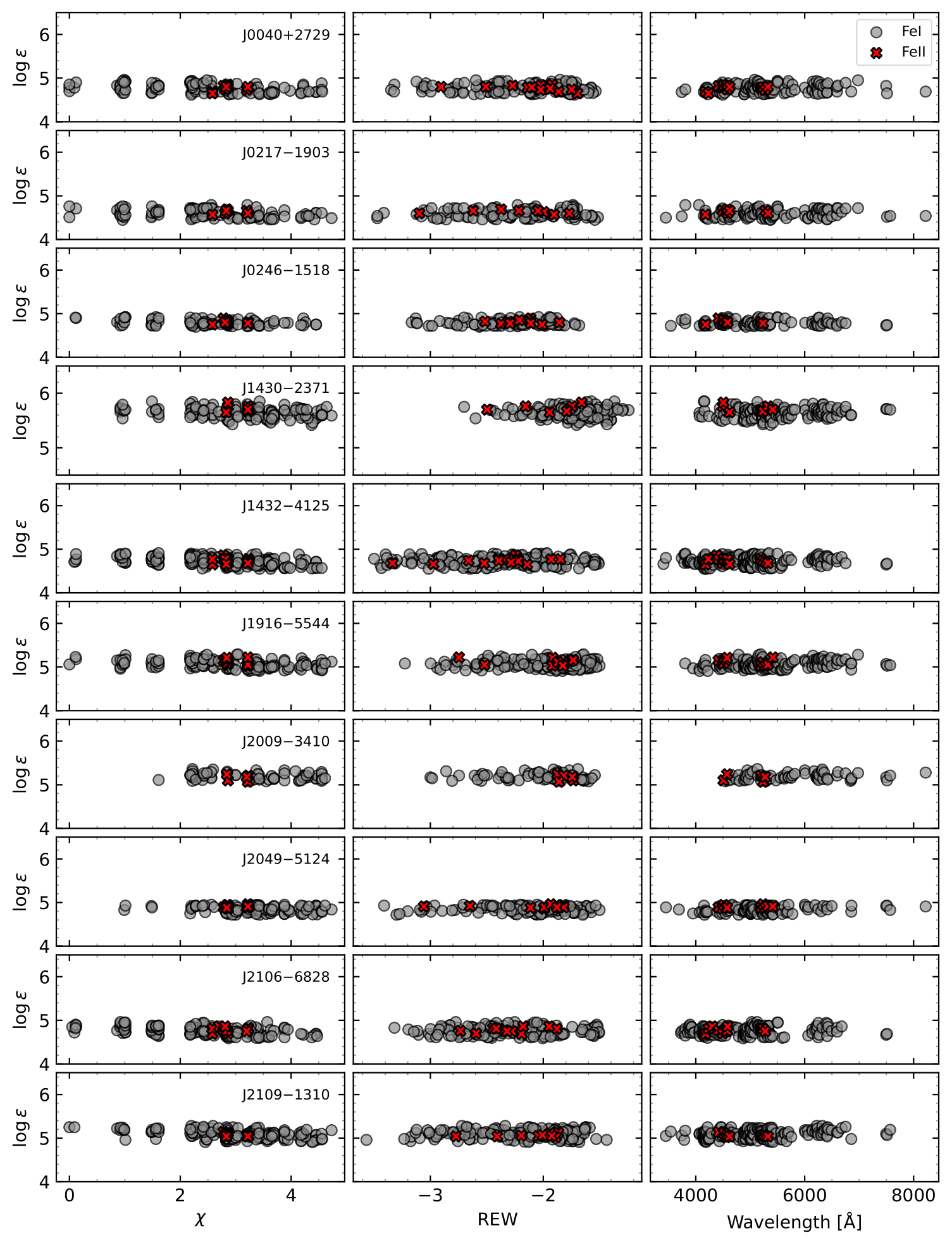}
    \captionof{figure}{Logarithmic abundances (\(\log \epsilon\)) of \ion{Fe}{I} and \ion{Fe}{II} lines used to determine the stellar parameters of the sample stars, shown as a function of excitation potential (left), reduced equivalent width (center), and wavelength (right).}
    \label{Ion_imbalance1}
\end{mdframed}
\end{center}

\clearpage

\section{Chemical abundances}
\begin{landscape}
\begin{table}[ht]

\centering
\tiny
\caption{Derived elemental abundances for the ten stars. 
}
\begin{tabular}{|l|rrrrrr|rrrrrr|rrrrrr|}
\hline
Element & \multicolumn{6}{c|}{J0040+2729} & \multicolumn{6}{c|}{J0217$$-$$1903} & \multicolumn{6}{c|}{J0246$$-$$1518} \\
 & $N$ & $\log\epsilon(X)$ & [X/H] & $\sigma_{[X/H]}$ & [X/Fe] & $\sigma_{[X/Fe]}$ & $N$ & $\log\epsilon(X)$ & [X/H] & $\sigma_{[X/H]}$ & [X/Fe] & $\sigma_{[X/Fe]}$ & $N$ & $\log\epsilon(X)$ & [X/H] & $\sigma_{[X/H]}$ & [X/Fe] & $\sigma_{[X/Fe]}$ \\
\hline
\ion{C-H}{} & -- & 5.55 & $-$2.81 & 0.18 & +0.02 & 0.17 & -- & 5.20 & $-$3.15 & 0.14 & $-$0.13 & 0.14 & -- & -- & -- & -- & -- & -- \\
\ion{C-N}{} & -- & -- & -- & -- & -- & -- & -- & 4.73 & $-$2.37 & 0.40 & +0.65 & 0.39 & -- & -- & -- & -- & -- & -- \\
\ion{$C_{corr}$}{} & -- & -- & -- & -- & +0.54 & -- & -- & -- & -- & -- & +0.38 & -- & -- & -- & -- & -- & -- & -- \\
\ion{N-H}{} & -- & -- & -- & -- & -- & -- & -- & 5.23 & $-$2.60 & 0.33 & +0.42 & 0.32 & -- & -- & -- & -- & -- & -- \\
\ion{OI}{} & 1 & 6.46 & $-$2.23 & 0.30 & +0.60 & 0.29 & 1 & 6.64 & $-$2.05 & 0.32 & +0.97 & 0.31 & 2 & 6.85 & $-$1.84 & 0.25 & +0.95 & 0.26 \\
\ion{NaI}{} & 2 & 3.67 & $-$2.56 & 0.31 & +0.26 & 0.31 & 4 & 3.52 & $-$2.69 & 0.13 & +0.33 & 0.13 & 4 & 3.63 & $-$2.60 & 0.11 & +0.19 & 0.12 \\
\ion{MgI}{} & 5 & 5.25 & $-$2.35 & 0.14 & +0.47 & 0.14 & 6 & 5.17 & $-$2.46 & 0.14 & +0.55 & 0.15 & 7 & 5.21 & $-$2.38 & 0.14 & +0.41 & 0.13 \\
\ion{AlI}{} & 2 & 3.04 & $-$3.37 & 0.25 & $-$0.55 & 0.25 & 1 & 3.06 & $-$3.39 & 0.32 & $-$0.37 & 0.32 & 1 & 2.80 & $-$3.65 & 0.66 & $-$0.85 & 0.64 \\
\ion{SiI}{} & 2 & 5.15 & $-$2.45 & 0.30 & +0.37 & 0.30 & 3 & 4.97 & $-$2.52 & 0.14 & +0.50 & 0.14 & 2 & 5.24 & $-$2.30 & 0.25 & +0.50 & 0.25 \\
\ion{KI}{} & 2 & 2.98 & $-$2.02 & 0.26 & +0.81 & 0.26 & 2 & 2.79 & $-$2.28 & 0.27 & +0.73 & 0.27 & 1 & 2.84 & $-$2.19 & 0.25 & +0.61 & 0.25 \\
\ion{CaI}{} & 18 & 3.90 & $-$2.43 & 0.14 & +0.40 & 0.14 & 17 & 3.78 & $-$2.54 & 0.14 & +0.47 & 0.13 & 25 & 3.94 & $-$2.39 & 0.14 & +0.40 & 0.13 \\
\ion{ScI}{} & 1 & 0.30 & $-$2.85 & 0.49 & $-$0.03 & 0.48 & -- & -- & -- & -- & -- & -- & -- & -- & -- & -- & -- & -- \\
\ion{ScII}{} & 9 & 0.45 & $-$2.72 & 0.15 & $-$0.05 & 0.12 & 9 & 0.35 & $-$2.75 & 0.14 & +0.04 & 0.16 & 13 & 0.45 & $-$2.57 & 0.14 & +0.11 & 0.15 \\
\ion{TiI}{} & 15 & 2.34 & $-$2.67 & 0.20 & +0.15 & 0.20 & 12 & 2.28 & $-$2.83 & 0.20 & +0.18 & 0.19 & 17 & 2.46 & $-$2.55 & 0.18 & +0.25 & 0.17 \\
\ion{TiII}{} & 25 & 2.46 & $-$2.35 & 0.17 & +0.32 & 0.13 & 17 & 2.40 & $-$2.37 & 0.22 & +0.42 & 0.17 & 31 & 2.53 & $-$2.33 & 0.15 & +0.34 & 0.15 \\
\ion{VI}{} & 3 & 1.02 & $-$2.91 & 0.14 & $-$0.09 & 0.14 & 2 & 0.90 & $-$3.03 & 0.24 & $-$0.01 & 0.24 & 2 & 1.17 & $-$2.76 & 0.24 & +0.04 & 0.24 \\
\ion{VII}{} & 8 & 1.40 & $-$2.49 & 0.11 & +0.18 & 0.13 & 7 & 1.28 & $-$2.62 & 0.12 & +0.18 & 0.16 & 10 & 1.35 & $-$2.53 & 0.12 & +0.15 & 0.14 \\
\ion{CrI}{} & 6 & 2.63 & $-$2.94 & 0.22 & $-$0.11 & 0.21 & 9 & 2.48 & $-$3.09 & 0.22 & $-$0.07 & 0.21 & 10 & 2.67 & $-$2.88 & 0.16 & $-$0.09 & 0.16 \\
\ion{CrII}{} & 4 & 2.94 & $-$2.70 & 0.14 & $-$0.03 & 0.12 & 2 & 2.76 & $-$2.82 & 0.25 & $-$0.03 & 0.24 & 2 & 2.88 & $-$2.76 & 0.25 & $-$0.08 & 0.24 \\
\ion{MnI}{} & 6 & 2.14 & $-$3.22 & 0.15 & $-$0.39 & 0.15 & 4 & 1.89 & $-$3.56 & 0.11 & $-$0.54 & 0.12 & 7 & 2.21 & $-$3.14 & 0.13 & $-$0.35 & 0.13 \\
\ion{MnII}{} & -- & -- & -- & -- & -- & -- & 4 & 1.88 & $-$3.54 & 0.14 & $-$0.74 & 0.16 & 3 & 2.44 & $-$2.99 & 0.13 & $-$0.31 & 0.14 \\
\ion{FeI}{} & 109 & 4.77 & $-$2.83 & 0.11 & 0.00 & 0.10 & 103 & 4.59 & $-$3.02 & 0.12 & 0.00 & 0.10 & 83 & 4.79 & $-$2.79 & 0.11 & 0.00 & 0.10 \\
\ion{FeII}{} & 14 & 4.78 & $-$2.67 & 0.14 & 0.00 & 0.10 & 8 & 4.63 & $-$2.79 & 0.16 & 0.00 & 0.10 & 10 & 4.81 & $-$2.68 & 0.14 & 0.00 & 0.10 \\
\ion{CoI}{} & 6 & 2.16 & $-$2.81 & 0.16 & +0.02 & 0.16 & 6 & 1.95 & $-$3.00 & 0.16 & +0.01 & 0.16 & 15 & 2.40 & $-$2.49 & 0.16 & +0.30 & 0.16 \\
\ion{NiI}{} & 15 & 3.52 & $-$2.80 & 0.15 & +0.03 & 0.16 & 15 & 3.28 & $-$2.99 & 0.14 & +0.03 & 0.14 & 12 & 3.46 & $-$2.71 & 0.14 & +0.09 & 0.13 \\
\ion{ZnI}{} & 2 & 2.02 & $-$2.55 & 0.25 & +0.27 & 0.25 & 2 & 1.95 & $-$2.62 & 0.26 & +0.39 & 0.26 & 3 & 2.19 & $-$2.38 & 0.16 & +0.42 & 0.16 \\
\ion{SrI}{} & -- & -- & -- & -- & -- & -- & 1 & 0.01 & $-$2.86 & 0.37 & +0.16 & 0.37 & -- & -- & -- & -- & -- & -- \\
\ion{SrII}{} & 2 & 0.32 & $-$2.53 & 0.26 & +0.15 & 0.27 & 3 & 0.63 & $-$2.12 & 0.20 & +0.68 & 0.21 & 3 & 0.65 & $-$2.20 & 0.20 & +0.48 & 0.19 \\
\ion{YII}{} & 21 & $-$0.36 & $-$2.54 & 0.11 & +0.14 & 0.14 & 22 & $-$0.15 & $-$2.38 & 0.11 & +0.42 & 0.16 & 21 & $-$0.11 & $-$2.25 & 0.14 & +0.43 & 0.15 \\
\ion{ZrII}{} & 15 & 0.35 & $-$2.20 & 0.11 & +0.48 & 0.14 & 17 & 0.55 & $-$1.99 & 0.11 & +0.80 & 0.16 & 19 & 0.54 & $-$2.02 & 0.12 & +0.66 & 0.14 \\
\ion{MoI}{} & 1 & $-$0.18 & $-$2.06 & 0.23 & +0.76 & 0.24 & 1 & $-$0.19 & $-$2.07 & 0.23 & +0.95 & 0.23 & 1 & $-$0.02 & $-$1.90 & 0.25 & +0.89 & 0.25 \\
\ion{RuI}{} & 1 & 0.14 & $-$1.61 & 0.34 & +1.22 & 0.34 & 3 & 0.12 & $-$1.52 & 0.14 & +1.50 & 0.14 & 2 & 0.42 & $-$1.16 & 0.42 & +1.64 & 0.41 \\
\ion{PdI}{} & -- & -- & -- & -- & -- & -- & 1 & $-$0.02 & $-$1.59 & 0.27 & +1.43 & 0.27 & -- & -- & -- & -- & -- & -- \\
\ion{BaII}{} & 5 & $-$0.19 & $-$2.30 & 0.19 & +0.38 & 0.16 & 6 & $-$0.03 & $-$2.14 & 0.16 & +0.65 & 0.17 & 5 & 0.32 & $-$1.90 & 0.20 & +0.78 & 0.18 \\
\ion{LaII}{} & 27 & $-$0.89 & $-$2.01 & 0.12 & +0.66 & 0.13 & 27 & $-$0.61 & $-$1.77 & 0.12 & +1.02 & 0.16 & 26 & $-$0.49 & $-$1.60 & 0.12 & +1.08 & 0.14 \\
\ion{CeII}{} & 35 & $-$0.57 & $-$2.15 & 0.11 & +0.53 & 0.14 & 30 & $-$0.28 & $-$1.83 & 0.11 & +0.97 & 0.16 & 43 & $-$0.16 & $-$1.75 & 0.11 & +0.93 & 0.14 \\
\ion{PrII}{} & 12 & $-$1.08 & $-$1.72 & 0.13 & +0.95 & 0.14 & 13 & $-$0.75 & $-$1.50 & 0.12 & +1.30 & 0.16 & 10 & $-$0.70 & $-$1.31 & 0.14 & +1.37 & 0.15 \\
\ion{NdII}{} & 64 & $-$0.44 & $-$1.86 & 0.10 & +0.81 & 0.14 & 59 & $-$0.15 & $-$1.54 & 0.10 & +1.25 & 0.16 & 61 & $-$0.06 & $-$1.48 & 0.11 & +1.20 & 0.14 \\
\ion{SmII}{} & 46 & $-$0.78 & $-$1.73 & 0.10 & +0.95 & 0.14 & 58 & $-$0.45 & $-$1.44 & 0.10 & +1.36 & 0.16 & 61 & $-$0.34 & $-$1.27 & 0.10 & +1.41 & 0.14 \\
\ion{EuII}{} & 6 & $-$1.10 & $-$1.67 & 0.15 & +1.00 & 0.13 & 8 & $-$0.73 & $-$1.22 & 0.11 & +1.57 & 0.16 & 7 & $-$0.66 & $-$1.25 & 0.18 & +1.43 & 0.16 \\
\ion{GdII}{} & 13 & $-$0.65 & $-$1.71 & 0.11 & +0.96 & 0.14 & 21 & $-$0.37 & $-$1.41 & 0.11 & +1.38 & 0.16 & 20 & $-$0.13 & $-$1.18 & 0.10 & +1.50 & 0.14 \\
\ion{TbII}{} & 3 & $-$1.40 & $-$1.47 & 0.16 & +1.21 & 0.13 & 4 & $-$1.14 & $-$1.36 & 0.22 & +1.43 & 0.19 & 2 & $-$0.86 & $-$1.07 & 0.26 & +1.61 & 0.25 \\
\ion{DyII}{} & 10 & $-$0.48 & $-$1.62 & 0.12 & +1.05 & 0.14 & 16 & $-$0.34 & $-$1.46 & 0.11 & +1.34 & 0.16 & 16 & 0.06 & $-$1.11 & 0.13 & +1.56 & 0.15 \\
\ion{HoII}{} & 4 & $-$1.23 & $-$1.74 & 0.22 & +0.94 & 0.16 & 6 & $-$1.03 & $-$1.45 & 0.16 & +1.34 & 0.17 & 6 & $-$0.71 & $-$1.23 & 0.17 & +1.45 & 0.16 \\
\ion{ErII}{} & 6 & $-$0.79 & $-$1.72 & 0.14 & +0.96 & 0.15 & 9 & $-$0.52 & $-$1.36 & 0.15 & +1.44 & 0.16 & 8 & $-$0.16 & $-$1.03 & 0.14 & +1.65 & 0.15 \\
\ion{TmII}{} & 4 & $-$1.54 & $-$1.53 & 0.19 & +1.14 & 0.16 & 4 & $-$1.42 & $-$1.46 & 0.20 & +1.34 & 0.18 & 5 & $-$0.96 & $-$1.08 & 0.11 & +1.60 & 0.14 \\
\ion{YbII}{} & 1 & $-$0.88 & $-$1.72 & 0.25 & +0.95 & 0.24 & 1 & $-$0.81 & $-$1.65 & 0.47 & +1.15 & 0.40 & 1 & $-$0.45 & $-$1.29 & 0.30 & +1.39 & 0.27 \\
\ion{LuII}{} & 1 & $-$1.50 & $-$1.60 & 0.32 & +1.07 & 0.31 & 1 & $-$0.96 & $-$1.06 & 0.27 & +1.73 & 0.24 & 2 & $-$0.90 & $-$1.02 & 0.35 & +1.66 & 0.31 \\
\ion{HfII}{} & 2 & $-$0.84 & $-$1.73 & 0.25 & +0.95 & 0.26 & 4 & $-$0.75 & $-$1.53 & 0.13 & +1.27 & 0.16 & 2 & $-$0.67 & $-$1.53 & 0.26 & +1.15 & 0.25 \\
\ion{OsI}{} & 1 & $-$0.20 & $-$1.60 & 0.25 & +1.22 & 0.26 & 1 & $-$0.01 & $-$1.41 & 0.27 & +1.61 & 0.26 & 1 & 0.22 & $-$1.18 & 0.26 & +1.61 & 0.27 \\
\ion{IrI}{} & 1 & $-$0.26 & $-$1.64 & 0.36 & +1.18 & 0.35 & 1 & $-$0.07 & $-$1.45 & 0.37 & +1.57 & 0.36 & 2 & 0.40 & $-$0.97 & 0.23 & +1.82 & 0.24 \\
\ion{ThII}{} & 1 & $-$1.52 & $-$1.54 & 0.28 & +1.14 & 0.25 & 3 & $-$1.24 & $-$1.23 & 0.11 & +1.57 & 0.16 & 1 & $-$1.04 & $-$1.06 & 0.26 & +1.61 & 0.24
\\
\hline
\end{tabular}
\tablefoot{For each element and star, we report: the number of spectral lines used ($N$), the logarithmic abundance $\log\epsilon(X)$, the abundance relative to hydrogen $[\mathrm{X}/\mathrm{H}]$, its uncertainty $\sigma_{[\mathrm{X}/\mathrm{H}]}$, the abundance relative to iron $[\mathrm{X}/\mathrm{Fe}]$, and the corresponding uncertainty $\sigma_{[\mathrm{X}/\mathrm{Fe}]}$. Reported values include both detections and upper limits, the latter indicated with ``<''. All abundances listed in this table are 1D and LTE.}
\label{tab:Abundances}
\end{table}
\end{landscape}

\begin{landscape}
\begin{table}[ht]
\centering
\tiny

    \begin{tabular}{|l|rrrrrr|rrrrrr|rrrrrr|}
    \multicolumn{2}{c}{\textbf{Table 6} Continue} \\
    \hline
    Element & \multicolumn{6}{c|}{J1430$$-$$2371} & \multicolumn{6}{c|}{J1432$$-$$4125} & \multicolumn{6}{c|}{J1916$$-$$5544} \\
 & $N$ & $\log\epsilon(X)$ & [X/H] & $\sigma_{[X/H]}$ & [X/Fe] & $\sigma_{[X/Fe]}$ & $N$ & $\log\epsilon(X)$ & [X/H] & $\sigma_{[X/H]}$ & [X/Fe] & $\sigma_{[X/Fe]}$ & $N$ & $\log\epsilon(X)$ & [X/H] & $\sigma_{[X/H]}$ & [X/Fe] & $\sigma_{[X/Fe]}$ \\
\hline
\ion{C-H}{} & --& 6.03 & $-$2.33 & 0.18 & $-$0.41 & 0.20 & --& 6.08 & $-$2.48 & 0.18 & +0.44 & 0.17 & --& 5.49 & $-$2.94 & 0.17 & $-$0.49 & 0.20 \\
\ion{C-N}{} & --& 6.10 & $-$1.73 & 0.36 & +0.20 & 0.36 & --& --& --& --& --& --& --& 5.86 & $-$1.98 & 0.45 & +0.54 & 0.46 \\
\ion{$C_{corr}$}{} & -- & -- & -- & -- & +0.19 & -- & -- & -- & -- & -- & +0.45 & -- & -- & -- & -- & -- & +0.26 & -- \\
\ion{N-H}{} & --& --& --& --& --& --& --& 4.64 & $-$3.15 & 0.34 & $-$0.23 & 0.34 & --& --& --& --& --& --\\
\ion{OI}{} & 4 & 7.62 & $-$1.10 & 0.15 & +0.82 & 0.16 & 2 & 7.19 & $-$1.50 & 0.25 & +1.42 & 0.26 & 3 & 7.22 & $-$1.47 & 0.14 & +1.04 & 0.14 \\
\ion{NaI}{} & 6 & 4.24 & $-$1.96 & 0.15 & $-$0.04 & 0.15 & 2 & 3.86 & $-$2.33 & 0.26 & +0.59 & 0.27 & 6 & 4.05 & $-$2.22 & 0.12 & +0.29 & 0.12 \\
\ion{MgI}{} & 4 & 6.25 & $-$1.29 & 0.19 & +0.63 & 0.19 & 7 & 5.31 & $-$2.26 & 0.14 & +0.66 & 0.13 & 3 & 5.86 & $-$1.78 & 0.15 & +0.73 & 0.16 \\
\ion{AlI}{} & 3 & 4.61 & $-$1.87 & 0.17 & +0.05 & 0.17 & 1 & 2.89 & $-$3.56 & 0.47 & $-$0.64 & 0.46 & --& --& --& --& --& --\\
\ion{SiI}{} & 10 & 6.05 & $-$1.47 & 0.12 & +0.45 & 0.13 & 3 & 5.41 & $-$2.05 & 0.11 & +0.88 & 0.12 & 7 & 5.63 & $-$1.90 & 0.12 & +0.61 & 0.12 \\
\ion{KI}{} & 2 & 3.78 & $-$1.42 & 0.40 & +0.50 & 0.39 & 1 & 2.98 & $-$2.05 & 0.25 & +0.87 & 0.25 & 1 & 3.16 & $-$1.87 & 0.31 & +0.64 & 0.31 \\
\ion{CaI}{} & 23 & 4.76 & $-$1.49 & 0.15 & +0.43 & 0.15 & 28 & 3.96 & $-$2.34 & 0.13 & +0.58 & 0.13 & 22 & 4.24 & $-$2.18 & 0.14 & +0.33 & 0.13 \\
\ion{ScI}{} & 4 & 0.87 & $-$2.30 & 0.19 & $-$0.37 & 0.19 & --& --& --& --& --& --& 3 & 0.29 & $-$2.82 & 0.15 & $-$0.30 & 0.14 \\
\ion{ScII}{} & 9 & 1.41 & $-$1.80 & 0.14 & $-$0.04 & 0.14 & 14 & 0.57 & $-$2.53 & 0.12 & +0.27 & 0.15 & 12 & 0.73 & $-$2.35 & 0.14 & +0.05 & 0.13 \\
\ion{TiI}{} & 8 & 3.17 & $-$1.63 & 0.25 & +0.30 & 0.25 & 19 & 2.53 & $-$2.50 & 0.18 & +0.43 & 0.17 & 16 & 2.55 & $-$2.49 & 0.18 & +0.03 & 0.18 \\
\ion{TiII}{} & 7 & 3.41 & $-$1.61 & 0.19 & +0.14 & 0.16 & 30 & 2.59 & $-$2.32 & 0.14 & +0.47 & 0.14 & 18 & 2.86 & $-$2.17 & 0.15 & +0.23 & 0.14 \\
\ion{VI}{} & 2 & 1.78 & $-$2.10 & 0.44 & $-$0.18 & 0.43 & 3 & 1.15 & $-$2.75 & 0.12 & +0.17 & 0.13 & 3 & 1.23 & $-$2.68 & 0.27 & $-$0.17 & 0.27 \\
\ion{VII}{} & 4 & 1.94 & $-$1.97 & 0.12 & $-$0.21 & 0.16 & 10 & 1.32 & $-$2.61 & 0.12 & +0.19 & 0.14 & 7 & 1.55 & $-$2.41 & 0.16 & $-$0.01 & 0.15 \\
\ion{CrI}{} & 8 & 3.66 & $-$2.04 & 0.37 & $-$0.12 & 0.36 & 10 & 2.72 & $-$2.89 & 0.14 & +0.04 & 0.14 & 11 & 2.99 & $-$2.62 & 0.14 & $-$0.10 & 0.13 \\
\ion{CrII}{} & 2 & 3.84 & $-$1.78 & 0.24 & $-$0.03 & 0.23 & 2 & 2.74 & $-$2.90 & 0.25 & $-$0.10 & 0.23 & 4 & 3.13 & $-$2.52 & 0.12 & $-$0.11 & 0.12 \\
\ion{MnI}{} & 6 & 3.28 & $-$2.23 & 0.22 & $-$0.30 & 0.22 & 8 & 2.16 & $-$3.21 & 0.12 & $-$0.28 & 0.13 & 6 & 2.39 & $-$3.06 & 0.18 & $-$0.55 & 0.18 \\
\ion{MnII}{} & --& --& --& --& --& --& 4 & 2.40 & $-$3.09 & 0.14 & $-$0.30 & 0.14 & 2 & 3.04 & $-$2.47 & 0.36 & $-$0.07 & 0.33 \\
\ion{FeI}{} & 131 & 5.64 & $-$1.92 & 0.11 & 0.00 & 0.10 & 174 & 4.73 & $-$2.92 & 0.12 & 0.00 & 0.10 & 137 & 5.07 & $-$2.51 & 0.11 & 0.00 & 0.10 \\
\ion{FeII}{} & 7 & 5.72 & $-$1.76 & 0.16 & 0.00 & 0.10 & 15 & 4.74 & $-$2.80 & 0.15 & 0.00 & 0.10 & 11 & 5.12 & $-$2.40 & 0.15 & 0.00 & 0.10 \\
\ion{CoI}{} & 6 & 2.67 & $-$2.38 & 0.20 & $-$0.46 & 0.20 & 19 & 2.47 & $-$2.55 & 0.13 & +0.37 & 0.13 & 7 & 2.46 & $-$2.48 & 0.20 & +0.03 & 0.19 \\
\ion{NiI}{} & 10 & 4.26 & $-$2.06 & 0.15 & $-$0.14 & 0.16 & 22 & 3.52 & $-$2.79 & 0.12 & +0.13 & 0.12 & 15 & 3.68 & $-$2.58 & 0.12 & $-$0.07 & 0.12 \\
\ion{CuI}{} & 1 & 1.88 & $-$2.31 & 0.29 & $-$0.39 & 0.29 & --& --& --& --& --& --& 1 & 1.14 & $-$3.05 & 0.28 & $-$0.54 & 0.27 \\
\ion{ZnI}{} & 2 & 2.68 & $-$1.88 & 0.25 & +0.04 & 0.26 & 3 & 2.10 & $-$2.46 & 0.16 & +0.46 & 0.16 & 2 & 2.10 & $-$2.47 & 0.25 & +0.04 & 0.25 \\
\ion{RbI}{} & 1 & 0.82 & $-$1.70 & 0.33 & +0.22 & 0.32 & --& --& --& --& --& --& --& --& --& --& --& --\\
\ion{SrI}{} & 1 & 0.56 & $-$2.31 & 0.37 & $-$0.39 & 0.37 & --& --& --& --& --& --& --& --& --& --& --& --\\
\ion{SrII}{} & 3 & 1.20 & $-$1.61 & 0.19 & +0.15 & 0.17 & 3 & 0.46 & $-$2.30 & 0.12 & +0.49 & 0.16 & 3 & 0.58 & $-$2.22 & 0.15 & +0.19 & 0.16 \\
\ion{YII}{} & 22 & 0.29 & $-$1.93 & 0.12 & $-$0.17 & 0.14 & 17 & $-$0.35 & $-$2.62 & 0.15 & +0.17 & 0.14 & 18 & $-$0.29 & $-$2.48 & 0.14 & $-$0.08 & 0.13 \\
\ion{ZrII}{} & 13 & 1.04 & $-$1.50 & 0.12 & +0.25 & 0.15 & 22 & 0.43 & $-$2.08 & 0.11 & +0.72 & 0.15 & 16 & 0.42 & $-$2.10 & 0.11 & +0.31 & 0.14 \\
\ion{MoI}{} & 2 & 0.32 & $-$1.54 & 0.34 & +0.38 & 0.34 & 1 & $-$0.19 & $-$2.07 & 0.28 & +0.85 & 0.29 & 2 & 0.10 & $-$2.05 & 0.25 & +0.46 & 0.25 \\
\ion{RuI}{} & 1 & 0.20 & $-$1.55 & 0.52 & +0.37 & 0.51 & 1 & 0.18 & $-$1.57 & 0.28 & +1.36 & 0.27 & 2 & 0.09 & $-$1.72 & 0.57 & +0.80 & 0.57 \\
\ion{RhI}{} & --& --& --& --& --& --& 1 & $-$0.24 & $-$1.16 & 0.35 & +1.77 & 0.34 & 1 & $-$0.31 & $-$1.22 & 0.37 & +1.29 & 0.37 \\
\ion{PdI}{} & 1 & 0.58 & $-$0.99 & 0.64 & +0.93 & 0.64 & 1 & $-$0.03 & $-$1.60 & 0.25 & +1.33 & 0.24 & --& <$-$0.01 & --& --& --& --\\
\ion{AgI}{} & --& --& --& --& --& --& 1 & $-$0.70 & $-$1.64 & 0.38 & +1.28 & 0.37 & --& --& --& --& --& --\\
\ion{BaII}{} & 3 & 0.47 & $-$1.71 & 0.23 & +0.05 & 0.19 & 3 & 0.32 & $-$1.94 & 0.27 & +0.86 & 0.22 & 4 & $-$0.20 & $-$2.29 & 0.29 & +0.11 & 0.21 \\
\ion{LaII}{} & 27 & $-$0.41 & $-$1.58 & 0.12 & +0.17 & 0.14 & 34 & $-$0.53 & $-$1.69 & 0.11 & +1.10 & 0.15 & 30 & $-$0.92 & $-$2.01 & 0.12 & +0.40 & 0.14 \\
\ion{CeII}{} & 28 & $-$0.09 & $-$1.70 & 0.11 & +0.06 & 0.15 & 43 & $-$0.18 & $-$1.80 & 0.10 & +0.99 & 0.15 & 37 & $-$0.61 & $-$2.24 & 0.11 & +0.16 & 0.14 \\
\ion{PrII}{} & 11 & $-$0.71 & $-$1.40 & 0.14 & +0.36 & 0.14 & 13 & $-$0.65 & $-$1.33 & 0.12 & +1.46 & 0.15 & 10 & $-$1.21 & $-$1.91 & 0.15 & +0.49 & 0.15 \\
\ion{NdII}{} & 61 & $-$0.02 & $-$1.46 & 0.11 & +0.30 & 0.15 & 68 & $-$0.02 & $-$1.47 & 0.11 & +1.33 & 0.15 & 61 & $-$0.49 & $-$1.90 & 0.11 & +0.50 & 0.14 \\
\ion{SmII}{} & 39 & $-$0.32 & $-$1.30 & 0.12 & +0.46 & 0.15 & 66 & $-$0.36 & $-$1.36 & 0.10 & +1.43 & 0.15 & 49 & $-$0.81 & $-$1.79 & 0.11 & +0.61 & 0.14 \\
\ion{EuII}{} & 9 & $-$0.65 & $-$1.14 & 0.14 & +0.62 & 0.14 & 7 & $-$0.70 & $-$1.31 & 0.20 & +1.49 & 0.16 & 11 & $-$1.04 & $-$1.59 & 0.12 & +0.81 & 0.14 \\
\ion{GdII}{} & 11 & $-$0.21 & $-$1.30 & 0.12 & +0.46 & 0.14 & 23 & $-$0.26 & $-$1.32 & 0.10 & +1.48 & 0.15 & 11 & $-$0.74 & $-$1.82 & 0.11 & +0.59 & 0.14 \\
\ion{TbII}{} & 3 & $-$1.26 & $-$1.46 & 0.27 & +0.30 & 0.24 & 2 & $-$1.03 & $-$1.31 & 0.25 & +1.49 & 0.24 & 3 & $-$1.45 & $-$1.81 & 0.20 & +0.60 & 0.16 \\
\ion{DyII}{} & 5 & 0.01 & $-$1.11 & 0.15 & +0.65 & 0.14 & 20 & $-$0.18 & $-$1.25 & 0.11 & +1.54 & 0.15 & 13 & $-$0.55 & $-$1.71 & 0.12 & +0.70 & 0.14 \\
\ion{HoII}{} & 3 & $-$0.97 & $-$1.36 & 0.19 & +0.40 & 0.17 & 6 & $-$0.87 & $-$1.33 & 0.17 & +1.46 & 0.16 & 4 & $-$1.40 & $-$1.90 & 0.15 & +0.50 & 0.13 \\
\ion{ErII}{} & 4 & $-$0.13 & $-$0.86 & 0.32 & +0.90 & 0.31 & 9 & $-$0.33 & $-$1.28 & 0.14 & +1.51 & 0.16 & 5 & $-$0.84 & $-$1.81 & 0.17 & +0.59 & 0.13 \\
\ion{TmII}{} & 2 & $-$1.44 & $-$1.44 & 0.36 & +0.32 & 0.32 & 6 & $-$1.07 & $-$1.18 & 0.11 & +1.61 & 0.15 & 3 & $-$1.79 & $-$1.78 & 0.22 & +0.62 & 0.17 \\
\ion{YbII}{} & 1 & $-$0.75 & $-$1.59 & 0.41 & +0.17 & 0.39 & 1 & $-$0.38 & $-$1.22 & 0.30 & +1.58 & 0.26 & 1 & $-$0.90 & $-$1.74 & 0.40 & +0.66 & 0.34 \\
\ion{LuII}{} & 1 & $-$0.95 & $-$1.05 & 0.28 & +0.71 & 0.26 & 2 & $-$1.19 & $-$1.37 & 0.31 & +1.43 & 0.27 & 1 & $-$1.54 & $-$1.64 & 0.25 & +0.77 & 0.27 \\
\ion{HfII}{} & 2 & $-$0.79 & $-$1.75 & 0.31 & +0.01 & 0.27 & 4 & $-$0.58 & $-$1.39 & 0.14 & +1.40 & 0.16 & 2 & $-$0.76 & $-$1.64 & 0.25 & +0.76 & 0.26 \\
\ion{OsI}{} & 1 & 0.23 & $-$1.18 & 0.42 & +0.75 & 0.41 & 1 & 0.20 & $-$1.20 & 0.28 & +1.73 & 0.29 & 1 & $-$0.29 & $-$1.69 & 0.31 & +0.82 & 0.30 \\
\ion{IrI}{} & 1 & $-$0.00 & $-$1.38 & 0.52 & +0.54 & 0.52 & 1 & 0.01 & $-$1.37 & 0.28 & +1.56 & 0.27 & 1 & $-$0.26 & $-$1.64 & 0.42 & +0.87 & 0.42 \\
\ion{PbI}{} & 1 & 0.13 & $-$1.62 & 0.29 & +0.30 & 0.29 & --& --& --& --& --& --& --& --& --& --& --& --\\
\ion{ThII}{} & 2 & $-$1.38 & $-$1.25 & 0.36 & +0.50 & 0.33 & 4 & $-$1.05 & $-$1.09 & 0.14 & +1.71 & 0.16 & 3 & $-$1.56 & $-$1.73 & 0.15 & +0.67 & 0.15 \\
    \hline
    \end{tabular}

\end{table}
\end{landscape}

\begin{landscape}
\begin{table}[ht]
\centering
\tiny

    \begin{tabular}{|l|rrrrrr|rrrrrr|rrrrrr|}
    \multicolumn{2}{c}{\textbf{Table 6} Continue} \\
    \hline
    Element & \multicolumn{6}{c|}{J2009$$-$$3410} & \multicolumn{6}{c|}{J2049$$-$$5124} & \multicolumn{6}{c|}{J2106$$-$$6828} \\
 & $N$ & $\log\epsilon(X)$ &[X/H] & $\sigma_{[X/H]}$ &  [X/Fe] & $\sigma_{[X/Fe]}$ & $N$ & $\log\epsilon(X)$& [X/H] & 
$\sigma_{[X/H]}$ & [X/Fe] & $\sigma_{[X/Fe]}$ & $N$ & $\log\epsilon(X)$ & [X/H] & $\sigma_{[X/H]}$ &  [X/Fe] & $\sigma_{[X/Fe]}$ \\
\hline
\ion{C-H}{} & -- & 5.37 & $-$3.01 & 0.13 & $-$0.67 & 0.15 & -- & 5.31 & $-$3.16 & 0.14 & $-$0.46 & 0.14 & -- & 6.07 & $-$2.36 & 0.14 & +0.54 & 0.15 \\
\ion{C-N}{} & -- & 6.01 & $-$1.82 & 0.31 & +0.52 & 0.32 & -- & 5.68 & $-$2.15 & 0.35 & +0.56 & 0.34 & -- & -- & -- & -- & -- & -- \\
\ion{$C_{corr}$}{} & -- & -- & -- & -- & +0.11 & -- & -- & -- & -- & -- & +0.29 & -- & -- & -- & -- & -- & +0.55 & -- \\
\ion{N-H}{} & -- & -- & -- & -- & -- & -- & -- & 5.52 & $-$2.31 & 0.27 & +0.40 & 0.26 & -- & -- & -- & -- & -- & -- \\
\ion{OI}{} & 3 & 7.14 & $-$1.56 & 0.117 & +0.78 & 0.13 & 3 & 6.95 & $-$1.74 & 0.12 & +0.96 & 0.12 & 3 & 7.04 & $-$1.67 & 0.16 & +1.23 & 0.17 \\
\ion{NaI}{} & 5 & 3.86 & $-$2.39 & 0.12 & $-$0.05 & 0.12 & 3 & 3.70 & $-$2.61 & 0.15 & +0.09 & 0.15 & 2 & 3.88 & $-$2.25 & 0.27 & +0.65 & 0.26 \\
\ion{MgI}{} & 5 & 5.84 & $-$1.72 & 0.13 & +0.62 & 0.12 & 3 & 5.57 & $-$2.06 & 0.14 & +0.65 & 0.14 & 8 & 5.29 & $-$2.29 & 0.14 & +0.61 & 0.13 \\
\ion{AlI}{} & -- & -- & -- & -- & -- & -- & 2 & 3.32 & $-$3.02 & 0.36 & $-$0.31 & 0.36 & 2 & 2.93 & $-$3.54 & 0.23 & $-$0.65 & 0.24 \\
\ion{SiI}{} & 8 & 5.72 & $-$1.77 & 0.12 & +0.57 & 0.12 & 5 & 5.47 & $-$2.07 & 0.12 & +0.64 & 0.12 & 1 & 5.23 & $-$2.28 & 0.33 & +0.62 & 0.32 \\
\ion{KI}{} & 1 & 3.28 & $-$1.75 & 0.37 & +0.59 & 0.36 & 1 & 3.02 & $-$2.01 & 0.21 & +0.70 & 0.28 & 2 & 2.88 & $-$2.15 & 0.26 & +0.75 & 0.25 \\
\ion{CaI}{} & 19 & 4.33 & $-$1.93 & 0.14 & +0.41 & 0.13 & 24 & 4.02 & $-$2.40 & 0.14 & +0.31 & 0.13 & 21 & 4.01 & $-$2.31 & 0.14 & +0.59 & 0.13 \\
\ion{ScI}{} & 1 & 0.35 & $-$2.80 & 0.49 & $-$0.46 & 0.48 & 3 & 0.37 & $-$2.78 & 0.14 & $-$0.08 & 0.14 & -- & -- & -- & -- & -- & -- \\
\ion{ScII}{} & 6 & 0.82 & $-$2.35 & 0.11 & $-$0.01 & 0.11 & 13 & 0.58 & $-$2.54 & 0.12 & +0.03 & 0.12 & 11 & 0.57 & $-$2.54 & 0.12 & +0.24 & 0.16 \\
\ion{TiI}{} & 11 & 2.72 & $-$2.12 & 0.16 & +0.22 & 0.17 & 15 & 2.44 & $-$2.59 & 0.20 & +0.11 & 0.20 & 17 & 2.52 & $-$2.51 & 0.18 & +0.39 & 0.17 \\
\ion{TiII}{} & 9 & 2.84 & $-$2.04 & 0.14 & +0.29 & 0.13 & 24 & 2.65 & $-$2.34 & 0.13 & +0.23 & 0.13 & 30 & 2.63 & $-$2.33 & 0.14 & +0.45 & 0.16 \\
\ion{VI}{} & 2 & 1.18 & $-$2.72 & 0.51 & $-$0.38 & 0.50 & 3 & 1.20 & $-$2.70 & 0.18 & +0.01 & 0.17 & 2 & 1.15 & $-$2.82 & 0.23 & +0.08 & 0.24 \\
\ion{VII}{} & 3 & 1.45 & $-$2.50 & 0.11 & $-$0.15 & 0.11 & 8 & 1.53 & $-$2.40 & 0.12 & +0.17 & 0.12 & 7 & 1.34 & $-$2.60 & 0.11 & +0.18 & 0.16 \\
\ion{CrI}{} & 8 & 3.13 & $-$2.31 & 0.23 & +0.03 & 0.23 & 11 & 2.81 & $-$2.72 & 0.17 & $-$0.02 & 0.17 & 11 & 2.74 & $-$2.82 & 0.16 & +0.08 & 0.15 \\
\ion{CrII}{} & 1 & 3.25 & $-$2.39 & 0.23 & $-$0.05 & 0.23 & 3 & 3.02 & $-$2.72 & 0.12 & $-$0.15 & 0.12 & 4 & 2.91 & $-$2.77 & 0.16 & +0.01 & 0.14 \\
\ion{MnI}{} & 6 & 2.67 & $-$2.84 & 0.20 & $-$0.51 & 0.19 & 7 & 2.35 & $-$2.99 & 0.13 & $-$0.29 & 0.13 & 5 & 2.08 & $-$3.27 & 0.16 & $-$0.37 & 0.16 \\
\ion{MnII}{} & 1 & 3.10 & $-$2.33 & 0.35 & +0.01 & 0.36 & 2 & 2.08 & $-$3.35 & 0.25 & $-$0.78 & 0.27 & 2 & 2.31 & $-$3.12 & 0.26 & $-$0.34 & 0.24 \\
\ion{FeI}{} & 65 & 5.19 & $-$2.34 & 0.11 & 0.00 & 0.10 & 99 & 4.84 & $-$2.70 & 0.10 & 0.00 & 0.10 & 149 & 4.76 & $-$2.90 & 0.12 & 0.00 & 0.10 \\
\ion{FeII}{} & 5 & 5.16 & $-$2.34 & 0.11 & 0.00 & 0.10 & 11 & 4.91 & $-$2.57 & 0.13 & 0.00 & 0.10 & 11 & 4.78 & $-$2.78 & 0.16 & 0.00 & 0.10 \\
\ion{CoI}{} & 4 & 2.25 & $-$2.60 & 0.14 & $-$0.26 & 0.14 & 8 & 2.39 & $-$2.56 & 0.14 & +0.14 & 0.13 & 18 & 2.47 & $-$2.58 & 0.16 & +0.32 & 0.16 \\
\ion{NiI}{} & 14 & 3.82 & $-$2.51 & 0.15 & $-$0.17 & 0.14 & 17 & 3.59 & $-$2.62 & 0.11 & +0.08 & 0.11 & 9 & 3.54 & $-$2.75 & 0.13 & +0.15 & 0.13 \\
\ion{CuI}{} & 1 & 1.29 & $-$2.90 & 0.38& $-$0.56 & 0.37 & 1 & 0.79 & $-$3.40 & 0.25 & $-$0.70 & 0.24 & -- & -- & -- & -- & -- & -- \\
\ion{ZnI}{} & 2 & 2.03 & $-$2.52 & 0.23 & $-$0.18 & 0.24 & 2 & 2.04 & $-$2.52 & 0.25 & +0.18 & 0.24 & 2 & 2.07 & $-$2.50 & 0.26 & +0.40 & 0.26 \\
\ion{SrI}{} & 1 & 0.33 & $-$2.54 & 0.49 & $-$0.20 & 0.48 & 1 & $-$0.31 & $-$3.18 & 0.26 & $-$0.48 & 0.26 & -- & -- & -- & -- & -- & -- \\
\ion{SrII}{} & 3 & 0.77 & $-$2.15 & 0.23 & +0.19 & 0.23 & 2 & $-$0.16 & $-$3.04 & 0.33 & $-$0.47 & 0.29 & 2 & 0.56 & $-$2.27 & 0.37 & +0.51 & 0.31 \\
\ion{YII}{} & 16 & $-$0.05 & $-$2.32 & 0.11 & +0.02 & 0.12 & 23 & $-$0.84 & $-$3.02 & 0.12 & $-$0.45 & 0.12 & 11 & $-$0.33 & $-$2.61 & 0.16 & +0.17 & 0.15 \\
\ion{ZrII}{} & 12 & 0.63 & $-$1.97 & 0.11 & +0.37 & 0.11 & 10 & $-$0.06 & $-$2.61 & 0.12 & $-$0.04 & 0.13 & 10 & 0.44 & $-$2.08 & 0.16 & +0.70 & 0.15 \\
\ion{MoI}{} & 2 & $-$0.14 & $-$1.97 & 0.48 & +0.37 & 0.48 & 1 & $-$0.46 & $-$2.33 & 0.23 & +0.37 & 0.23 & -- & -- & -- & -- & -- & -- \\
\ion{RuI}{} & 2 & 0.06 & $-$1.71 & 0.55 & +0.63 & 0.55 & 1 & $-$0.14 & $-$1.89 & 0.41 & +0.81 & 0.41 & -- & -- & -- & -- & -- & -- \\
\ion{RhI}{} & -- & <0.48 & -- & -- & -- & -- &-- & -- & -- & -- & -- & -- & -- & -- & -- & -- & -- & -- \\
\ion{PdI}{} & 1 & $-$0.17 & $-$1.74 & 0.58 & +0.60 & 0.57 & 1 & $-$0.38 & $-$1.95 & 0.44 & +0.75 & 0.44 & -- & -- & -- & -- & -- & -- \\
\ion{BaII}{} & 4 & 0.05 & $-$2.15 & 0.13 & +0.18 & 0.12 & 5 & $-$0.61 & $-$2.71 & 0.21 & $-$0.14 & 0.17 & 5 & 0.01 & $-$2.25 & 0.19 & +0.52 & 0.18 \\
\ion{LaII}{} & 26 & $-$0.58 & $-$1.72 & 0.10 & +0.62 & 0.11 & 21 & $-$1.38 & $-$2.51 & 0.12 & +0.06 & 0.13 & 7 & $-$0.92 & $-$2.08 & 0.16 & +0.70 & 0.14 \\
\ion{CeII}{} & 28 & $-$0.34 & $-$1.90 & 0.10 & +0.44 & 0.11 & 26 & $-$1.05 & $-$2.62 & 0.11 & $-$0.05 & 0.13 & 5 & $-$0.57 & $-$2.21 & 0.17 & +0.56 & 0.16 \\
\ion{PrII}{} & 12 & $-$0.85 & $-$1.55 & 0.12 & +0.79 & 0.12 & 4 & $-$1.51 & $-$2.16 & 0.13 & +0.41 & 0.13 & 1 & $-$1.05 & $-$1.77 & 0.27 & +1.00 & 0.26 \\
\ion{NdII}{} & 56 & $-$0.16 & $-$1.58 & 0.11 & +0.76 & 0.11 & 52 & $-$0.94 & $-$2.37 & 0.10 & +0.20 & 0.13 & 17 & $-$0.42 & $-$1.83 & 0.12 & +0.94 & 0.16 \\
\ion{SmII}{} & 48 & $-$0.43 & $-$1.33 & 0.11 & +1.01 & 0.11 & 39 & $-$1.22 & $-$2.20 & 0.11 & +0.37 & 0.13 & 8 & $-$0.76 & $-$1.78 & 0.14 & +0.99 & 0.15 \\
\ion{EuII}{} & 10 & $-$0.57 & $-$1.14 & 0.11 & +1.20 & 0.11 & 6 & $-$1.61 & $-$2.18 & 0.14 & +0.39 & 0.13 & 5 & $-$1.17 & $-$1.68 & 0.12 & +1.10 & 0.16 \\
\ion{GdII}{} & 10 & $-$0.25 & $-$1.28 & 0.11 & +1.06 & 0.11 & 10 & $-$1.05 & $-$2.10 & 0.11 & +0.47 & 0.13 & 4 & $-$0.61 & $-$1.71 & 0.12 & +1.06 & 0.16 \\
\ion{TbII}{} & 3 & $-$1.24 & $-$1.57 & 0.22 & +0.77 & 0.22 & 2 & $-$1.92 & $-$2.08 & 0.25 & +0.49 & 0.23 & 2 & $-$1.09 & $-$1.40 & 0.21 & +1.38 & 0.22 \\
\ion{DyII}{} & 6 & $-$0.01 & $-$1.13 & 0.12 & +1.21 & 0.12 & 10 & $-$0.99 & $-$2.12 & 0.12 & +0.45 & 0.13 & 8 & $-$0.44 & $-$1.53 & 0.11 & +1.25 & 0.16 \\
\ion{HoII}{} & 4 & $-$0.86 & $-$1.35 & 0.12 & +0.99 & 0.13 & 3 & $-$1.85 & $-$2.36 & 0.14 & +0.21 & 0.13 & 3 & $-$1.18 & $-$1.73 & 0.19 & +1.05 & 0.16 \\
\ion{ErII}{} & 5 & $-$0.26 & $-$1.21 & 0.12 & +1.13 & 0.13 & 4 & $-$1.35 & $-$2.29 & 0.13 & +0.28 & 0.13 & 3 & $-$0.63 & $-$1.47 & 0.20 & +1.31 & 0.17 \\
\ion{TmII}{} & 5 & $-$1.44 & $-$1.39 & 0.20 & +0.95 & 0.20 & 2 & $-$2.15 & $-$2.29 & 0.24 & +0.28 & 0.24 & -- & -- & -- & -- & -- & -- \\
\ion{YbII}{} & 1 & $-$0.80 & $-$1.64 & 0.37 & +0.70 & 0.37 & 1 & $-$1.52 & $-$2.37 & 0.27 & +0.20 & 0.24 & 1 & $-$0.87 & $-$1.71 & 0.28 & +1.07 & 0.25 \\
\ion{LuII}{} & 1 & $-$1.04 & $-$1.14 & 0.28 & +1.20 & 0.28 & 1 & $-$1.90 & $-$2.00 & 0.33 & +0.57 & 0.29 & -- & -- & -- & -- & -- & -- \\
\ion{HfII}{} & 2 & $-$0.88 & $-$1.75 & 0.25 & +0.59 & 0.25 & 2 & $-$1.24 & $-$2.11 & 0.30 & +0.46 & 0.28 & -- & -- & -- & -- & -- & -- \\
\ion{OsI}{} & 1 & 0.17 & $-$1.23 & 0.48 & +1.11 & 0.48 & 1 & $-$0.68 & $-$2.08 & 0.43 & +0.62 & 0.43 & 1 & $-$0.18 & $-$1.58 & 0.40 & +1.32 & 0.39 \\
\ion{IrI}{} & 1 & 0.25 & $-$1.13 & 0.41 & +1.21 & 0.41 & 1 & $-$0.61 & $-$2.00 & 0.35 & +0.71 & 0.35 & -- & -- & -- & -- & -- & -- \\
\ion{PbI}{} & 1 & $-$0.03 & $-$1.78 & 0.38 & +0.56 & 0.38 & -- & -- & -- & -- & -- & -- & -- & -- & -- & -- & -- & -- \\
\ion{ThII}{} & 3 & $-$1.52 & $-$1.55 & 0.33 & +0.79 & 0.33 & 2 & $-$1.93 & $-$2.10 & 0.34 & +0.47 & 0.32 & -- & -- & -- & -- & -- & -- \\
    \hline
    \end{tabular}
\end{table}
\end{landscape}

\begin{landscape}
\begin{table}[ht]
\centering
\tiny
    \begin{tabular}{|l|rrrrrr|}
    \multicolumn{2}{c}{\textbf{Table 6} Continue} \\
    \hline
    Element & \multicolumn{6}{c|}{J2109$$-$$1310} \\
 & $N$ & $\log\epsilon(X)$ &[X/H] & $\sigma_{[X/H]}$ &  [X/Fe] & $\sigma_{[X/Fe]}$ \\
\hline
\ion{C-H}{} & -- & 6.28 & $-$2.14 & 0.16 & +0.39 & 0.16 \\
\ion{C-N}{} & -- & 6.01 & $-$1.75 & 0.52 & +0.78 & 0.51 \\
\ion{N-H}{} & -- & 5.86 & $-$1.97 & 0.31 & +0.56 & 0.30 \\
\ion{$C_{corr}$}{} & -- & -- & -- & -- & +0.74 & --  \\
\ion{OI}{} & 4 & 7.08 & $-$1.61 & 0.12 & +0.92 & 0.13 \\
\ion{NaI}{} & 4 & 4.21 & $-$2.13 & 0.18 & +0.41 & 0.18 \\
\ion{MgI}{} & 7 & 5.63 & $-$1.98 & 0.14 & +0.56 & 0.13 \\
\ion{AlI}{} & 2 & 3.29 & $-$3.01 & 0.34 & $-$0.47 & 0.34 \\
\ion{SiI}{} & 3 & 5.36 & $-$2.06 & 0.20 & +0.48 & 0.20 \\
\ion{KI}{} & 2 & 3.17 & $-$1.80 & 0.25 & +0.73 & 0.25 \\
\ion{CaI}{} & 29 & 4.25 & $-$2.14 & 0.12 & +0.39 & 0.12 \\
\ion{ScI}{} & 2 & 0.55 & $-$2.56 & 0.26 & $-$0.03 & 0.26 \\
\ion{ScII}{} & 14 & 0.75 & $-$2.48 & 0.14 & 0.00 & 0.15 \\
\ion{TiI}{} & 17 & 2.72 & $-$2.24 & 0.17 & +0.29 & 0.16 \\
\ion{TiII}{} & 28 & 2.74 & $-$2.11 & 0.16 & +0.37 & 0.16 \\
\ion{VI}{} & 2 & 1.48 & $-$2.49 & 0.25 & +0.04 & 0.25 \\
\ion{VII}{} & 8 & 1.60 & $-$2.32 & 0.13 & +0.15 & 0.14 \\
\ion{CrI}{} & 11 & 3.19 & $-$2.37 & 0.15 & +0.16 & 0.14 \\
\ion{CrII}{} & 5 & 3.23 & $-$2.38 & 0.14 & +0.09 & 0.13 \\
\ion{MnI}{} & 9 & 2.95 & $-$2.42 & 0.12 & +0.11 & 0.12 \\
\ion{MnII}{} & 3 & 3.04 & $-$2.43 & 0.12 & +0.05 & 0.16 \\
\ion{FeI}{} & 157 & 5.10 & $-$2.53 & 0.11 & 0.00 & 0.10 \\
\ion{FeII}{} & 10 & 5.07 & $-$2.48 & 0.16 & 0.00 & 0.10 \\
\ion{CoI}{} & 12 & 2.68 & $-$2.40 & 0.14 & +0.14 & 0.15 \\
\ion{NiI}{} & 20 & 3.92 & $-$2.33 & 0.11 & +0.20 & 0.12 \\
\ion{ZnI}{} & 3 & 2.31 & $-$2.27 & 0.15 & +0.26 & 0.15 \\
\ion{SrI}{} & 1 & 0.14 & $-$2.73 & 0.36 & $-$0.19 & 0.35 \\
\ion{SrII}{} & 3 & 0.52 & $-$2.25 & 0.22 & +0.22 & 0.21 \\
\ion{YII}{} & 29 & $-$0.22 & $-$2.39 & 0.11 & +0.08 & 0.16 \\
\ion{ZrII}{} & 26 & 0.52 & $-$2.07 & 0.10 & +0.41 & 0.16 \\
\ion{MoI}{} & 1 & 0.07 & $-$1.81 & 0.31 & +0.72 & 0.30 \\
\ion{RuI}{} & 2 & 0.38 & $-$1.38 & 0.24 & +1.16 & 0.24 \\
\ion{RhI}{} & 1 & $-$0.14 & $-$1.05 & 0.31 & +1.48 & 0.31 \\
\ion{PdI}{} & 2 & 0.08 & $-$1.43 & 0.34 & +1.11 & 0.34 \\
\ion{BaII}{} & 5 & 0.16 & $-$1.96 & 0.22 & +0.51 & 0.19 \\
\ion{LaII}{} & 36 & $-$0.61 & $-$1.73 & 0.11 & +0.75 & 0.16 \\
\ion{CeII}{} & 50 & $-$0.32 & $-$1.90 & 0.10 & +0.57 & 0.16 \\
\ion{PrII}{} & 14 & $-$0.79 & $-$1.44 & 0.12 & +1.03 & 0.15 \\
\ion{NdII}{} & 77 & $-$0.18 & $-$1.58 & 0.10 & +0.90 & 0.16 \\
\ion{SmII}{} & 66 & $-$0.48 & $-$1.47 & 0.10 & +1.00 & 0.16 \\
\ion{EuII}{} & 8 & $-$0.83 & $-$1.42 & 0.16 & +1.06 & 0.15 \\
\ion{GdII}{} & 26 & $-$0.31 & $-$1.38 & 0.10 & +1.09 & 0.16 \\
\ion{TbII}{} & 4 & $-$1.11 & $-$1.39 & 0.11 & +1.08 & 0.16 \\
\ion{DyII}{} & 26 & $-$0.18 & $-$1.29 & 0.10 & +1.19 & 0.16 \\
\ion{HoII}{} & 6 & $-$0.98 & $-$1.39 & 0.20 & +1.09 & 0.16 \\
\ion{ErII}{} & 12 & $-$0.46 & $-$1.40 & 0.12 & +1.08 & 0.16 \\
\ion{TmII}{} & 6 & $-$1.18 & $-$1.34 & 0.11 & +1.14 & 0.16 \\
\ion{YbII}{} & 1 & $-$0.70 & $-$1.54 & 0.32 & +0.93 & 0.26 \\
\ion{LuII}{} & 1 & $-$1.10 & $-$1.20 & 0.29 & +1.27 & 0.26 \\
\ion{HfII}{} & 4 & $-$0.63 & $-$1.50 & 0.12 & +0.97 & 0.16 \\
\ion{OsI}{} & 1 & 0.22 & $-$1.18 & 0.30 & +1.36 & 0.31 \\
\ion{IrI}{} & 1 & 0.11 & $-$1.27 & 0.25 & +1.26 & 0.25 \\
\ion{ThII}{} & 4 & $-$1.02 & $-$1.09 & 0.20 & +1.39 & 0.16\\
    \hline
    \end{tabular}
\end{table}
\end{landscape}

\begin{table*}[h!]
\centering
\caption{NLTE abundance corrections.}
\begin{tabular}{cccccccccc}
\hline\hline
Star & $\mathrm{[\ion{Sr}{ii}/Fe]}$ & $\mathrm{[\ion{Ba}{ii}/Fe]}$& $\mathrm{[\ion{Eu}{ii}/Fe]}$ & $\mathrm{[\ion{Sr}{ii}/Fe]}$ & $\mathrm{[\ion{Ba}{ii}/Fe]}$ & $\mathrm{[\ion{Eu}{ii}/Fe]}$ & $\mathrm{[\ion{Sr}{ii}/Fe]}$ & $\mathrm{[\ion{Ba}{ii}/Fe]}$ & $\mathrm{[\ion{Eu}{ii}/Fe]}$ \\
 & LTE & LTE & LTE & $_\mathrm{NLTE,corr}$ & $_\mathrm{NLTE,corr}$ & $_\mathrm{NLTE,corr}$ & $_\mathrm{corrected}$ & $_\mathrm{corrected}$  & $_\mathrm{corrected}$ \\
\hline
J0040+2729 & 0.15 & 0.38 & 1.00 & -0.06 & -0.28 & 0.01 & 0.09 & 0.10 & 1.01 \\
J0217$-$1903 & 0.68 & 0.65 & 1.57 & -- & -0.19 & -- & -- & 0.46 & -- \\
J0246$-$1518 & 0.48 & 0.78 & 1.43 & -0.07 & -0.14 & -0.01 & 0.41 & 0.64 & 1.42 \\
J1430$-$2317 & 0.15 & 0.05 & 0.62 & -- & -0.14 & -- & -- & -0.09 & -- \\
J1432$-$4125 & 0.49 & 0.86 & 1.49 & -- & -0.19 & -0.02 & -- & 0.67 & 1.47 \\
J1916$-$5544 & 0.19 & 0.11 & 0.81 & 0.08 & -0.20 & 0.02 & 0.27 & -0.09 & 0.83 \\
J2009$-$3410 & 0.19 & 0.18 & 1.20 & -0.07 & -0.15 & -- & 0.12 & 0.03 & -- \\
J2049$-$5124 & -0.47 & -0.14 & 0.39 & -- & -0.25 & 0.09 & -- & -0.39 & 0.48 \\
J2106$-$6828 & 0.51 & 0.52 & 1.10 & 0.03 & -0.24 & -- & 0.54 & 0.28 & -- \\
J2109$-$1310 & 0.22 & 0.51 & 1.06 & 0.03 & -0.34 & -0.01 & 0.25 & 0.17 & 1.05 \\

\hline
\end{tabular}
\tablefoot{Corrections were calculated for \ion{Sr}{ii} \citep{Mashonkina2022}, \ion{Ba}{ii} \citep{Mashonkina2019} and \ion{Eu}{ii} \citep{Mashonkina2000}. Column 1 lists the star identifier. Columns 2–4 give the LTE abundances $\mathrm{[\ion{Sr}{ii}/Fe]}$, $\mathrm{[\ion{Ba}{ii}/Fe]}$, and $\mathrm{[\ion{Eu}{ii}/Fe]}$. Columns 5–7 list the NLTE corrections applied to Sr, Ba, and Eu. Columns 8–10 show the resulting NLTE-corrected abundances. These corrections 
are based on literature values for stars where NLTE effects have been quantified.}
\label{tab:abundances_corrected}
\end{table*}

\clearpage
\section{The ten \emph{r}-process patterns}

\begin{center}
\begin{mdframed}[linewidth=0.pt,roundcorner=5pt,innermargin=0pt,outermargin=0pt]
    \centering
    \includegraphics[width=1.9\linewidth]{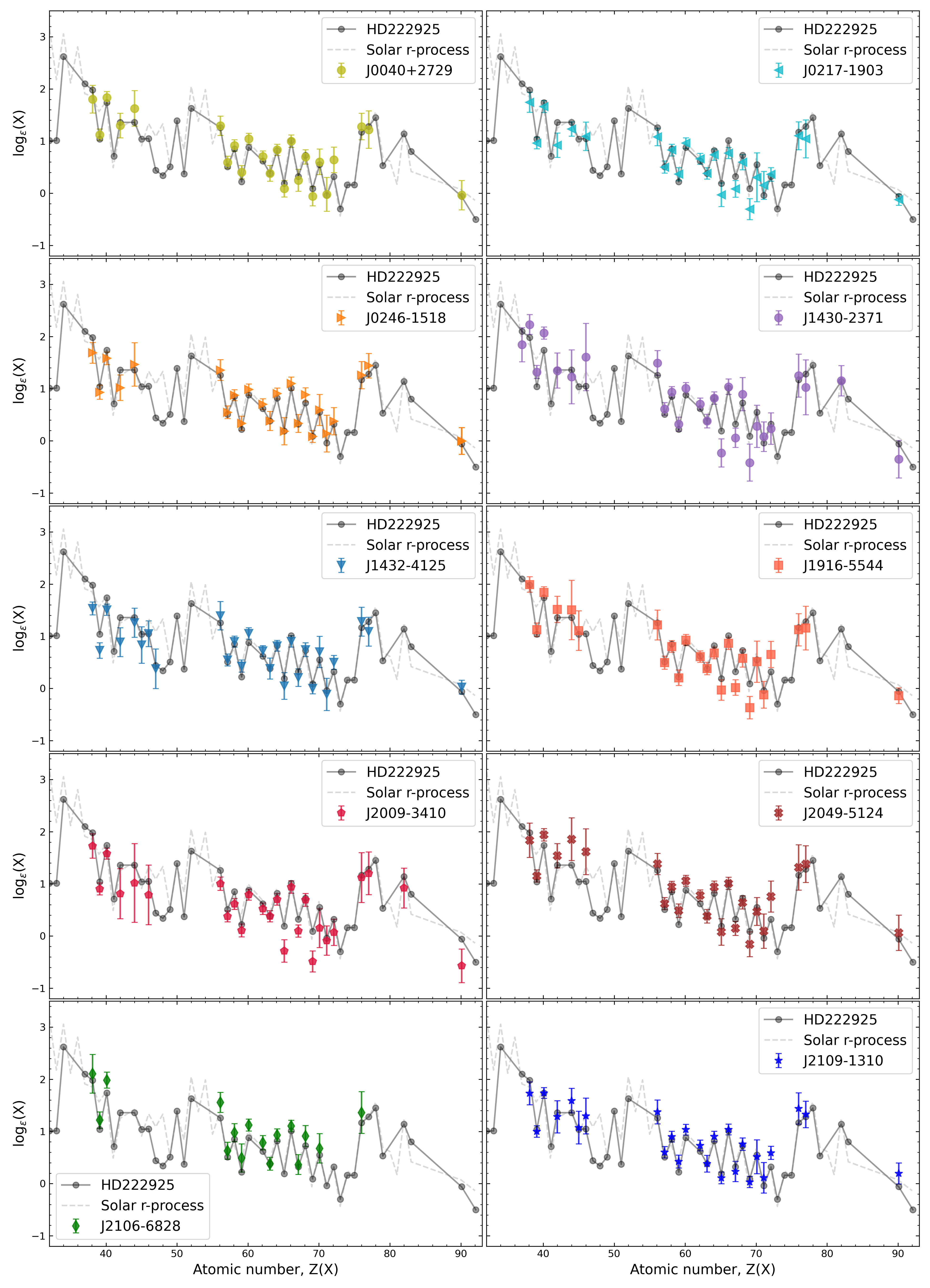}
    \captionof{figure}{\emph{r}-process patterns of the ten stars in the sample (colored markers), each rescaled to the $\log\epsilon$(Eu) value of HD~222925 (solid grey line). The dashed grey line shows the solar \emph{r}-process pattern from \cite{sneden2008}.}
    \label{fig:rescaled_Eu_10stars}
\end{mdframed}
\end{center}

\clearpage
\section{Atomic data}
\begin{table}[ht!]
\centering
\caption{Equivalent width, atomic species, excitation potential (EP), oscillator strength (loggf), equivalent widths (EQW), abundances (log($\epsilon$)), and references for every line analyzed in J0040+2729.}
\label{tab:lines}
\begin{tabular}{ccrrrrc}
\hline
$\lambda$& Species & $\chi$ & loggf & EQW    & log($\epsilon$) & Ref \\
 (\Angstrom) & & (eV) &  & (m\Angstrom) & (dex) & \\  
\hline
6300.30  &    \ion{O}{i}   &    0.00  &  -9.69   &     3.70    & 6.46   &    1 \\
5889.95  &    \ion{Na}{i}  &    0.00  &  +0.11   &     188.07  & 3.77   &    1 \\
5895.92  &    \ion{Na}{i}  &    0.00  &  -0.19   &     155.48  & 3.58   &    1 \\
4167.27  &    \ion{Mg}{i}  &    4.35  &  -0.74   &     58.40   & 5.30   &    1 \\
4702.99  &    \ion{Mg}{i}  &    4.33  & -0.44    &     70.15   & 5.10   &    1 \\
5183.60  &    \ion{Mg}{i}  &    2.72  & -0.17    &     254.51  & 5.24   &    2 \\
5528.40  &    \ion{Mg}{i}  &    4.35  & -0.55    &     77.67   & 5.30   &    2 \\
5711.09  &    \ion{Mg}{i}  &    4.35  & -1.84    &     9.74    & 5.32   &    2 \\
3961.52  &    \ion{Al}{i}  &    0.01  & -0.33    &     146.44  & 3.06   &    1 \\
\hline
\end{tabular}
\tablebib{
(1) NIST \cite{kramida2018}; 
(2) \citet{pehlivan2017}; 
(3) \citet{aldenius2009}; 
(4) \citet{lawler1989} using HFS from \citet{kurucz1995}; 
(5) \citet{lawler2013}; 
(6) \citet{wood2013}; 
(7) \citet{lawler2014a} for log($gf$) values and HFS; 
(8) \citet{wood2014b} for log($gf$) values and HFS, when available; 
(9) \citet{sobeck2007}; 
(10) \citet{lawler2017}; 
(11) \citet{denhartog2011} for log($gf$) values and HFS; 
(12) \citet{obrian1991}; 
(13) \citet{denhartog2014}; 
(14) \citet{belmonte2017}; 
(15) \citet{ruffoni2014}; 
(16) \citet{lawler2015} for log($gf$) values and HFS; 
(17) \citet{wood2014a}; 
(18) NIST \cite{kramida2018}, using HFS/IS from \citet{kurucz1995}; 
(19) \citet{roedererlawler2012}; 
(20) \citet{morton2000}; 
(21) \citet{biemont2011}; 
(22) \citet{ljung2006}; 
(23) \citet{nilsson2008}; 
(24) \citet{wickliffe1994}; 
(25) \citet{duquette1985}; 
(26) \citet{hansen2012} for log($gf$) value and HFS/IS; 
(27) NIST \cite{kramida2018}, using HFS/IS from \citet{mcwilliam1998}; 
(28) \citet{lawler2001a} using HFS from \citet{ivans2006}; 
(29) \citet{lawler2009}; 
(30) \citet{li2007} using HFS from \citet{Sneden_2009}; 
(31) \citet{ivarsson2001} using HFS from \citet{Sneden_2009}; 
(32) \citet{denhartog2003} using HFS/IS from \citet{roederer2008} when available; 
(33) \citet{lawler2006} using HFS/IS from \citet{roederer2008}; 
(34) \citet{lawler2001c} using HFS/IS from \citet{ivans2006}; 
(35) \citet{denhartog2006}; 
(36) \citet{lawler2001b} using HFS from \citet{lawler2001d}; 
(37) \citet{wickliffe2000}; 
(38) \citet{lawler2004} using HFS from \citet{Sneden_2009}; 
(39) \citet{lawler2008}; 
(40) \citet{wickliffe1997} using HFS from \citet{Sneden_2009}; 
(41) \citet{Sneden_2009} for log($gf$) value and HFS/IS; 
(42) \citet{lawler2009} for log($gf$) values and HFS; 
(43) \citet{lawler2007}; 
(44) \citet{quinet2006}; 
(45) \citet{xu2007} using HFS/IS from \citet{cowan2005}; 
(46) \citet{biemont2000} using HFS/IS from \citet{roederer2012b}; 
(47) \citet{nilsson2002a}; 
(48) \citet{nilsson2002b}. }

\tablefoot{The complete version of this table and the full table for the other nine stars are available at the CDS.}    

\label{tab:linelist}
\end{table}

\end{appendix}

\listofobjects
\end{document}